\documentclass[manuscript]{aastex}

\usepackage{epsfig}
\usepackage{epstopdf}
\usepackage{float}
\usepackage{amssymb,amsmath}
\usepackage{graphicx}
\usepackage{tabularx}
\usepackage{mathtools}
\begin{document}

\title{On ultra high energy cosmic rays and their resultant gamma rays}

\author{Eyal Gavish and David Eichler}
\affil{Physics Department, Ben-Gurion University, Be'er-Sheva 84105, Israel}
%\shorttitle{Ultra High Energy Cosmic Rays and Resultant Gamma Rays}
\shortauthors{Gavish & Eichler}

\begin{abstract}
The Fermi Large Area Telescope (Fermi LAT) collaboration has recently reported on 50 months  of measurements of the isotropic Extragalactic Gamma Ray Background (EGRB) spectrum between  $100\mbox{MeV}$ and $820\mbox{GeV}$. Ultrahigh Energy Cosmic Ray (UHECR) protons interact with the Cosmic Microwave Background (CMB) photons and produce cascade photons of energies $10\mbox{MeV}\mbox{-}1\mbox{TeV}$ that contribute to the EGRB flux.
We examine seven possible evolution models for UHECRs and find that UHECR sources that evolve as the Star Formation Rate (SFR), medium low luminosity active galactic nuclei type-1 ($L = 10^{43.5}\mbox{erg sec}^{-1}$ in the $[0.5\mbox{-}2]\mbox{KeV}$ band), and BL Lacertae objects (BL Lacs) are the most acceptable given the constraints imposed by the observed EGRB. Other possibilities produce too much secondary $\gamma$-radiation. In all cases, the decaying dark matter contribution improves the fit at high energy, but the contribution of still unresolved blazars, which would leave the smallest role for decaying dark matter, may yet provide an alternative improvement.

The possibility that the entire EGRB can be fitted with resolvable but not-yet-resolved blazars, as recently claimed by  \citet{2015ApJ...800L..27A}, would  leave little room in the EGRB to accommodate $\gamma$-rays from extragalactic UHECR production, even for many source evolution rates that would otherwise be acceptable.  We find that, under the assumption of UHECRs being mostly protons, there is not enough room for producing extragalactic UHECRs with AGN, GRB, or even SFR  source evolution. Sources that evolve as BL Lacs on the other hand, would produce much less secondary $\gamma$-radiation and would remain a viable source of UHECRs, provided that they dominate.

\end{abstract}


\section{Introduction}
The Cosmic Ray (CR) spectrum has been measured to unprecedented accuracy between the eneries of $\sim10^9\mbox{eV}$ and $\sim10^{20}\mbox{eV}$.  Two features in the spectrum where there is a change in the CR spectral shape, the second Knee at the energy $\sim0.6\mbox{EeV}$ \citep[e.g.][]{2007APh....27...76A, 2012APh....36...31B, 2014APh....53..120B} and the Ankle at $\sim 5\mbox{EeV}$ \citep[e.g.][]{2000NuPhS..87..345W, 2005PhRvD..72h1301D}, have been considered as the transition from a Galactic dominated spectrum to an extragalactic dominated spectrum \citep[see e.g.][]{2007APh....27...61A, 2012APh....39..129A}. 

At the transition, a change in the composition is not unexpected \citep[e.g.][]{1993A&A...274..902S}. The High Resolution Fly’s Eye (HiRes) collaboration reported \citep{2010PhRvL.104p1101A} that the CR composition is dominated by protons above $1.6\mbox{EeV}$. This result can indicate on a transition below the Ankle. The Telescope Array collaboration (TA) measurements \citep{2012JPhCS.404a2037J} are consistent with those of HiRes for a predominately protonic composition. These results are consistent with an earlier report by HiRes \citep{2008PhRvL.100j1101A} that claims to observe the Greisen-Zatsepin-Kuzmin (GZK) cutoff. The GZK cutoff predicted by \citet{1966PhRvL..16..748G} and \citet{1966JETPL...4...78Z} independently, is an upper limit of $E\sim 50\mbox{EeV}$ to the CR spectrum due to interactions of Ultra High Energy Cosmic Ray (UHECR) protons ($E\gtrsim 1\mbox{EeV}$) on the Cosmic Microwave Background (CMB) photons via pion photoproduction process. If UHECRs are dominated by heavier nuclei, the steepening of the spectrum is not as sharp as the GZK and it occurs at lower energies \citep{2013APh....41...73A, 2013APh....41...94A}. For this reason, the GZK cutoff, assuming that it is pinpointed with adequate energy resolution, is considered to be a signature of a proton dominated composition of UHECRs.

The Pierre Auger collaborations results \citep{2010PhRvL.104i1101A, 2014NIMPA.742...22B} are inconsistent with those of HiRes and TA, and show a gradual increase in the average mass of UHECRs with energy. This raises the possibility that the highest energy UHECRs are not protons, but it  is consistent with their being partly ($\sim 1/2$) protons, whereas mixed  composition models predict that at the highest energies there are no significant number of light nuclei.  Moreover,  there are difficulties with fitting the UHECR spectrum with an admixture of heavy elements \citep{2014JCAP...10..020A}. In this paper we therefore consider the UHECR to be protons, or to be half protons by composition, bearing in mind that it is only a hypothesis. The issue we consider is whether the secondary $\gamma$-rays that they produce are consistent with the extragalactic diffuse gamma ray background. If not, this can be taken as further evidence a) that they are mostly heavy nuclei  (which typically requires them to have a harder spectrum than $E^{-2.0}$) \citep{2011A&A...535A..66D,2014JCAP...10..020A}, or b) that they are not extragalactic.

UHECRs are widely thought to be accelerated at astrophysical shocks \citep[for a review see e.g.][]{1987PhR...154....1B}. The energy spectrum of the accelerated particles is assumed to be a power law $N(E) \propto E^{-\alpha}$ with a spectral index of $\alpha \gtrsim 2$. Observations confirm particles of energies $E>10^{20}\mbox{eV}$ and even an $E=3\times 10^{20}\mbox{eV}$ event \citep{1993PhRvL..71.3401B} has been detected. Possible sources that might be able to accelerate CRs up to energies $E\gtrsim 10^{20}\mbox{eV}$, among others are Active Galactic Nuclei (AGNs) and radio galaxies \citep[see e.g.][]{1995ApJ...454...60N, 1997JPhG...23....1B, 2000PhR...327..109B, 2004RPPh...67.1663T}. 

UHECR protons propagating in space are interacting on the CMB photons and initiating an electromagnetic cascade. The result is observable diffuse $\gamma$-rays \citep{1972JPhA....5.1419W}. The two main interactions of protons on the CMB are pair production $p + \gamma_{_{CMB}} \rightarrow p + e^+ + e^-$, at energies $2.4 \mbox{EeV} \lesssim E_p \lesssim 50\mbox{EeV}$,  and pion photoproduction $p + \gamma_{_{CMB}} \rightarrow n + \pi^+$, $p + \gamma_{_{CMB}} \rightarrow p + \pi^0$ at higher energies. The neutral pions ultimately decay into high energy photons while the positive pions decay into high energy photons, positrons, and neutrinos. The electrons and positrons, that emerge from the decays, interact with the background photons via inverse Compton process $e + \gamma_b \rightarrow e' + \gamma$. High energy photons are produced with a mean energy  of \citep{1970RvMP...42..237B}  $\varepsilon_{\gamma} = 4/3 (E_e/m_ec^2)^2 \varepsilon_b $, where $E_e$ is the energy of the incoming electron, $m_e$ is the rest mass of the electron, $c$ is the speed of light, and $\varepsilon_b$ is the energy of the background photon. The high energy photons interact with the Extragalactic Background Light (EBL) via pair production process $\gamma + \gamma_{_{EBL}} \rightarrow e^+ + e^-$, producing a pair of electron and positron with energy of $E_e=\varepsilon_{\gamma}/2$ each. These two processes, inverse Compton and pair production, drive the development of an electromagnetic cascade. The cascade develops until the energy of the photons drops bellow the pair creation threshold $\varepsilon_{th} = (m_ec^2)^2/\varepsilon_b$ . At this stage the photons stop interacting, while electrons continue losing energy and producing photons via inverse Compton. The cascade results in photons of energy below $\thicksim$ 1TeV that contribute to the isotropic diffuse $\gamma$-ray emission.

The isotropic diffuse $\gamma$-ray emission, also known as Extragalactic Gamma Ray Background (EGRB), was first detected by the SAS 2 satellite \citep{1977ApJ...217L...9F, 1978ApJ...222..833F} and  interpreted as of extragalactic origin. Later on, \citet{1998ApJ...494..523S} confirmed the existence of the EGRB by analyzing the EGRET data. In this work, we fit the recently reported  \citep{2015ApJ...799...86A} EGRB spectrum by Fermi Large Area Telescope (Fermi LAT) collaboration between $100\mbox{MeV}$ and $820\mbox{GeV}$.

\section{UHECR spectrum calculations}
In this section we follow \citet{2006PhRvD..74d3005B} and calculate UHECR spectrum under the assumptions of a pure proton composition, a homogeneous distribution of sources between redshift $z=0$ to the maximal redshift $z_{max}$ and continuous energy losses. For the energy loss rates of UHECR protons in interactions on the CMB we use the calculations made by \citet{2006PhRvD..74d3005B} as well.

The differential equation describing the energy loss rate of a  UHECR proton at redshift $z$ is
\begin{equation}
-\frac{dE}{dt} = E H(z) + b(E,t)
\end{equation}
where $E$ is the proton energy at epoch t, $b(E,t)$ is energy losses of UHECR protons of energy $E$ at epoch $t$ to pair production and to pion photoproduction, and $H(z) = H_0(\Omega_m(1+z)^3 + \Omega_{\Lambda})^{1/2}$ is the Hubble constant at redshift $z$ with the parameters $\Omega_m = 0.27$, $\Omega_{\Lambda} = 0.73$ and $H_0 = 70 \mbox{km} \  \mbox{sec}^{-1}\mbox{Mpc}^{-1}$. Changing variable $dt/dz =-1/(H(z)(1+z))$ and integrating we obtain
\begin{equation}
\label{eq:gen_energy}
E(E_{z_0},z) = E_{z_0} + \int_{z_0}^z dz' \frac{E}{1+z'} + \int_{z_0}^z dz' \frac{b(E,z')}{H(z')(1+z')}
\end{equation}
Equation (\ref{eq:gen_energy}) describes the energy of a  CR proton at redshift $z$ where its energy at redshift $z=z_0$ is $E_{z_0}$. The first integral from the left describes the energy that a CR proton is losing due to the expansion of the universe. The second integral describes the energy that a CR proton is losing due to its interactions on the CMB photons.
Assuming a power law distribution for UHECR protons, the production rate of the particles at redshift $z$ per unit energy per unit comoving volume is
\begin{equation}
\label{eq:np_z}
Q_p(E,z) = K_1F(z) E(E_0,z)^{-\alpha}
\end{equation}
where $K_1$ is constant, $\alpha$ is the power law index, $F(z) = \mbox{const}\times(1+z)^m$ is the density evolution of the UHECR sources as a function of the redshift, and $m$ is called the evolution index.

We assume that the number of protons is conserved. The interactions of protons on the CMB always involve a proton, except for the positive pion production. In this case, the outgoing neutron beta decays very fast to proton, electron, and anti neutrino electron. So eventually the total number of protons is conserved. Then, the number of particles per comoving volume at redshift $z$ is calculated as
 \begin{equation}
\label{eq:num_particles}
n_p(E_{z_0},z_0)dE_{z_0} = K_1\int_{z_0}^{z_{max}}\frac{dt}{dz} dzF(z) E^{-\alpha}dE
\end{equation}
The diffuse flux of UHECRs at the present time would be
\begin{equation}
\label{eq:cr_flux}
J_p(E_0) = K_1\frac{c}{4\pi}\int_0^{z_{max}}\frac{dz}{H(z)(1+z)}F(z) \ E^{-\alpha} \  \frac{dE}{dE_0}
\end{equation}
The UHECR spectrum is normalized to the experimental data through $K_1$. The spectrum in equation (\ref{eq:cr_flux}) is determined by the four parameters: $\alpha$, $m$, $z_{max}$, and $ E_{max}$. $E_{max}$ is the maximal energy that a CR proton can be accelerated to. So in equation (\ref{eq:cr_flux}) the energy of a CR particle is limited by $E(E_0,z)\leq E_{max}$. For each set of these four parameters, a different UHECR spectrum can be calculated and normalized to the experimental data.

\subsection{Energy density of photons resulted from UHECR interactions}
The energy density of photons originating from UHECR interactions, can be calculated by integrating over all energy losses of UHECR protons to the electromagnetic cascade. In the pion photoproduction process, UHECR protons are losing energy to the electromagnetic cascade and to the production of neutrinos. The fraction of energy that goes into the electromagnetic cascade in this process is $\sim 0.6$ \citep{2001PhRvD..64i3010E}. We denote by $b_{em}$ the relative energy losses of UHECR protons to the electromagnetic cascade.

 The amount of energy lost to the electromagnetic cascade, by a proton of energy $E$, propagating from redshift $z+dz$ to redshift $z$, is $b_{em}(E,z)dz/(H(z)(1+z))$.  Multiplying this amount of energy by the number of protons of energy $E$ per unit volume at redshift $z$, given in equation (\ref{eq:num_particles}), and integrating over all proton energies, we get the total energy density of photons produced by UHECR protons at redshift $z$
\begin{equation}
\label{eq:omega(z)}
\omega_c(z)dz = dz\int_0^{E_{max}}dE\frac{ b_{em}(E,z)}{H(z)(1+z)}n_p(E,z)(1+z)^3
\end{equation} 
where $n_p(E,z)(1+z)^3$ is the proper density of the protons.
Integrating over all redshifts and deviding by $(1+z)^4$ (as photons lose energy as $1/(1+z)$ and a unit volume expands by a factor of $(1+z)^3$) we get
\begin{equation}
\omega_c = \int_0^{E_{max}}dE\int_0^{z_{max}}dz\frac{ b_{em}(E,z)}{H(z)(1+z)^2}n_p(E,z)
\label{eq:energy_density_0}
\end{equation}
This energy density depends on the UHECR parameters $\alpha$, $m$, $z_{max}$, and $E_{max}$.

\section{The spectrum of $\gamma$-rays originating from UHECR interactions}
\label{sec:EGRB}
The generation rate of $\gamma$-ray photons originating from UHECR interactions at redshift $z$ per unit energy per unit volume is calculated as
\begin{equation}
Q_{\gamma}(\varepsilon,z) = K_2(z) \left \{ \begin{array}{lcl}
\left(\frac{\varepsilon}{\varepsilon_{\chi}}\right)^{-3/2} & \mbox{for} & \varepsilon < \varepsilon_{\chi} \\
\left(\frac{\varepsilon}{\varepsilon_{\chi}}\right)^{-2} & \mbox{for} & \varepsilon_{\chi} \leq \varepsilon \leq \varepsilon_a 
\end{array}\right.
\label{eq:gen_rate_photons}
\end{equation}
The spectral indexes were found by \citet{1975Ap&SS..32..461B}. The normalization factor $K_2(z)$ is constant in energy but depends on the redshift. For convenience we write equation (\ref{eq:gen_rate_photons}) in the following way
\begin{equation}
Q_{\gamma}(\varepsilon,z) = K_2(z)\mathcal{Q}_{\gamma}(\varepsilon,z)
\label{eq:Q_def}
\end{equation}
 $\varepsilon_a = (m_ec^2)^2/\varepsilon_{_{EBL}}$ is the threshold energy for pair production by a photon scattering on the EBL. Suppose a photon of energy $\varepsilon_a$ is interacting on the EBL. This photon will produce an electron and a positron of energy $\varepsilon_a/2$ each. This electron (or positron) will interact via inverse Compton on background photons, producing a photon of energy $\varepsilon_{\chi}= 1/3(\varepsilon_a/m_ec^2)^2\varepsilon_{b}$. So, photons of energies $\varepsilon_{\chi}\leq\varepsilon\leq \varepsilon_a$ do not interact with the background photons, but electrons continue to produce high energy photons of energies in this range. At energies below $\varepsilon_{\chi}$, photons are created by electrons of energies below $\varepsilon_a/2$.

The $\gamma$-ray spectrum at the present time is calculated as
\begin{equation}
J_{\gamma}(\varepsilon) = \frac{c}{4\pi} \int_0^{z_{max}} \left(\frac{dt}{dz}\right)dz \frac{K_2(z) \mathcal{Q}_{tot}(\varepsilon (1+z),z)}{(1+z)^2}\exp\left(-\tau(\varepsilon,z)\right)
\label{equ:gamma_flux}
\end{equation}
with
\begin{equation}
K_2(z) = \frac{\omega_c(z)H(z)(1+z)}{\int_0^{\infty}\mathcal{Q}_{tot} \varepsilon d\varepsilon \exp\left(-\tau\left(\varepsilon/(1+z),z\right)\right)}
\label{eq:norm_photons}
\end{equation}
where $\mathcal{Q}_{tot}$ is the total contribution at redshift $z$ from all photons in the EBL spectrum. $\tau(\varepsilon,z)$ is the optical depth for pair production for a cascade photon propagating through the EBL from redshift $z$ to redshift $z=0$, observed at the present time with energy $\varepsilon$, given by
\begin{equation}
\tau_{\gamma \gamma}(\varepsilon,z) = \int_0^z dz' \frac{dl}{dz'} \int_{-1}^1 d\mu \frac{1-\mu}{2}  \int_{E_{th}}^\infty d\varepsilon_b \  n \left( \varepsilon_b,z' \right) \  \sigma_{\gamma \gamma} \left(\varepsilon \left(1+z'\right),\varepsilon_b, \theta \right)
\label{eq:tau}
\end{equation}
where $dl/dz = cdt/dz$ is the cosmological line element, $\theta$ is the angle between the interacting photons, $\mu=\cos(\theta)$, $\varepsilon_b$ is the energy of an EBL photon, and $n(\varepsilon_b,z)$ is the number of photons of energy $\varepsilon_b$ at redshift $z$ per unit volume per unit energy. $E_{th}$ is the threshold energy for the pair production  process given by
\begin{equation}
E_{th} = \frac{2(m_ec^2)^2}{\varepsilon (1+z) (1-\cos(\theta))}
\end{equation}
The pair production cross section $\sigma_{\gamma \gamma}$ is given by \citep{1955jauch, 1967PhRv..155.1404G}
\begin{equation}
\sigma_{\gamma \gamma}(\varepsilon_1,\varepsilon_2,\theta) = \frac{3\sigma_T}{16}(1-\beta^2) \left[2\beta(\beta^2-2) + (3-\beta^4) \ln \left(\frac{1+\beta}{1-\beta} \right) \right]
\end{equation}
where $\sigma_T$ is the Thomson cross section and
\begin{equation}
\beta = \sqrt{1 - \frac{2(m_ec^2)^2}{\varepsilon_1 \varepsilon_2 (1-\cos(\theta))}}
\end{equation}
For the calculations of the $\gamma$-ray spectrum at redshift $z$, we use the best fit model in \citet{2004A&A...413..807K} as an EBL model.

\section{Fitting the Fermi LAT data}
\label{sec:fit}
In this section we fit the EGRB measured by Fermi LAT. $\gamma$-rays from UHECRs and from Star Forming Galaxies (SFGs) cannot explain the most energetic data points of Fermi LAT. A high energy $\gamma$-ray flux is required. We consider here a possible $\gamma$-ray flux from Dark Matter (DM) decay as the highest energy contribution to the EGRB.

\subsection{Components}
For calculating the contribution from SFGs, we use the $\gamma$-ray spectrum of our Galaxy from \citet{2015ApJ...799...86A}. We use the Sum of all modeled components in the right panel of Figure 4 (Model A) in \citet{2015ApJ...799...86A} and subtract the total EGB (Model A) in \citet{2015ApJ...799...86A}, Figure 8. We assume that each SFG in the universe produces this $\gamma$-ray spectrum.  We assume that the mass density of SFGs in the universe is half of the mass density of the universe, i.e $\sim 5\times 10 ^{-31} \mbox{gr}/ \mbox{cm}^{3}$. Further, we assume that SFGs evolve in time as the Star Formation Rate (SFR) (equation \ref{equ:SFR}). Lastly, we assume that the $\gamma$-rays from the SFGs are attenuated by the EBL when propagating in space (see Section \ref{sec:EGRB} for details). Under these assumptions we can calculate the $\gamma$-ray spectrum from SFGs in the universe.

As a high energy contribution to the EGRB, we use $\gamma$-rays from DM of mass $mc^2=3\mbox{TeV}$, decaying into bosons ($W^+W^-$). The spectrum we use is taken from Figure 7 in \citet{2012JCAP...10..043M}. The DM lifetime that was used by \citet{2012JCAP...10..043M} in Figure 7 is $\tau=  1.2\times 10^{27}\mbox{sec}$. We adjust this lifetime to optimize the fit. The reason for using the $W^+W^-$ channel is the improvement of the fit at high energies. Other decay channels, such as $\mbox{DM} \rightarrow \mu^+\mu^-$ or $\mbox{DM} \rightarrow b\bar{b}$, do not give such a good fit at high energies. As an example, we also show a fit using the $\mu^+\mu^-$ decay channel.

We examine four possibilities for the evolution of UHECR sources: SFR, Gamma Ray Bursts (GRBs), AGNs type-1, and BL Lacertae objects (BL Lacs).
\newline
The SFR function is taken from \citet{2008ApJ...683L...5Y}

\begin{equation}
F_{_{SFR}}(z) \propto \left\{ \begin{array}{lcl}
(1+z)^{3.4} & \mbox{for} & z \leq 1 \\
(1+z)^{-0.3} & \mbox{for} & 1 < z \leq 4 \\
(1+z)^{-3.5} & \mbox{for} & 4 < z 
\end{array}\right.
\label{equ:SFR}
\end{equation}
\newline
As suggested by \citet{2007PhRvD..75h3004Y}, we take the GRB evolution function to be $F_{_{GRB}}(z) \propto  F_{_{SFR}}(z)^{1.4}$. So we get

\begin{equation}
F_{_{GRB}}(z) \propto \left\{ \begin{array}{lcl}
(1+z)^{4.8} & \mbox{for} & z \leq 1 \\
(1+z)^{1.1} & \mbox{for} & 1 < z \leq 4 \\
(1+z)^{-2.1} & \mbox{for} & 4 < z 
\end{array}\right.
\label{equ:GRB}
\end{equation}
\newline
There is a significant difference in the evolution functions of different luminosity AGNs.  \citet{2005A&A...441..417H} calculated the evolution functions of AGNs of four different luminosities in the soft X-ray band ($0.5\mbox{-}2\mbox{KeV}$): Low Luminosity AGNs (LLAGNs) $L_X=10^{42.5} \mbox{erg sec}^{-1}$, Medium Low Luminosity AGNs (MLLAGNs) $L_X= 10^{43.5} \mbox{erg sec}^{-1}$, Medium High Luminosity AGNs (MHLAGNs)  $L_X= 10^{44.5} \mbox{erg sec}^{-1}$, and High Luminosity AGNs (HLAGNs) $L_X= 10^{45.5} \mbox{erg sec}^{-1}$. LLAGNs are considered as not being able to accelerate CRs to ultra high energies  \citep[see e.g.][]{2004NJPh....6..140W}. We thus discuss in this work the possibility that MLLAGNs, MHLAGNs, or HLAGNs are the sources of UHECRs:
\begin{equation}
F_{_{MLLAGN}}(z) \propto \left\{ \begin{array}{lcl}
(1+z)^{3.4} & \mbox{for} & z \leq 1.2 \\
10^{0.32(1.2-z)} & \mbox{for} & 1.2 < z 
\end{array}\right.
\label{equ:MLLAGN}
\end{equation}

\begin{equation}
F_{_{MHLAGN}}(z) \propto \left\{ \begin{array}{lcl}
(1+z)^{5} & \mbox{for} & z \leq 1.7 \\
2.7^5 & \mbox{for} & 1.7 < z \leq 2.7 \\
10^{0.43(2.7-z)} & \mbox{for} & 2.7 < z 
\end{array}\right.
\label{equ:AGN}
\end{equation}

\begin{equation}
F_{_{HLAGN}}(z) \propto \left\{ \begin{array}{lcl}
(1+z)^{7.1} & \mbox{for} & z \leq 1.7 \\
2.7^{7.1} & \mbox{for} & 1.7 < z \leq 2.7 \\
10^{0.43(2.7-z)} & \mbox{for} & 2.7 < z 
\end{array}\right.
\label{equ:AGN}
\end{equation}
\newline
For BL Lacs we use two different evolution functions. Various studies have found BL Lacs to evolve very slowly, or not evolve at all \citep[see e.g.][]{2002ApJ...566..181C,2007ApJ...662..182P}. Thus, the first evolution function we use is corresponding to a no evolution scenario:
\begin{equation}
F_{_{NoEvBL}} \propto (1+z)^0
\label{equ:HSP}
\end{equation}
\newline
The second function is related to a subclass of BL Lacs - High Synchrotron Peaked (HSP) objects. \citet{2014ApJ...780...73A} used a set of 211 BL Lac objects detected by Fermi LAT during the first year of operation \citep{2010ApJ...715..429A}. \citet{2014ApJ...780...73A} have found that the number density of HSP BL Lacs is strongly increasing with time (i.e. with decreasing $z$)  and that the number density of the 211 BL Lacs sample is almost entirely driven by this population at $z\leq1$. For these objects, the evolution function can be described roughly by:
\begin{equation}
F_{_{HSP}} \propto (1+z)^{-6}
\label{equ:BL_Lac_NoEvo}
\end{equation}
\newline
Most of the UHECR spectra are calculated with two power law indexes, defined as:
\begin{equation}
\alpha = \left\{ \begin{array}{rcl}
\alpha_1 & \mbox{for} & E \leq E_{br} \\
\alpha_2 & \mbox{for} & E_{br} < E 
\end{array}\right.
\end{equation}

\subsubsection{Blazars}
Blazars have been considered \citep[e.g][]{2015ApJ...800L..27A,2015MNRAS.450.2404G} as possible sources of $\gamma$-rays that explain the EGRB measurements, since most of the resolved sources are blazars. We show that the joint $\gamma$-ray flux from blazars and from UHECRs evolving as AGNs, GRBs, or SFR is too high and in most cases violates the limits imposed by the Fermi LAT data even without any additional contribution from SFGs or DM. UHECRs that evolve as BL Lacs, on the other hand, have enough room to fit the Fermi LAT data together with blazars. For a blazars contribution, we use the spectrum reported in \citet{2015ApJ...800L..27A}. This spectrum includes the resolved blazar sources. In order to get the unresolved blazar spectrum, we subtract the Fermi LAT resolved sources \citep{2015ApJ...799...86A} from the blazars spectrum.

\subsection{Results}

\begin{figure}
\centering
\includegraphics[width=0.49\textwidth]{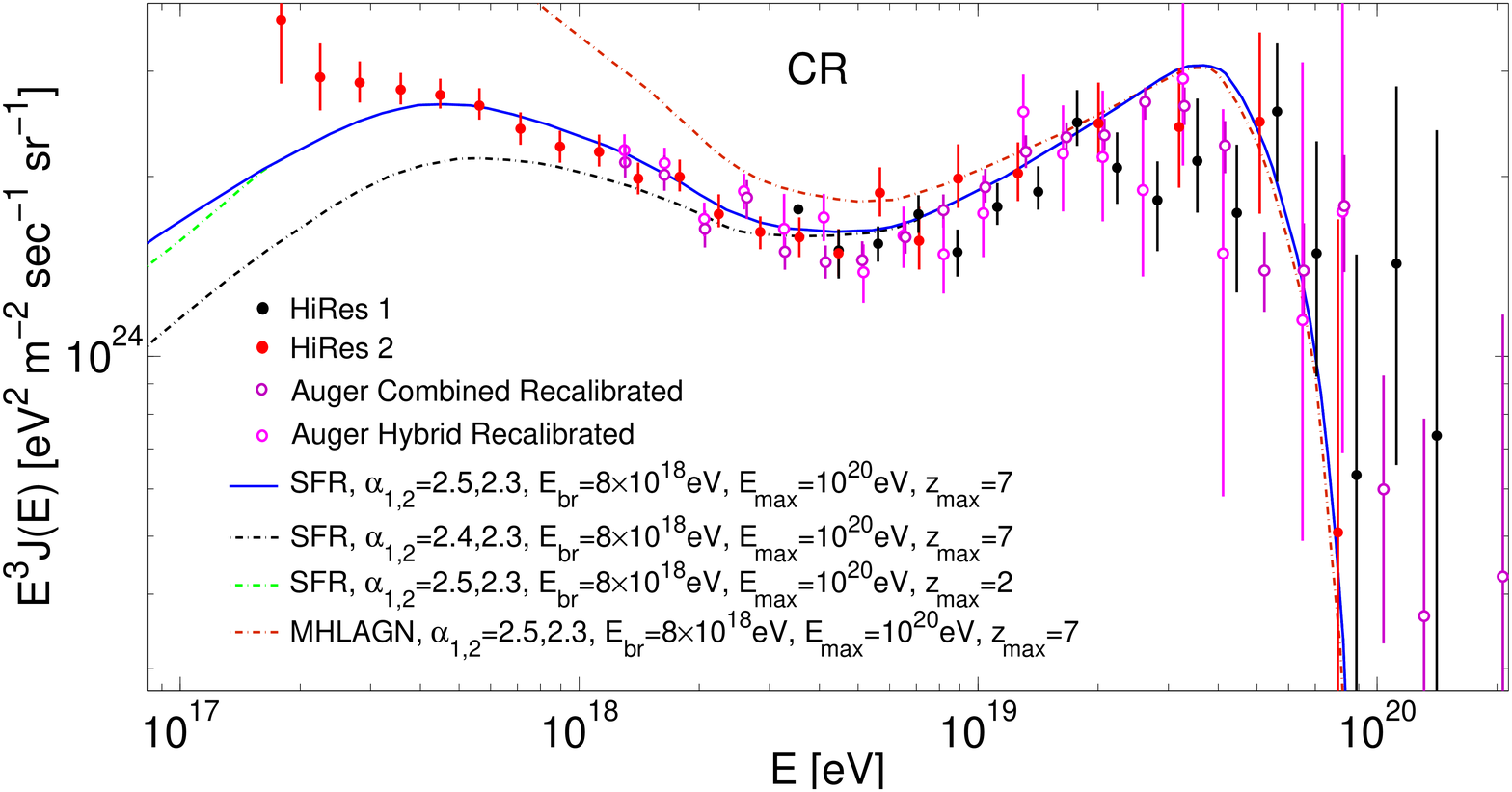}
\includegraphics[width=0.49\textwidth]{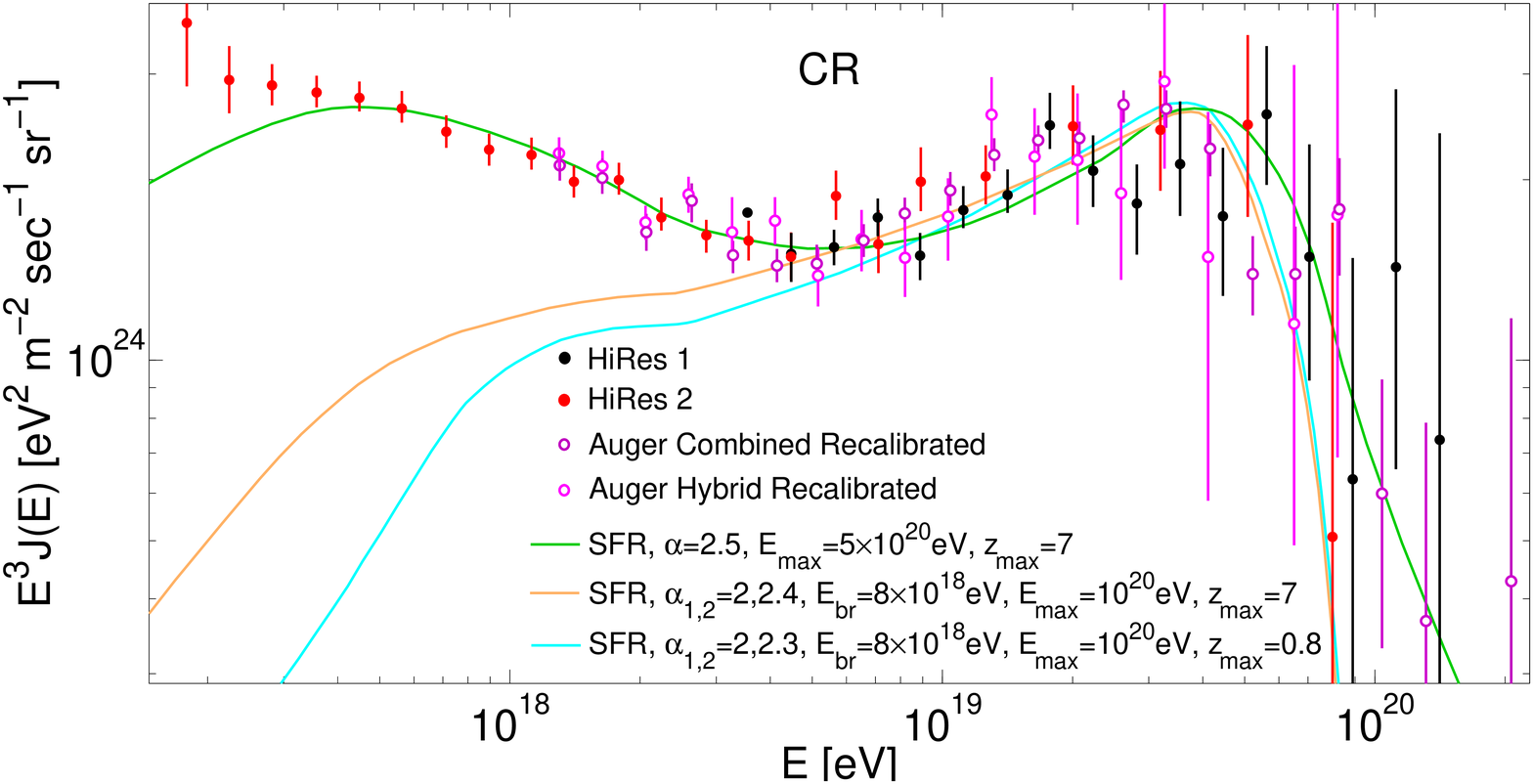}
\includegraphics[width=0.49\textwidth]{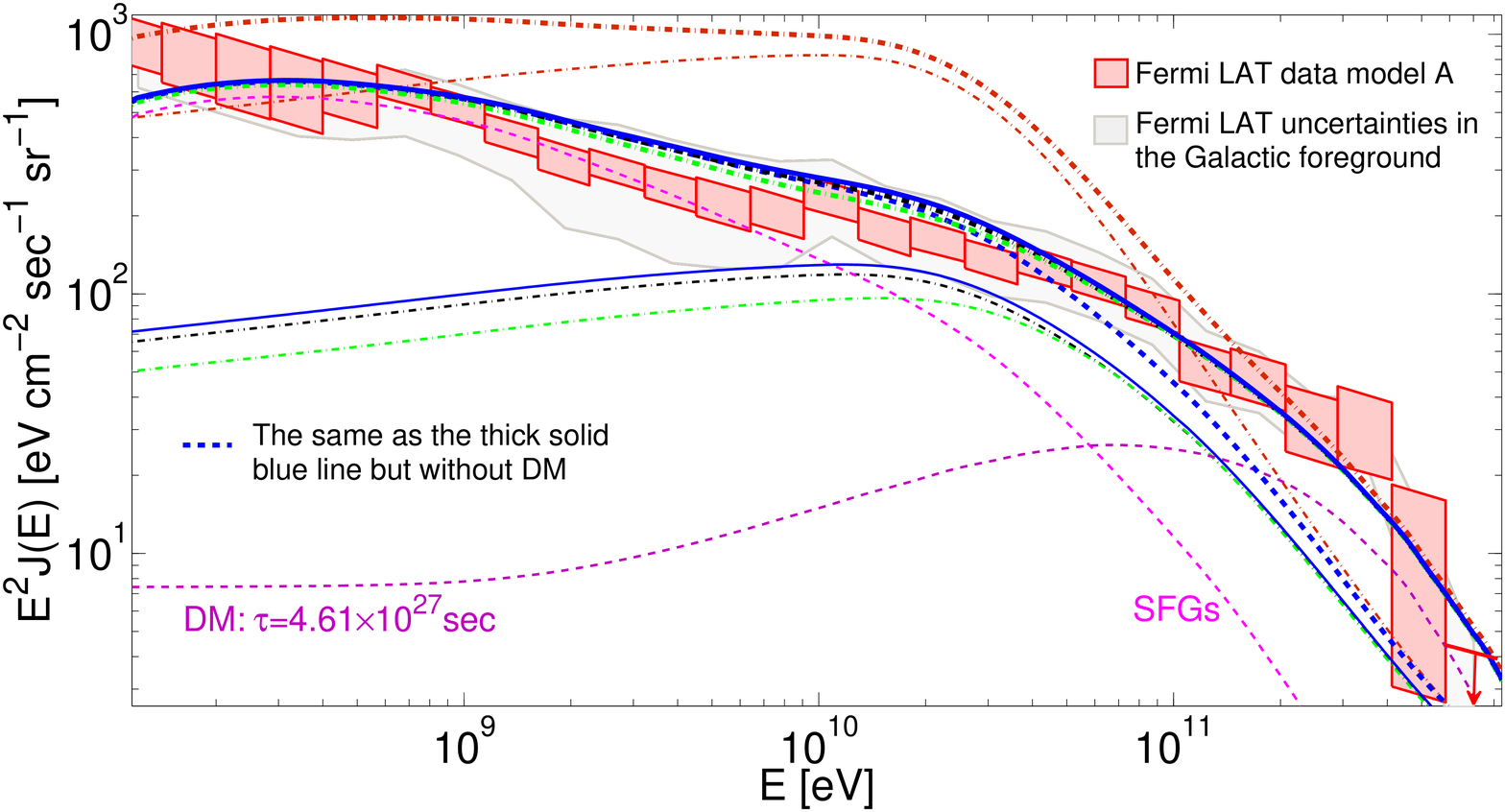}
\includegraphics[width=0.49\textwidth]{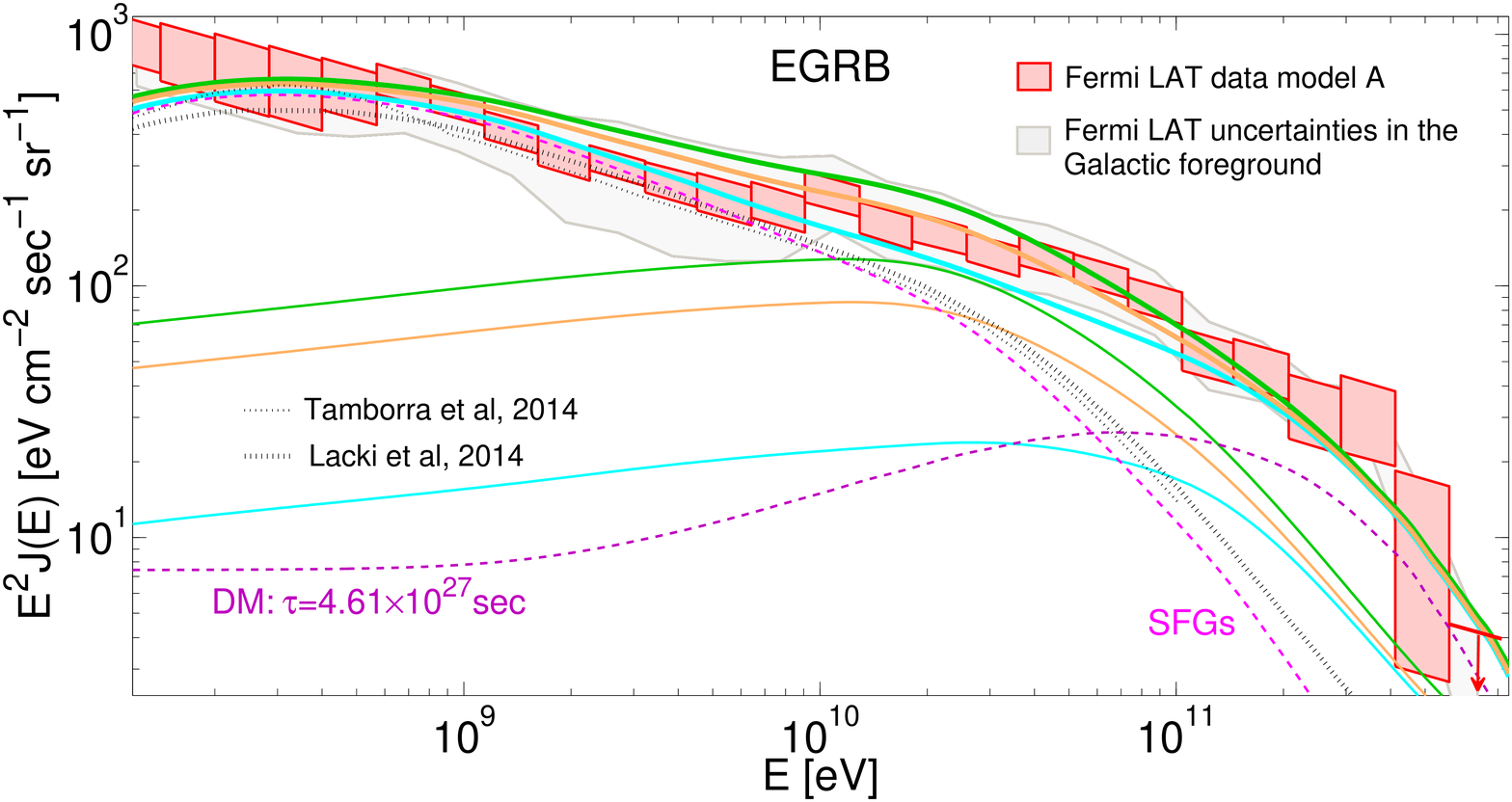}
\caption{\textbf{Upper panels}: UHECR spectra for different sets of parameters, normalized to the HiRes and the recalibrated Auger data (the parameters are written in the plots). All spectra are corresponding to the SFR evolution, except the dashed-dotted dark red curve which is corresponding to the MHLAGNs model. The dashed-dotted lines in the left panels differ from the solid blue line in one parameter.
\textbf{Lower panels}: The $\gamma$-ray fluxes corresponding to the different UHECR spectra in the upper panels are represented in the lower panels by the same lines as in the upper panels. The SFGs contribution and the DM ($W^+W^-$ channel, $mc^2=3\mbox{TeV}$) contribution with a lifetime of $\tau=4.61\times10^{27}\mbox{sec}$ are shown. The totals of the three components (SFGs, UHECRs, and DM) are represented by thick lines (same color and style as the thin lines for the same parameters). The thick dashed blue line in the lower left panel is the same as the thick solid blue line but without the contribution of DM. The thin dotted black line and the thick dotted black line in the lower right panel are SFGs spectra calculated by  \citet{2014JCAP...09..043T} and by \citet{2014ApJ...786...40L} respectively.}
\label{fig:SFR_HiRes}
\end{figure}

\begin{figure}
\centering
\includegraphics[width=0.95\textwidth]{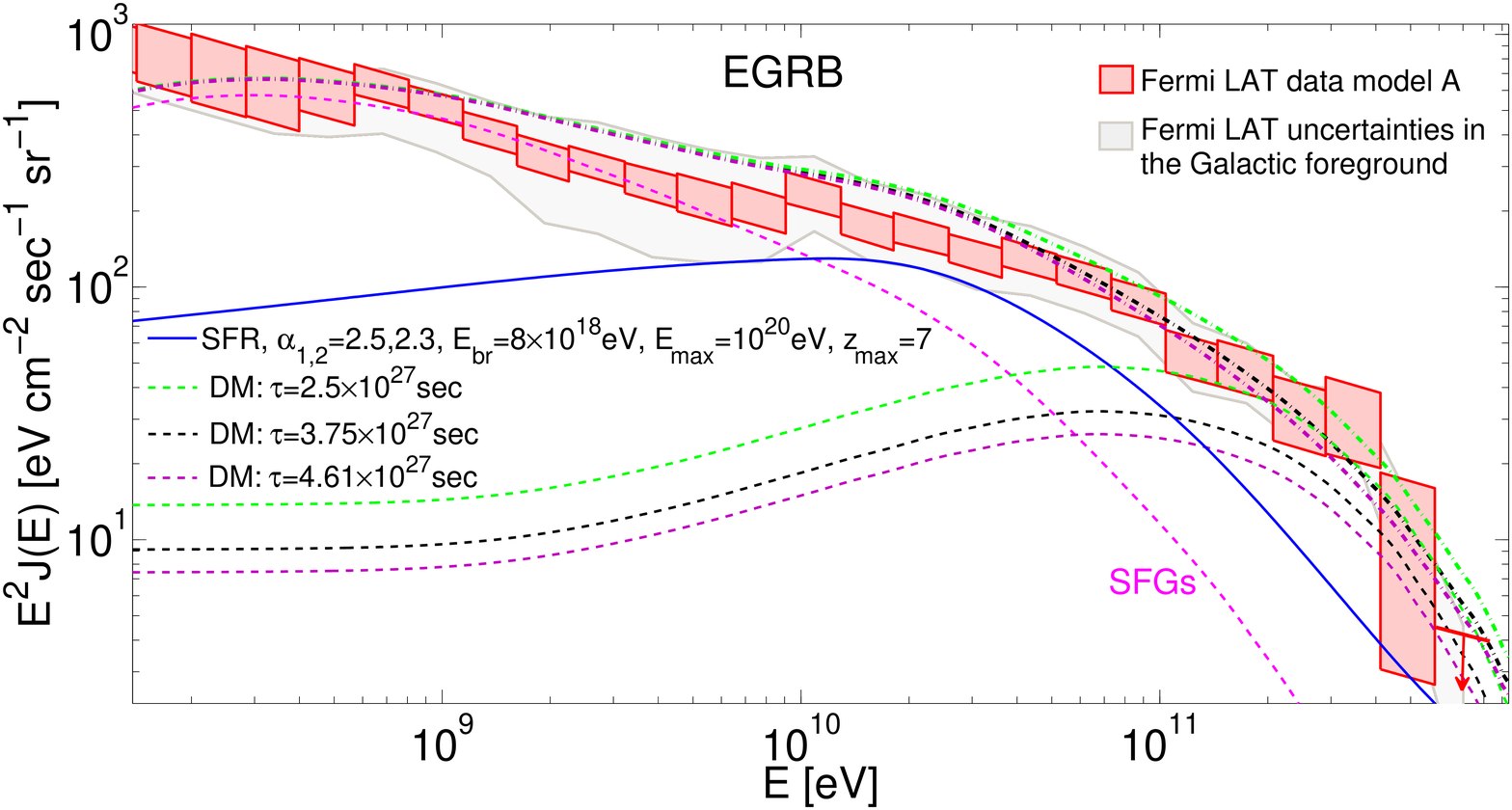}
\caption{Total contribution of $\gamma$-rays from SFGs, UHECRs, and DM for different DM lifetimes. The solid blue line is the same as the solid blue line in Figure \ref{fig:SFR_HiRes}. Dashed violet, dashed black, and dashed green lines are the contributions from DM for lifetimes of $4.61\times 10^{27},3.75\times 10^{27},2.5 \times 10^{27}\mbox{sec}$ respectively. Dashed magenta line is the SFGs contribution. Dashed-Dotted lines are the totals of SFGs, UHECRs, and DM.}
\label{fig:EGRB_DM_limits}
\end{figure}

\begin{figure}
\centering
\includegraphics[width=0.85\textwidth]{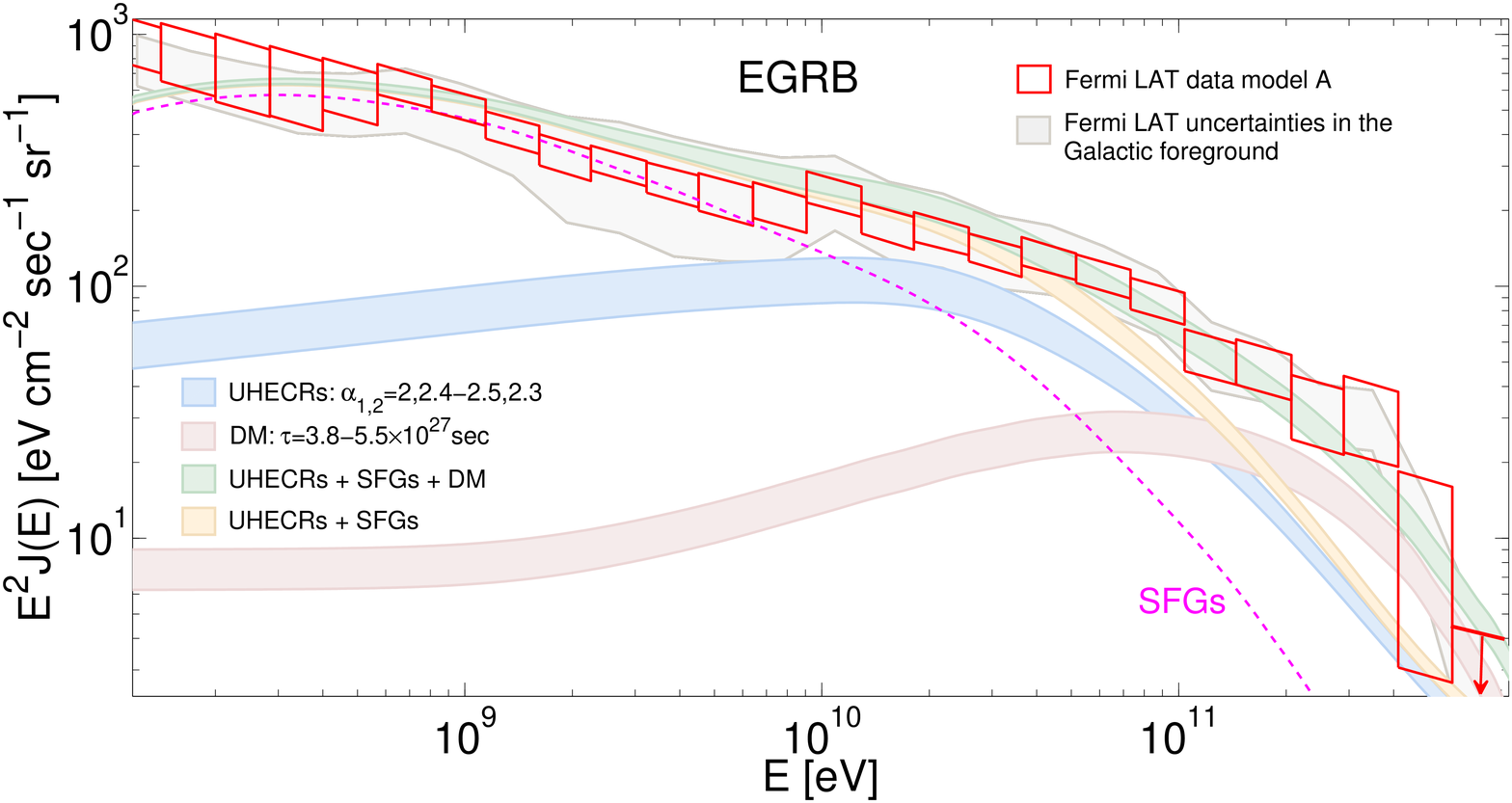}
\includegraphics[width=0.85\textwidth]{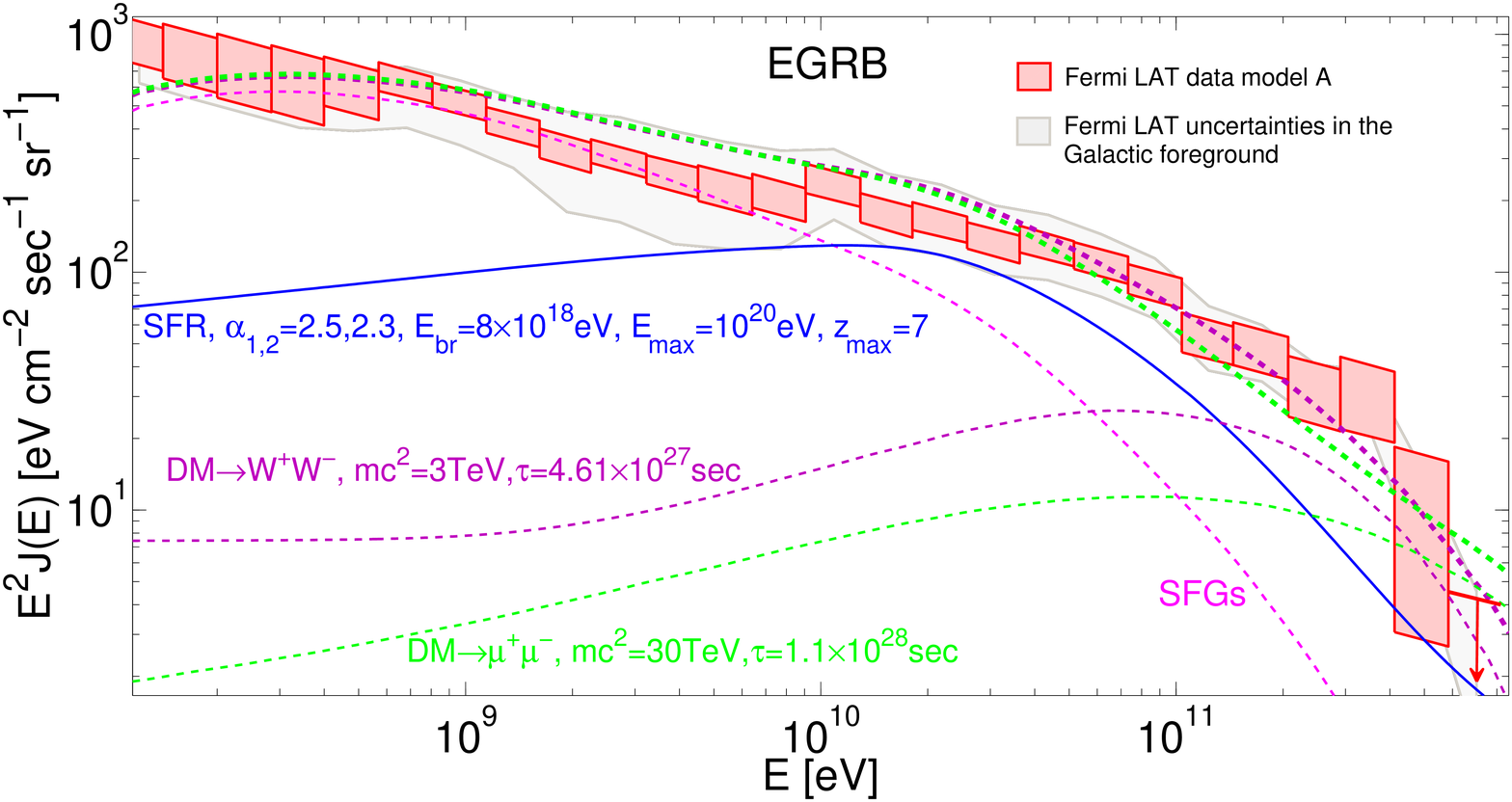}
\caption{\textbf{Upper panel:} Uncertainties in the total $\gamma$-ray flux from SFGs, UHECRs evolving as SFR, and DM. The UHECR band (blue) is corresponding to the area between the thin orange solid line and the thin blue solid line in the lower panels of Figure \ref{fig:SFR_HiRes}. This band reflects the uncertainties in $\gamma$-ray spectra from UHECRs that evolve in time as SFR, have a maximal acceleration energy of $E_{max}=10^{20}\mbox{eV}$, with maximal redshift of $z_{max}=7$, and are adjusted to the HiRes and recalibrated Auger data. The DM band (violet) is corresponding to lifetimes of $3.8\mbox{-}5.5\times10^{27}\mbox{sec}$. The yellow band is the total $\gamma$-ray flux from UHECRs and SFGs (without the DM). The green band is the total of UHECRs, SFGs, and DM.
\textbf{Lower Panel:} Comparison of a fit to the Fermi LAT data using the $W^+W^-$ decay channel (thick dashed violet line) to a fit using the $\mu^+\mu^-$ channel (thick green dashed line). The DM contributions are shown in thin magenta and green dashed lines. The contributions from UHECRs and SFGs are also shown.}
\label{fig:shading}
\end{figure}

\begin{figure}
\centering
\includegraphics[width=0.49\textwidth]{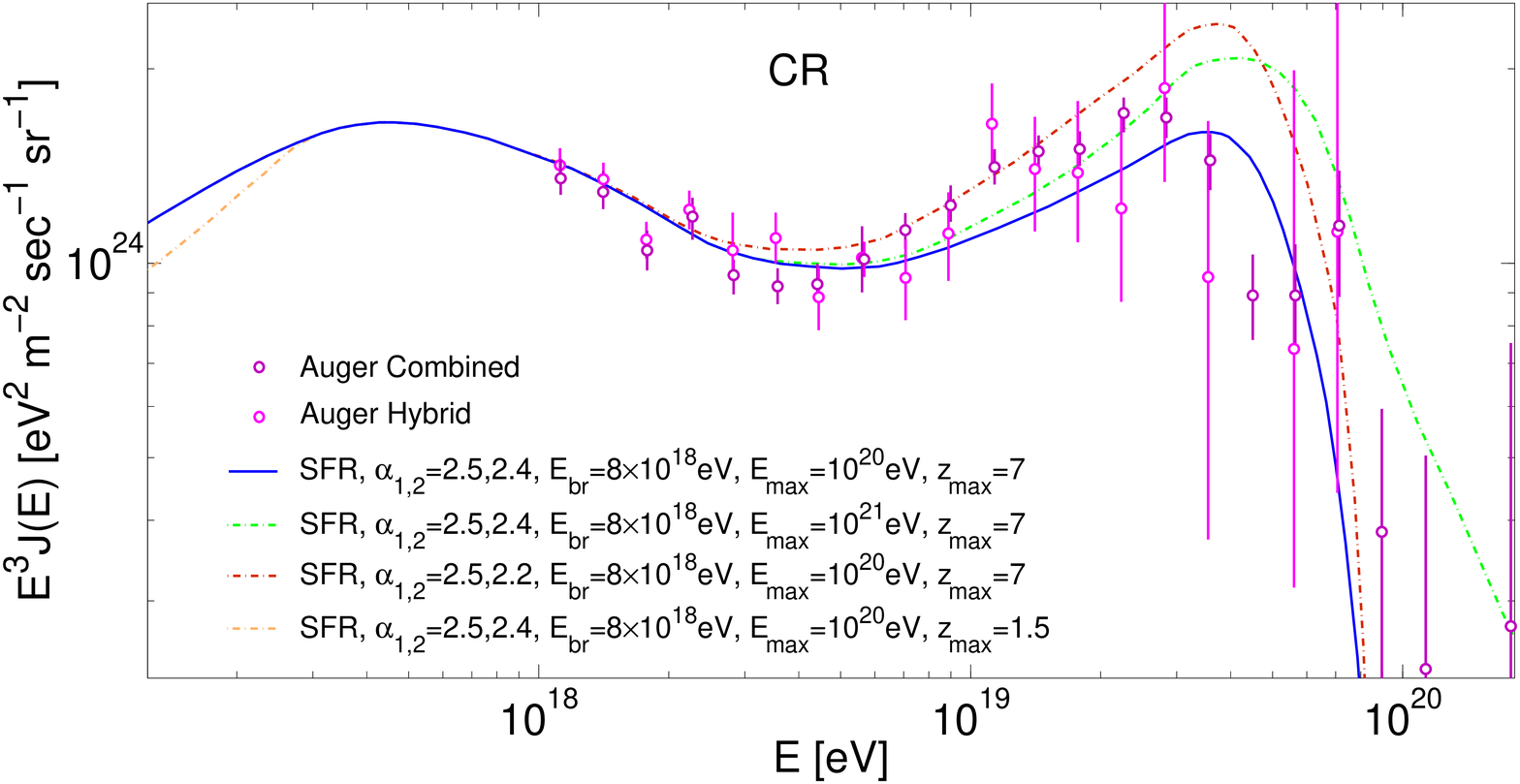}
\includegraphics[width=0.49\textwidth]{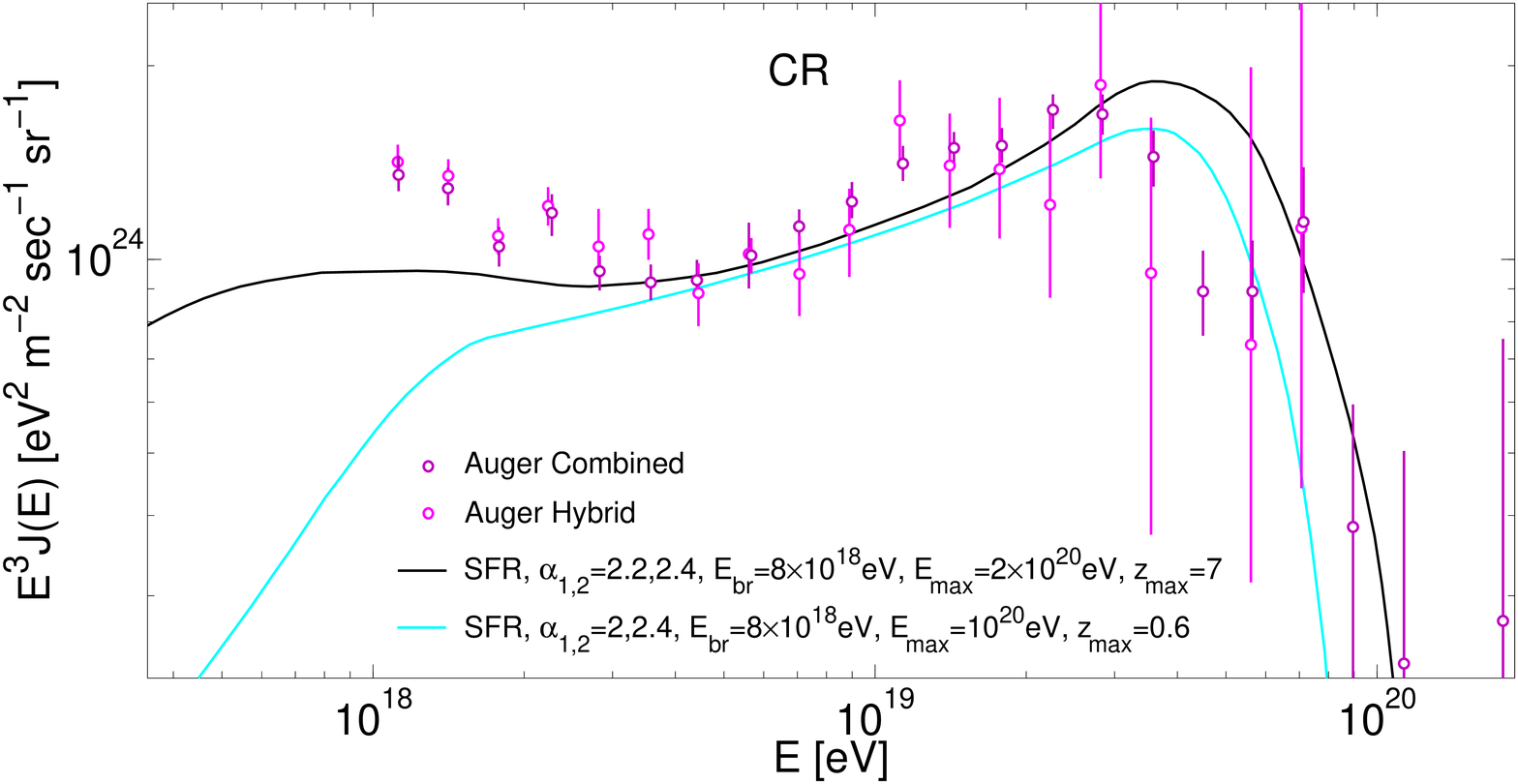}
\includegraphics[width=0.49\textwidth]{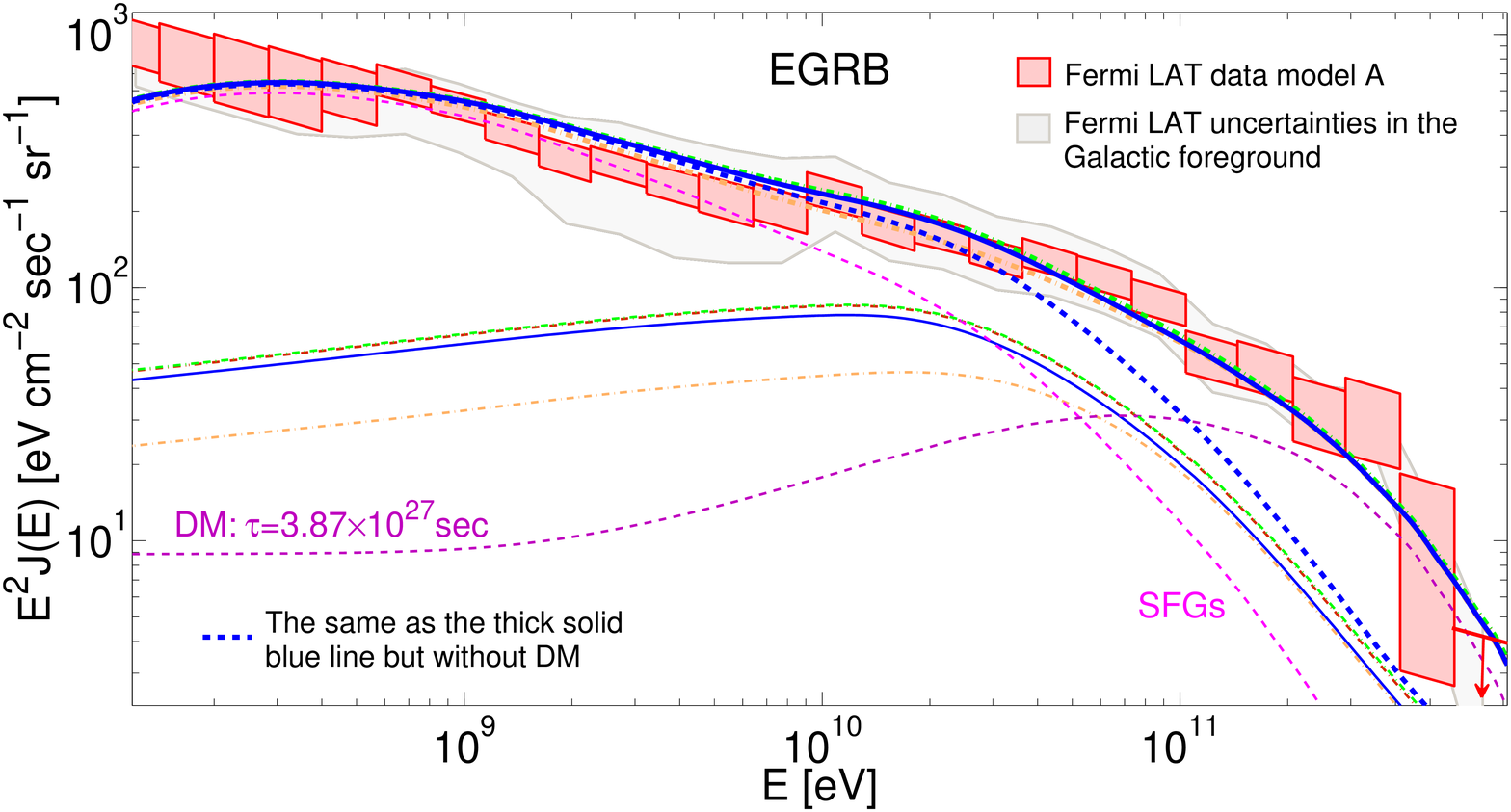}
\includegraphics[width=0.49\textwidth]{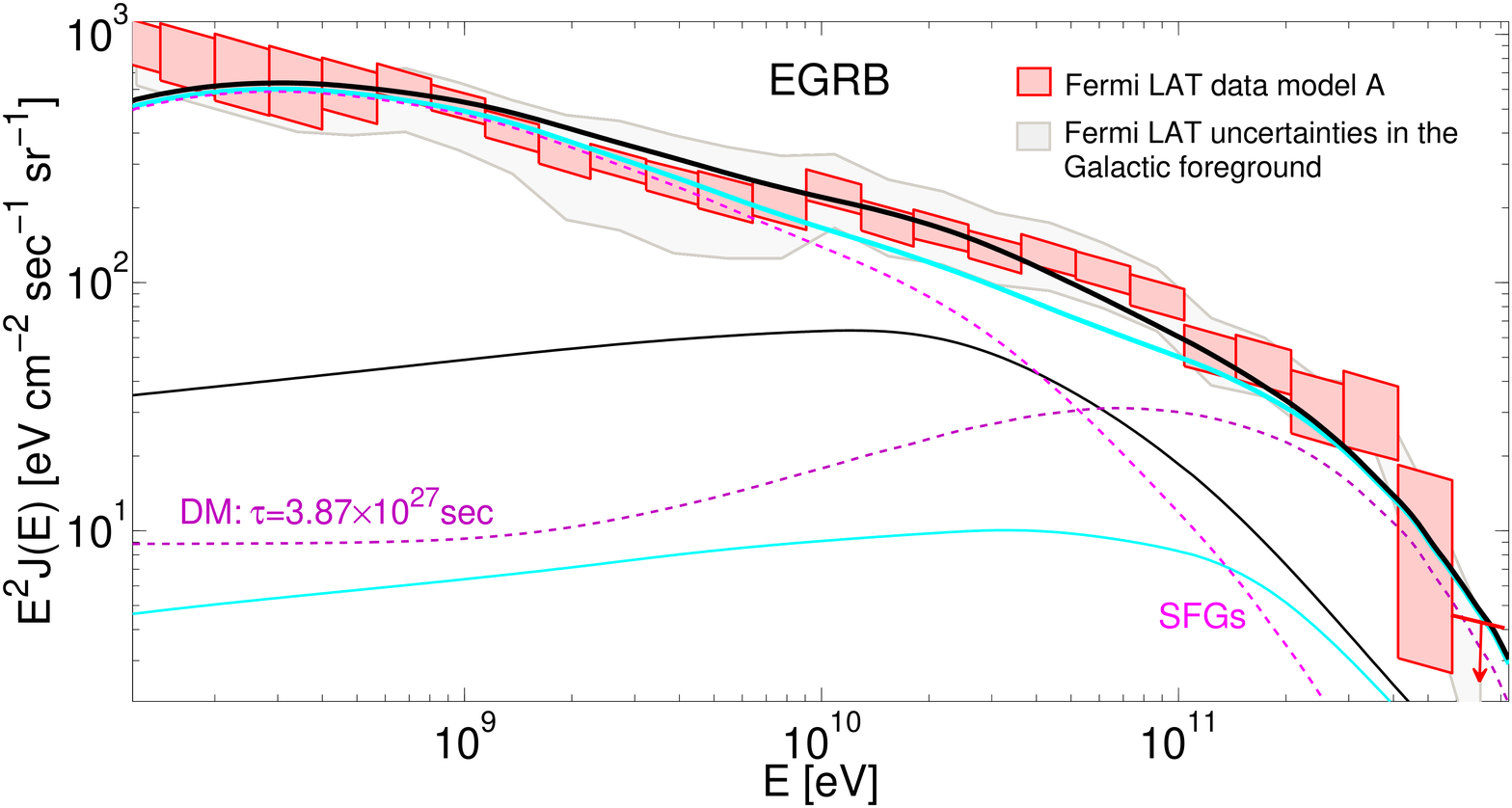}
\caption{The same as in Figure \ref{fig:SFR_HiRes}, but for different UHECR parameters and the UHECRs are normalized to the unrecalibrated Auger data. The evolution scenario here is SFR. The DM lifetime in this case is $\tau=3.87\times10^{27}\mbox{sec}$.}
\label{fig:SFR_Auger}
\end{figure}

\begin{figure}
\centering
\includegraphics[width=1\textwidth]{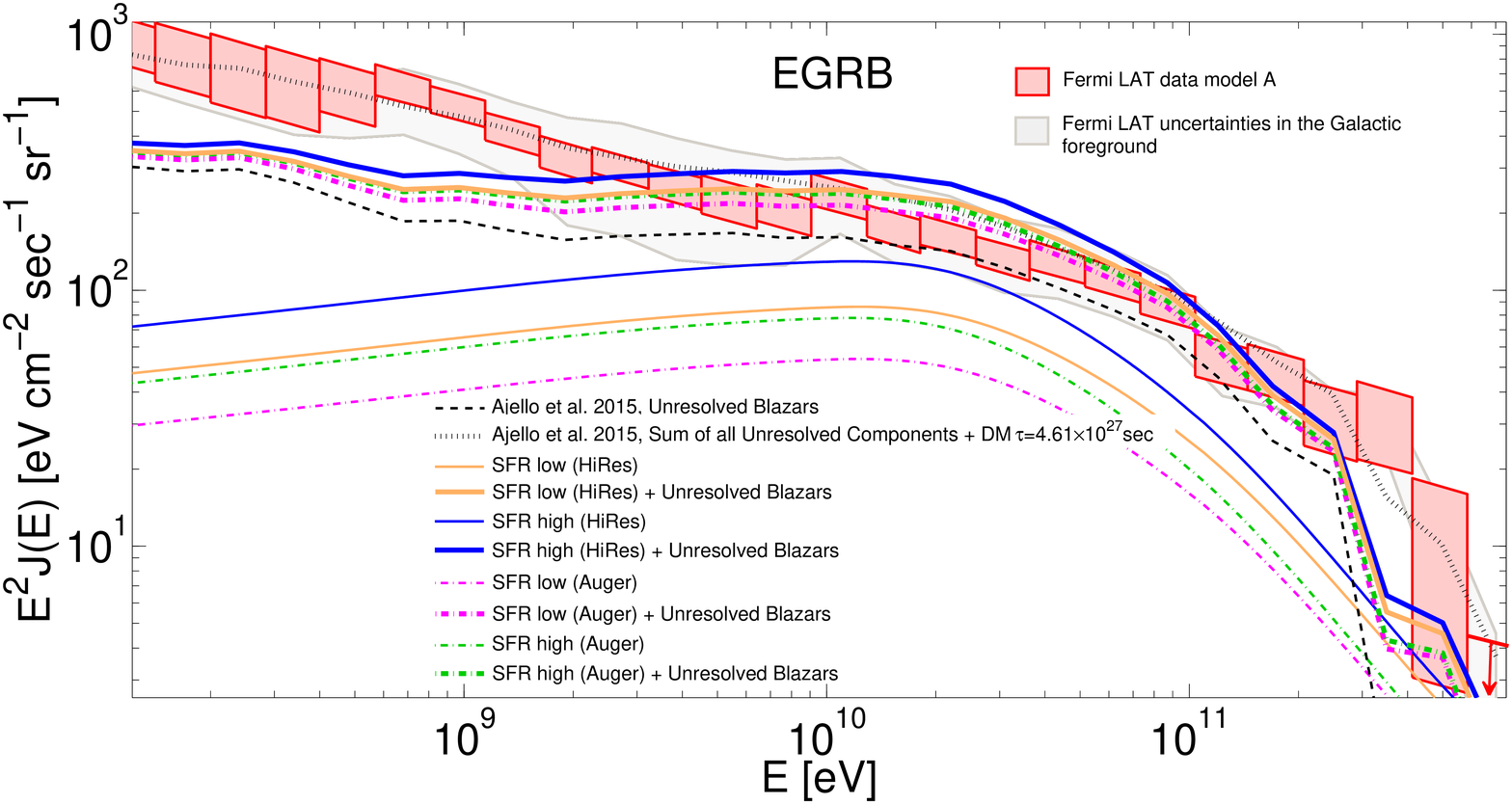}
\caption{The total flux of $\gamma$-rays originating from UHECRs evolving as SFR and from blazars. The thin dashed black line is the unresolved blazar spectrum as extrapolated by \citet{2015ApJ...800L..27A} from the resolved blazar data. The thick dotted black line is the sum of all unresolved components (blazars, SFGs, and radio galaxies) in \citet{2015ApJ...800L..27A}, Figure 3 (resolved point sources have been removed) with an additional contribution from $3\mbox{TeV}$ $W^+W^-$ decay DM with lifetime of $4.61\times10^{27}\mbox{sec}$. The thin solid orange and the thin solid blue lines represent the minimum and the maximum $\gamma$-ray fluxes available from UHECRs evolving as SFR, adjusted to the HiRes data, with $E_{max}=10^{20}\mbox{eV}$, and $z_{max}=7$. For the thin orange line: $\alpha_{1,2}=2,2.4$ and $E_{br}=8\times10^{18}\mbox{eV}$. For the thin blue line: $\alpha_{1,2}=2.5,2.3$ and $E_{br}=8\times10^{18}\mbox{eV}$. The thin dashed-dotted magenta and the thin dashed-dotted green lines represent the minimum and the maximum $\gamma$-ray fluxes available from UHECRs evolving as SFR, adjusted to the unrecalibrated Auger data, with $E_{max}=10^{20}\mbox{eV}$, and $z_{max}=7$. The thin dashed-dotted magenta line has the same parameters as the thin orange line. The thin dashed-dotted green line has the same parameters as the thin solid blue line but with $\alpha_2=2.4$. The thick lines are the totals of blazars and UHECRs.}
\label{fig:blazars}
\end{figure}

\begin{figure}
\centering
\includegraphics[width=0.49\textwidth]{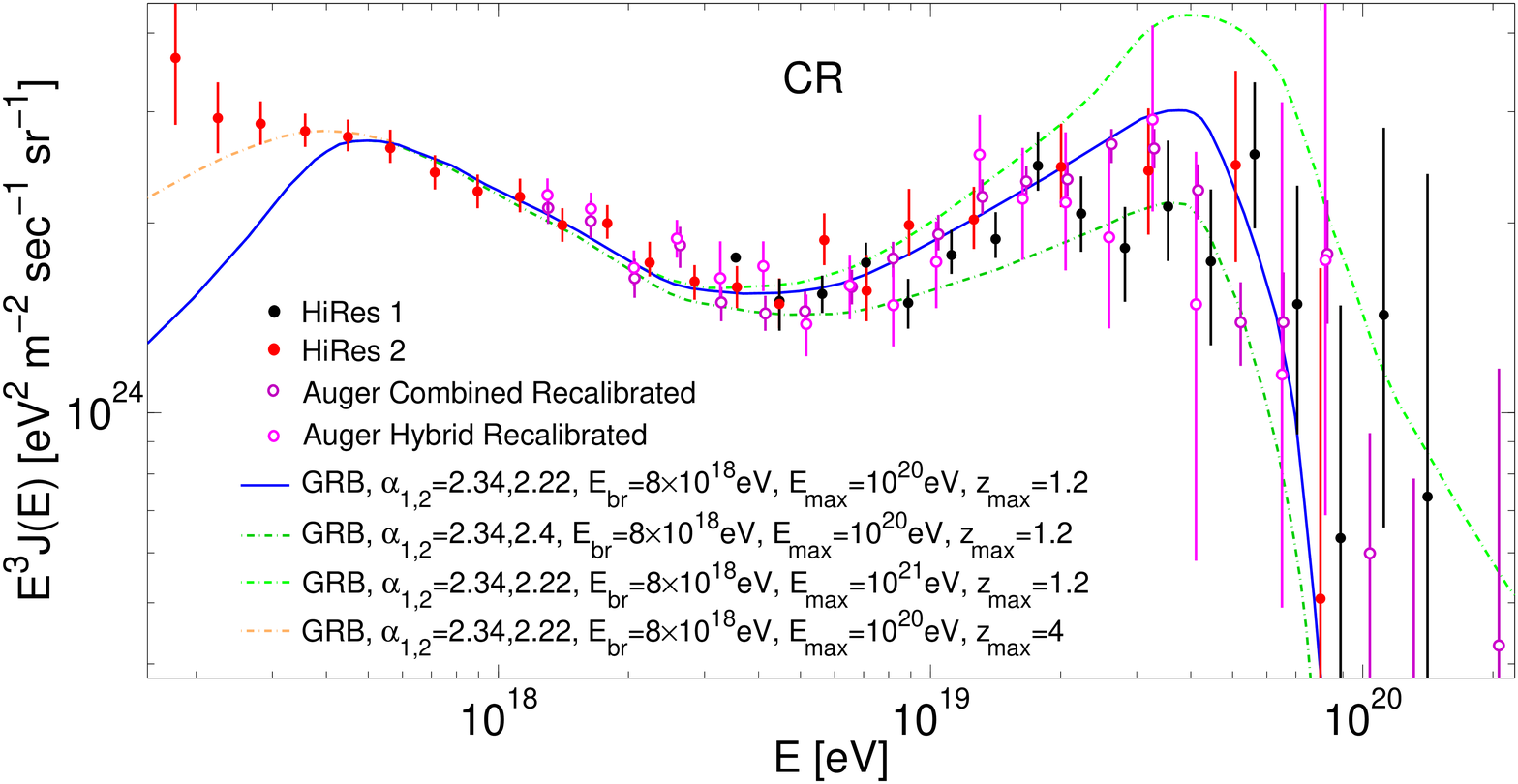}
\includegraphics[width=0.49\textwidth]{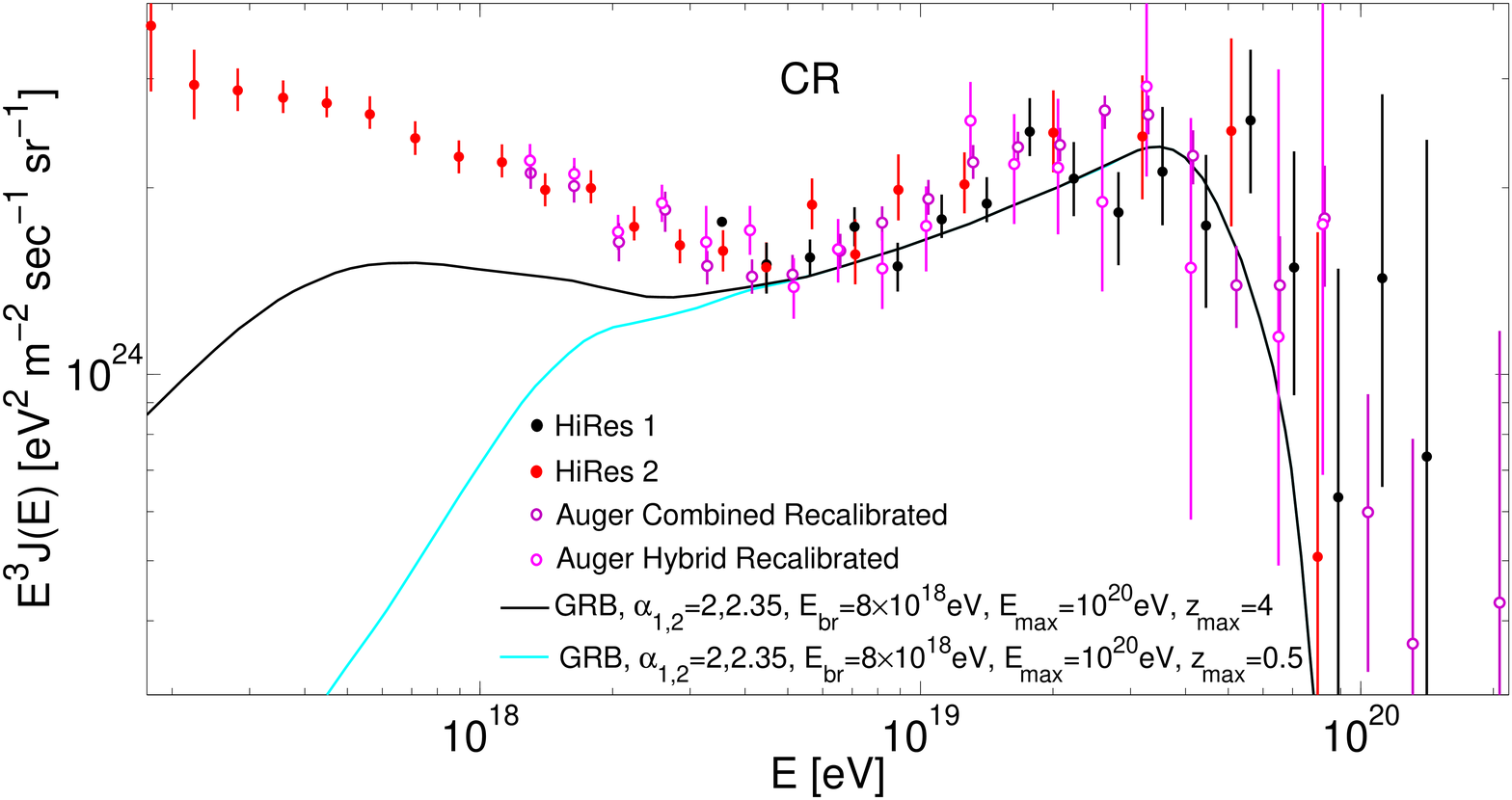}
\includegraphics[width=0.49\textwidth]{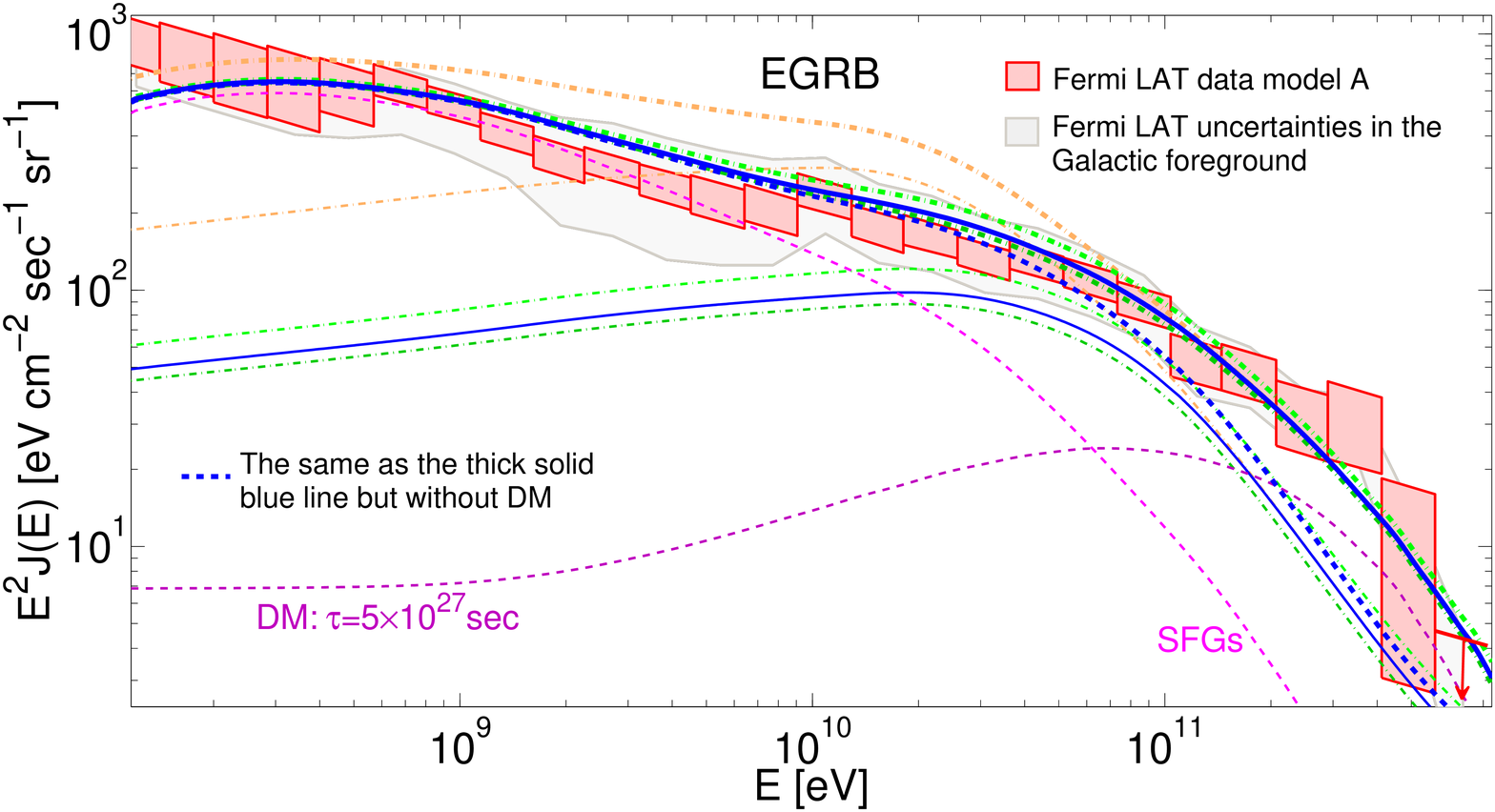}
\includegraphics[width=0.49\textwidth]{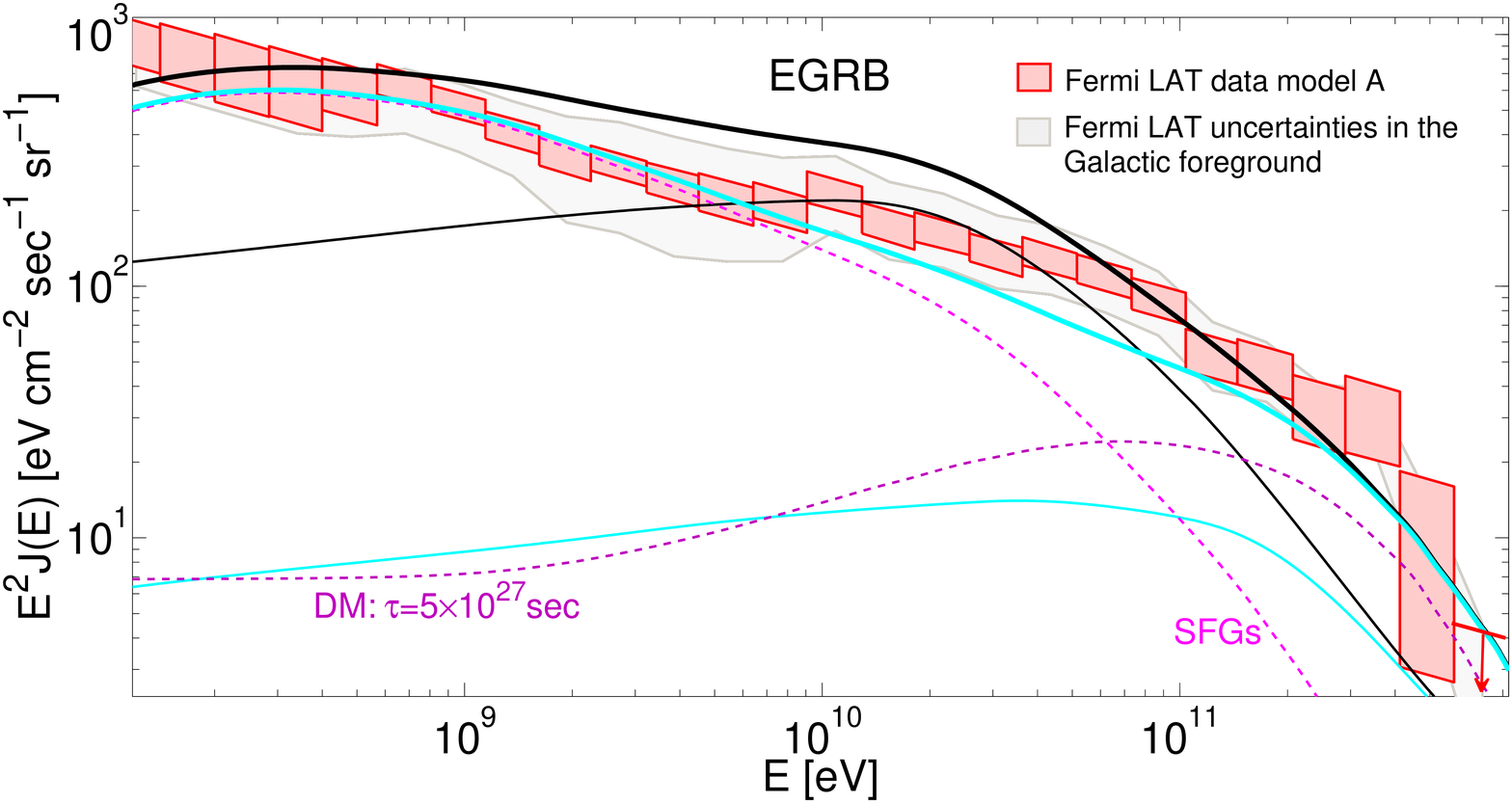}
\caption{The same as in Figure \ref{fig:SFR_HiRes}, but for UHECRs evolving as GRBs. The DM lifetime here is $\tau= 5\times10^{27}\mbox{sec}$.}
\label{fig:GRB_HiRes}
\end{figure}

\begin{figure}
\centering
\includegraphics[width=0.49\textwidth]{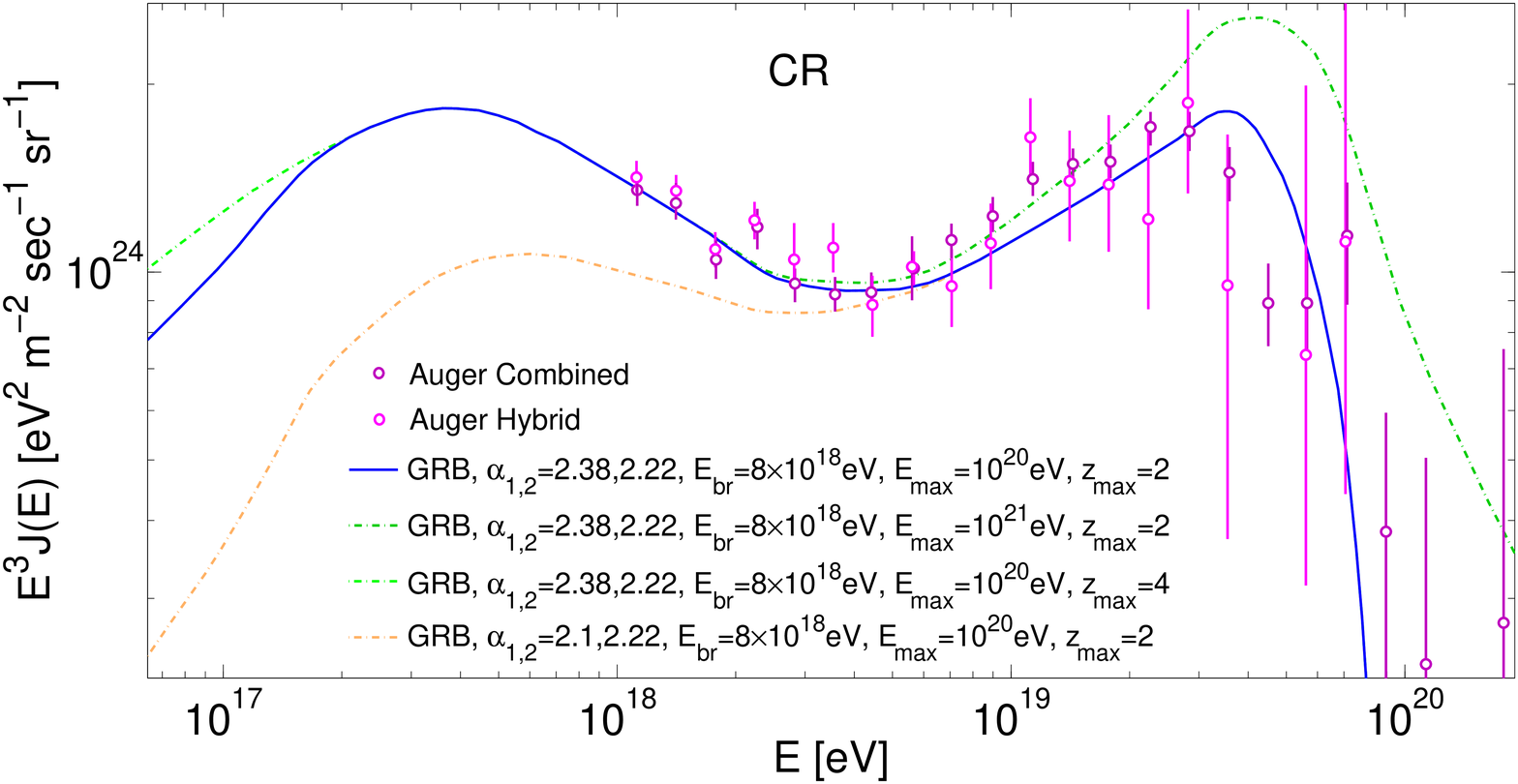}
\includegraphics[width=0.49\textwidth]{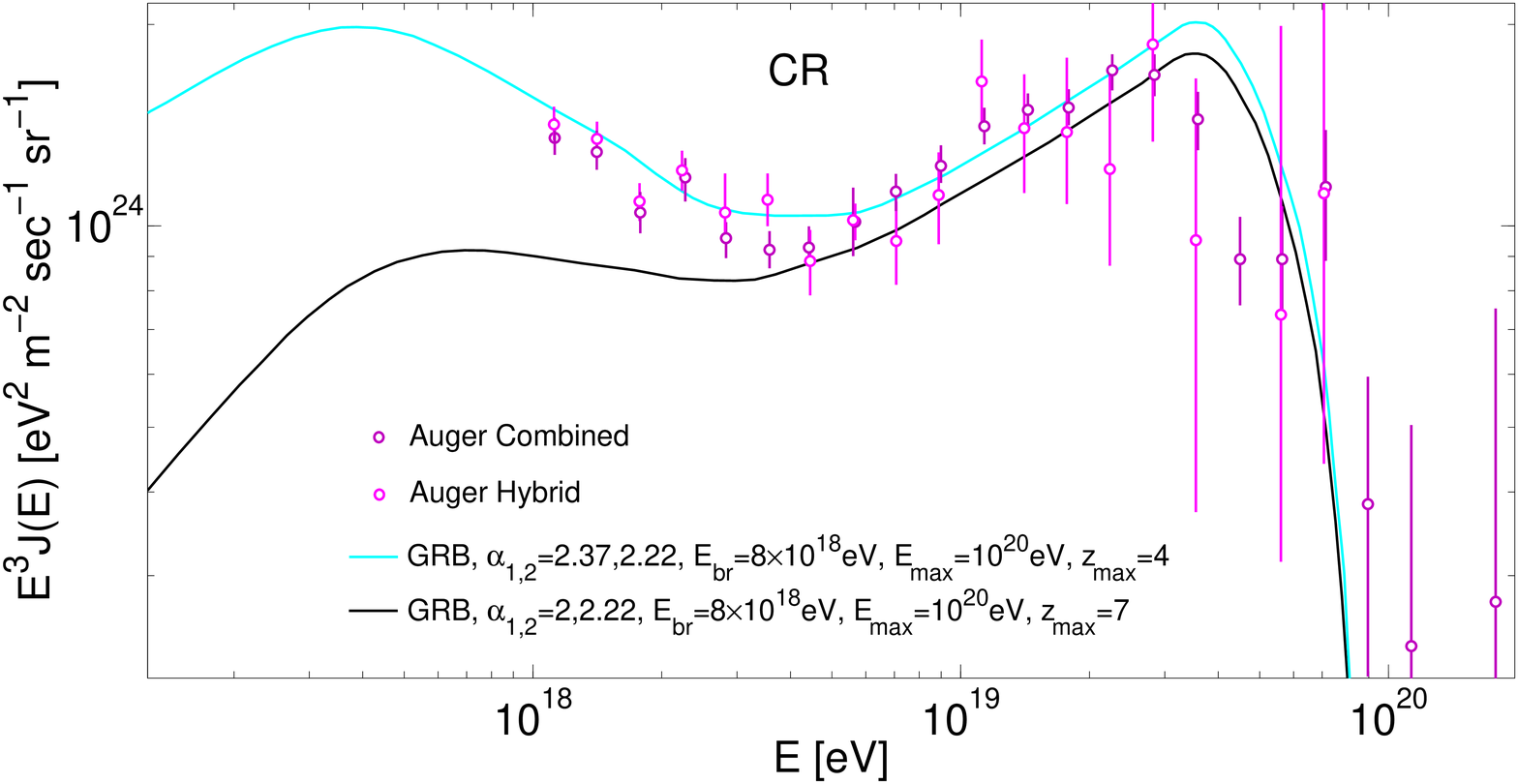}
\includegraphics[width=0.49\textwidth]{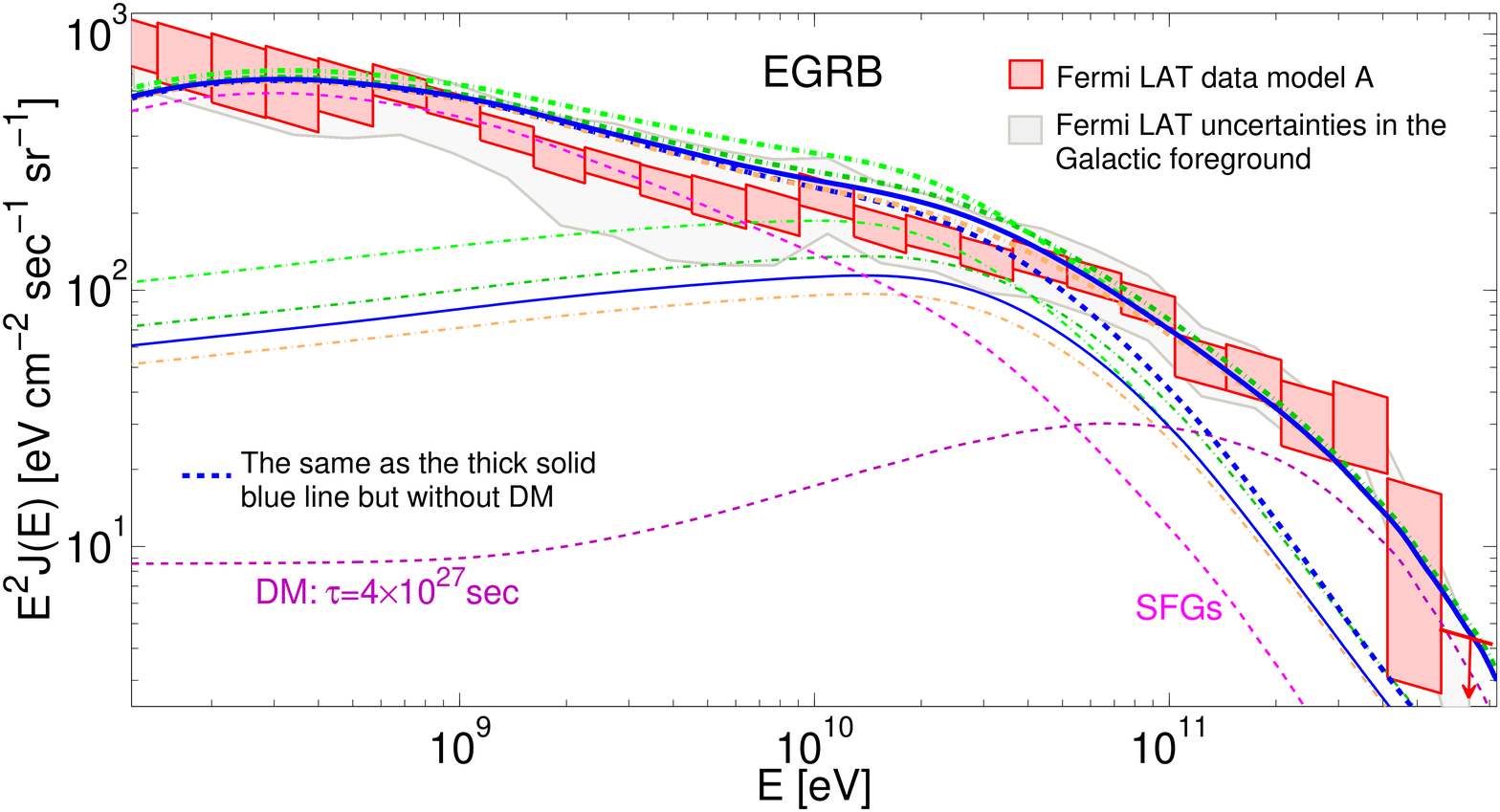}
\includegraphics[width=0.49\textwidth]{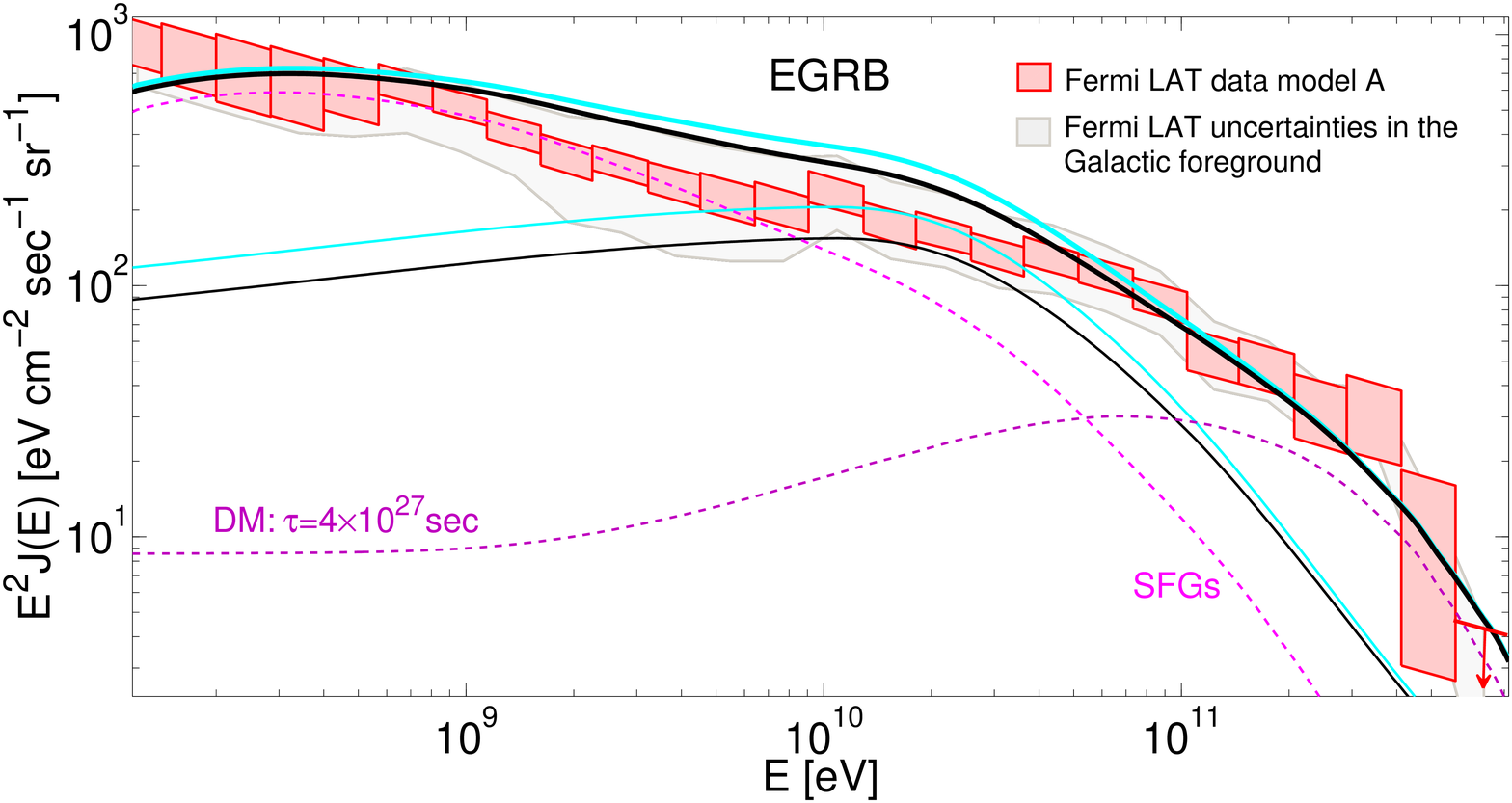}
\caption{The same as in Figure \ref{fig:SFR_Auger}, but for UHECRs evolving as GRBs. The DM lifetime here is $\tau= 4\times10^{27}\mbox{sec}$.}
\label{fig:GRB_Auger}
\end{figure}

\begin{figure}
\centering
\includegraphics[width=0.9\textwidth]{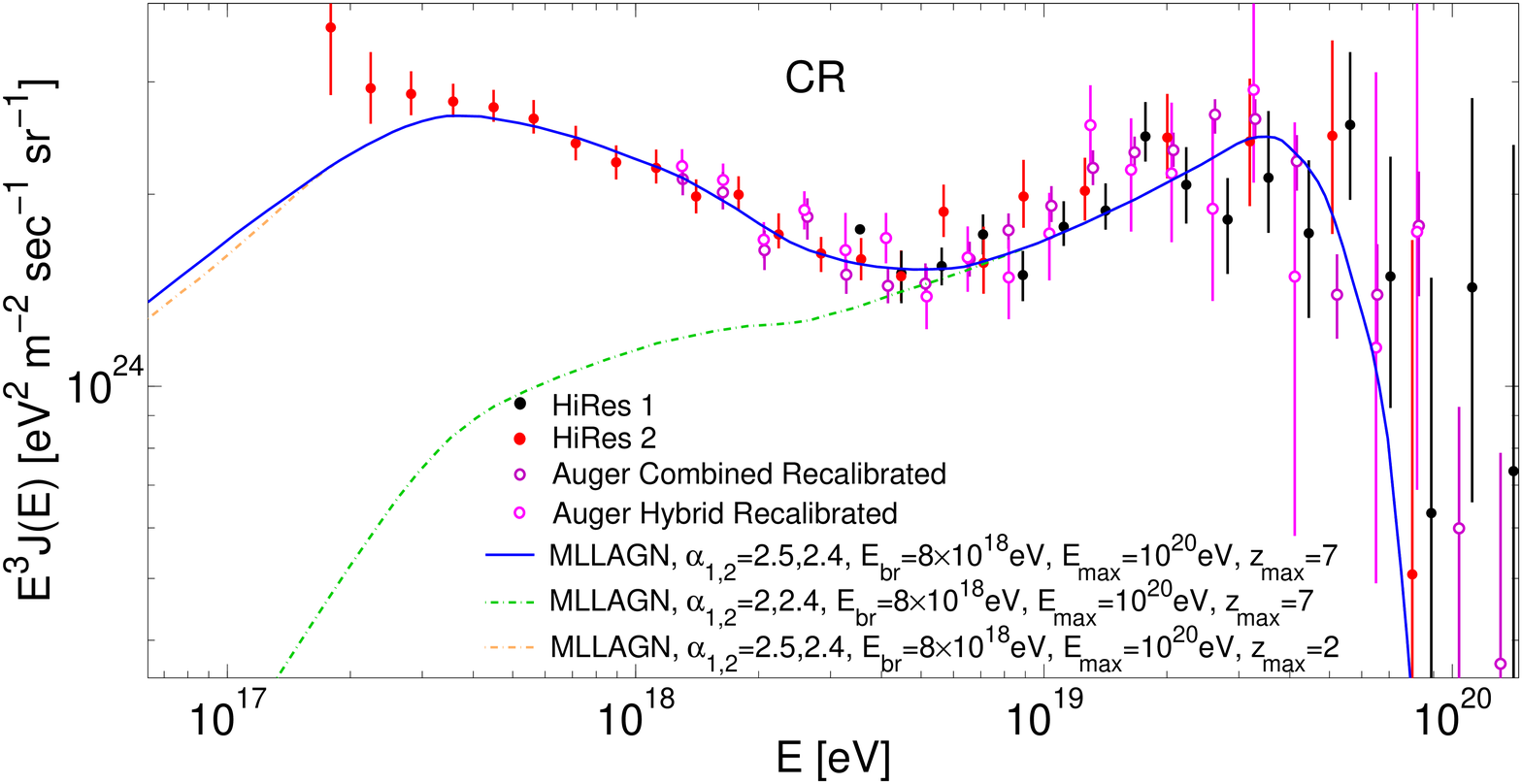}
\includegraphics[width=0.9\textwidth]{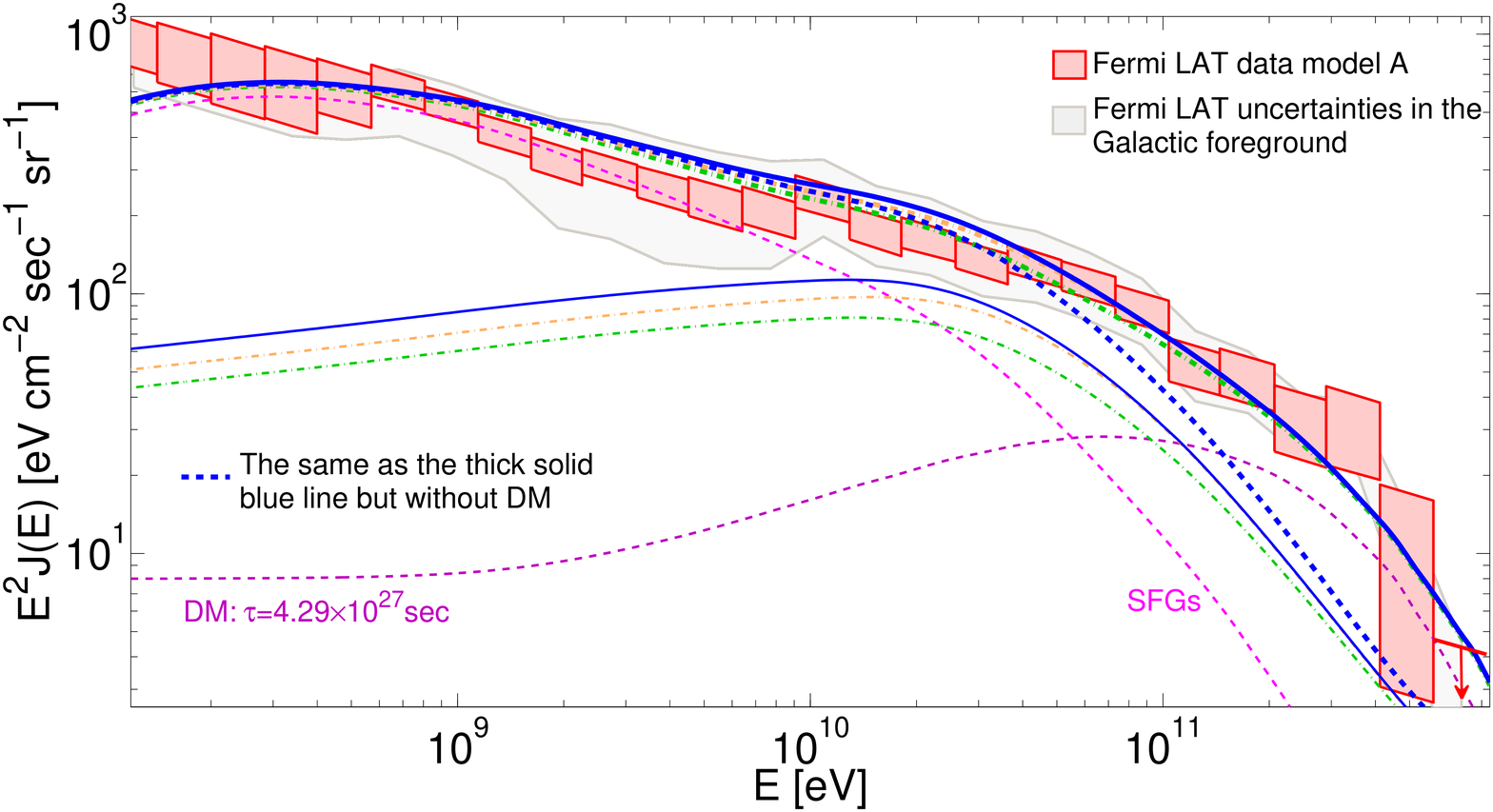}
\caption{UHECRs evolving as MLLAGNs and normalized to the HiRes data. \textbf{Upper Panel}: UHECR spectra. The dashed-dotted lines differ from the solid blue line in one parameter. \textbf{Lower Panel}: $\gamma$-ray fluxes corresponding to the UHECR spectra in the upper panel. Thick lines are the sum of the three components: SFGs, UHECRs, and DM. The thick dashed blue line is the same as the thick solid blue line but without DM contribution. The DM lifetime here is $\tau= 4.29\times10^{27}\mbox{sec}$.}
\label{fig:MLLAGN}
\end{figure}

\begin{figure}
\centering
\includegraphics[width=0.9\textwidth]{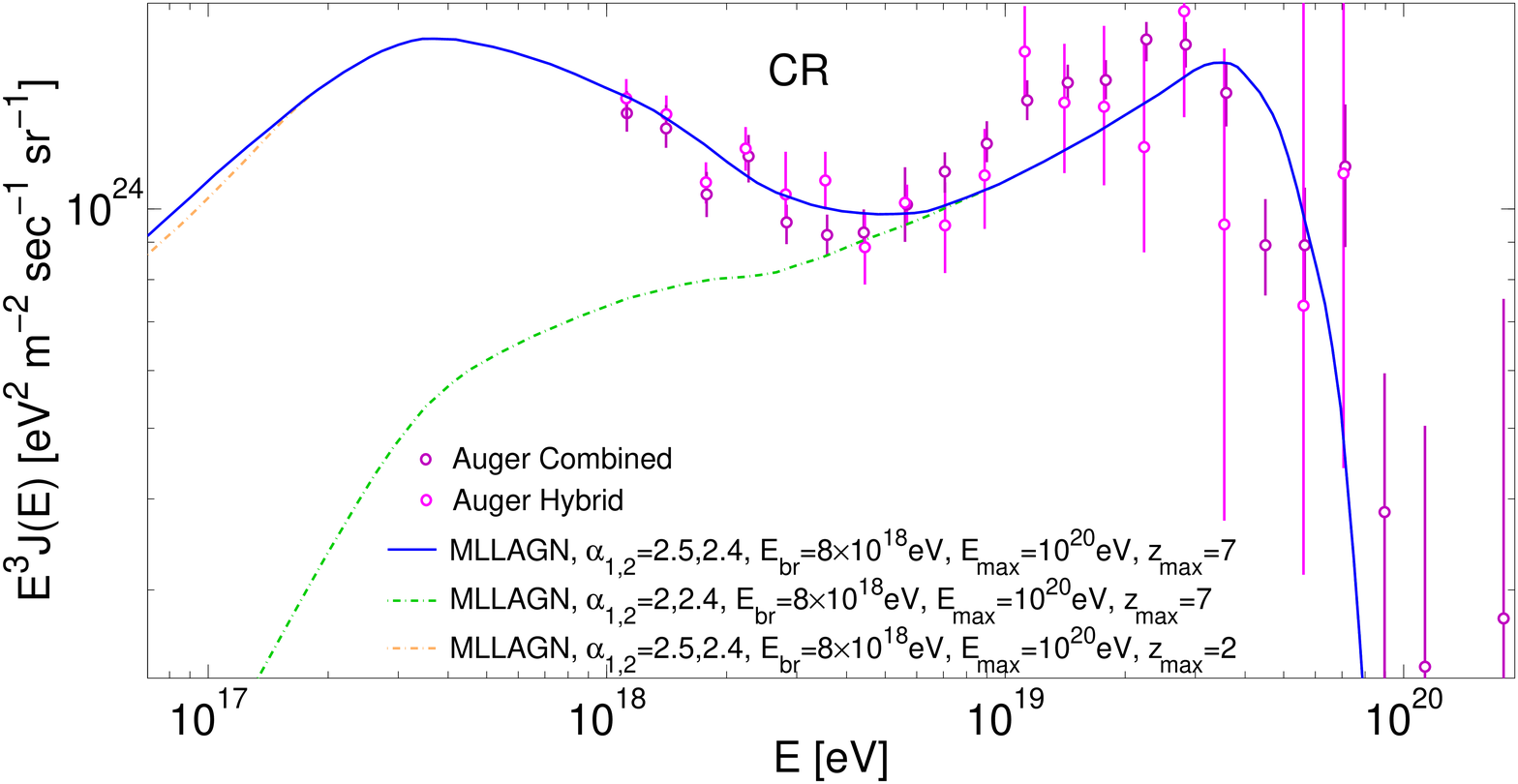}
\includegraphics[width=0.9\textwidth]{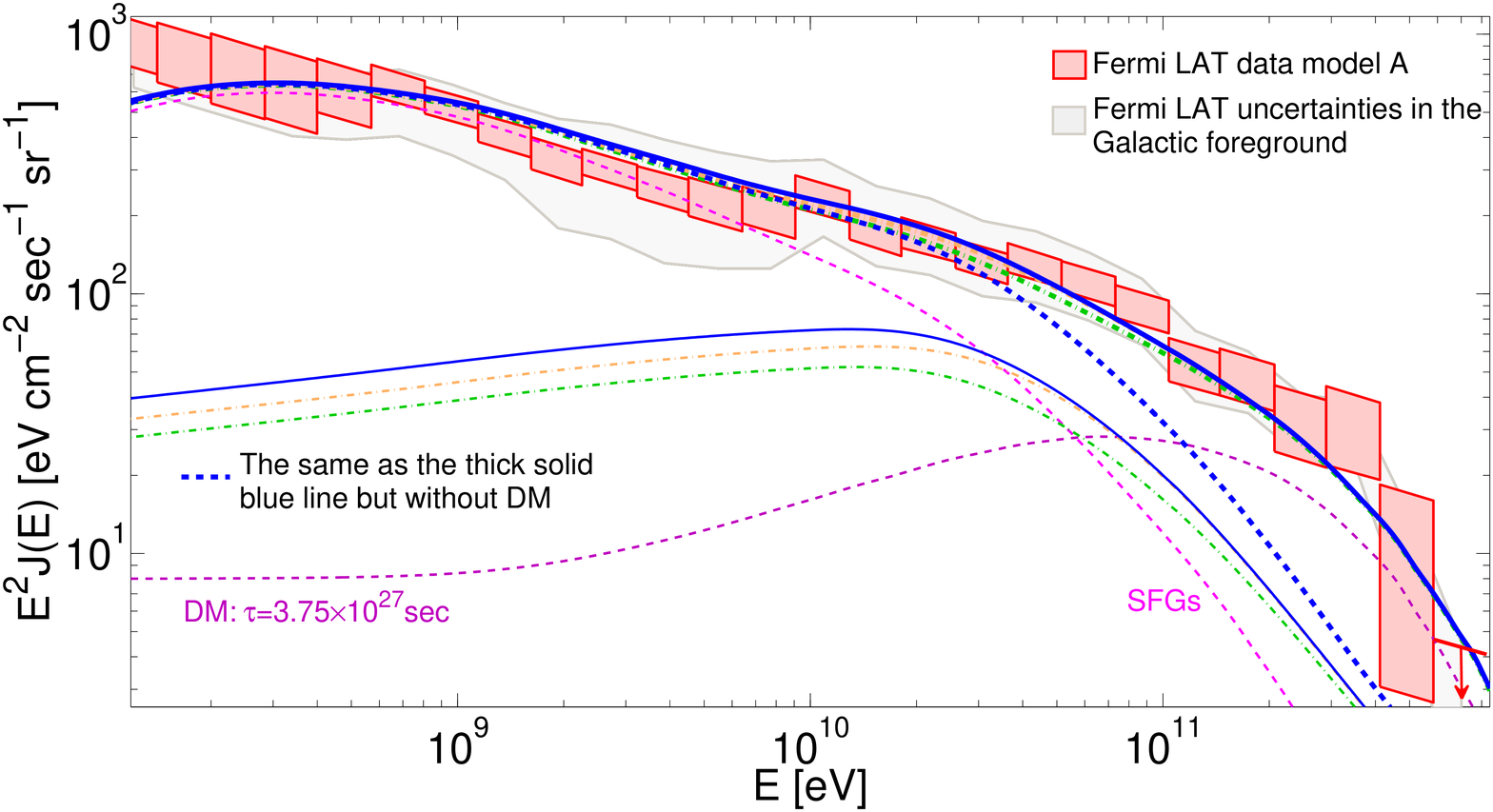}
\caption{The same as in Figure \ref{fig:MLLAGN}, but for UHECR spectra that are normalized to the Auger data. The DM lifetime here is $\tau= 3.75\times10^{27}\mbox{sec}$.}
\label{fig:MLLAGN_Auger}
\end{figure}

\begin{figure}
\centering
\includegraphics[width=0.49\textwidth]{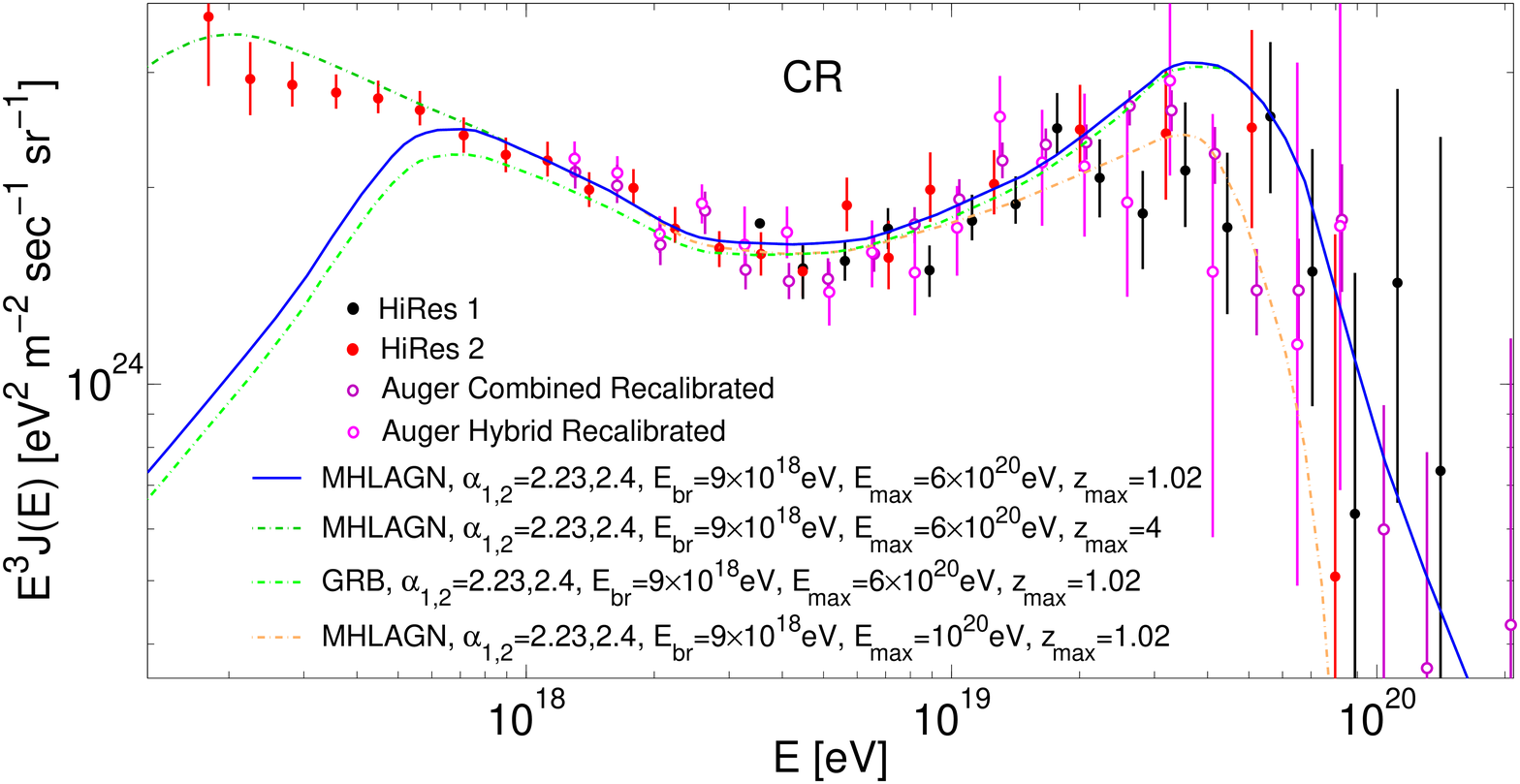}
\includegraphics[width=0.49\textwidth]{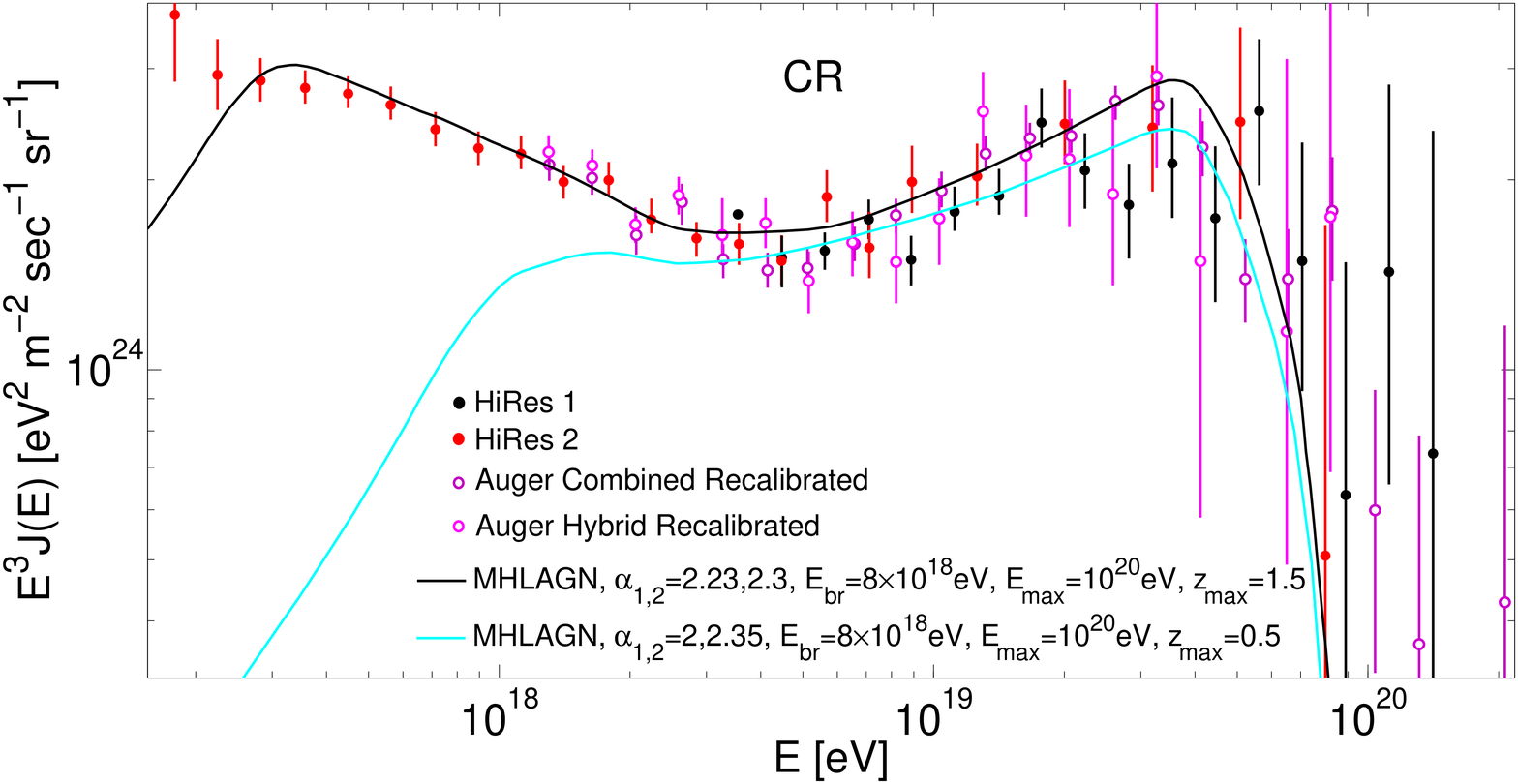}
\includegraphics[width=0.49\textwidth]{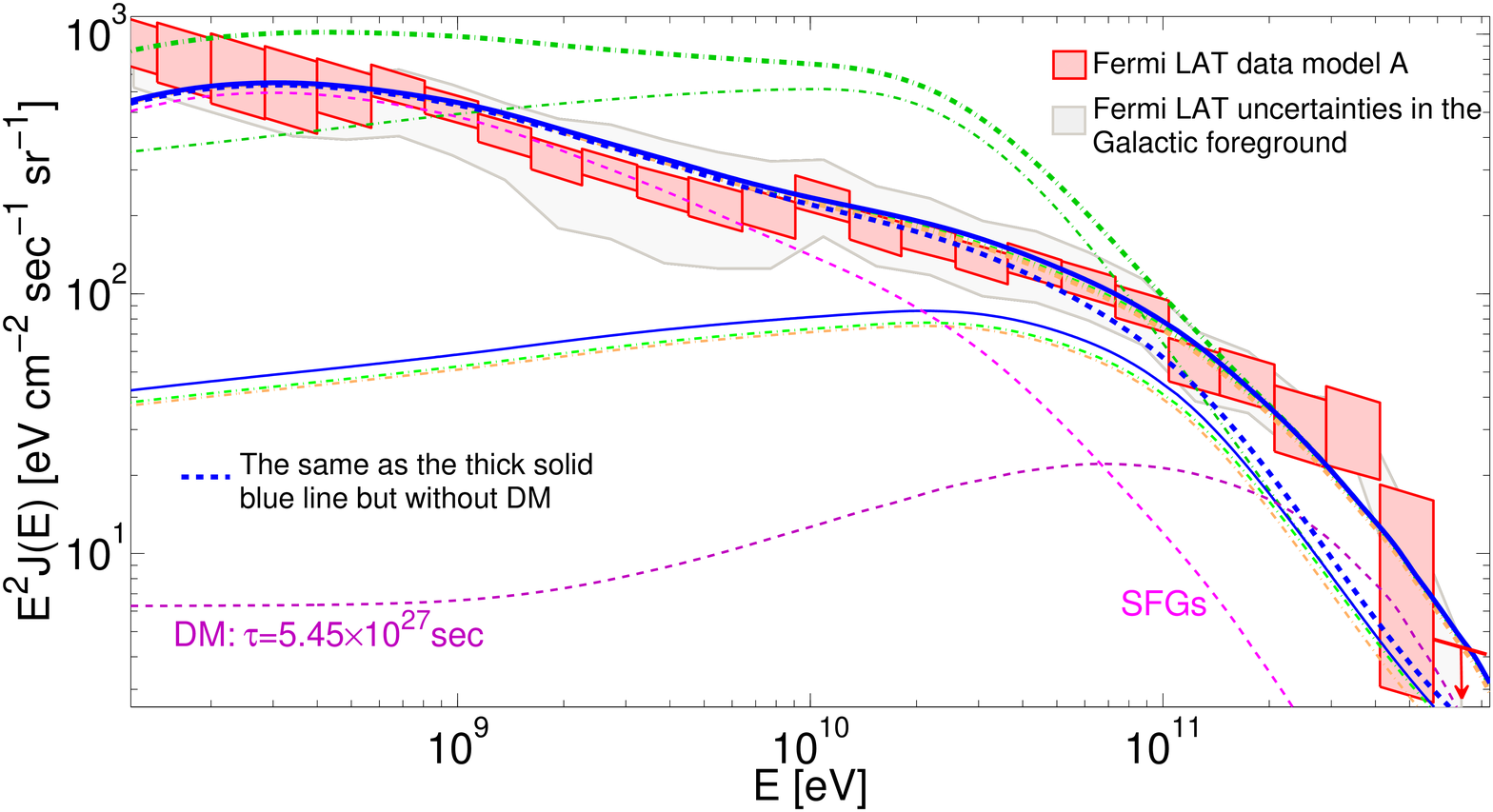}
\includegraphics[width=0.49\textwidth]{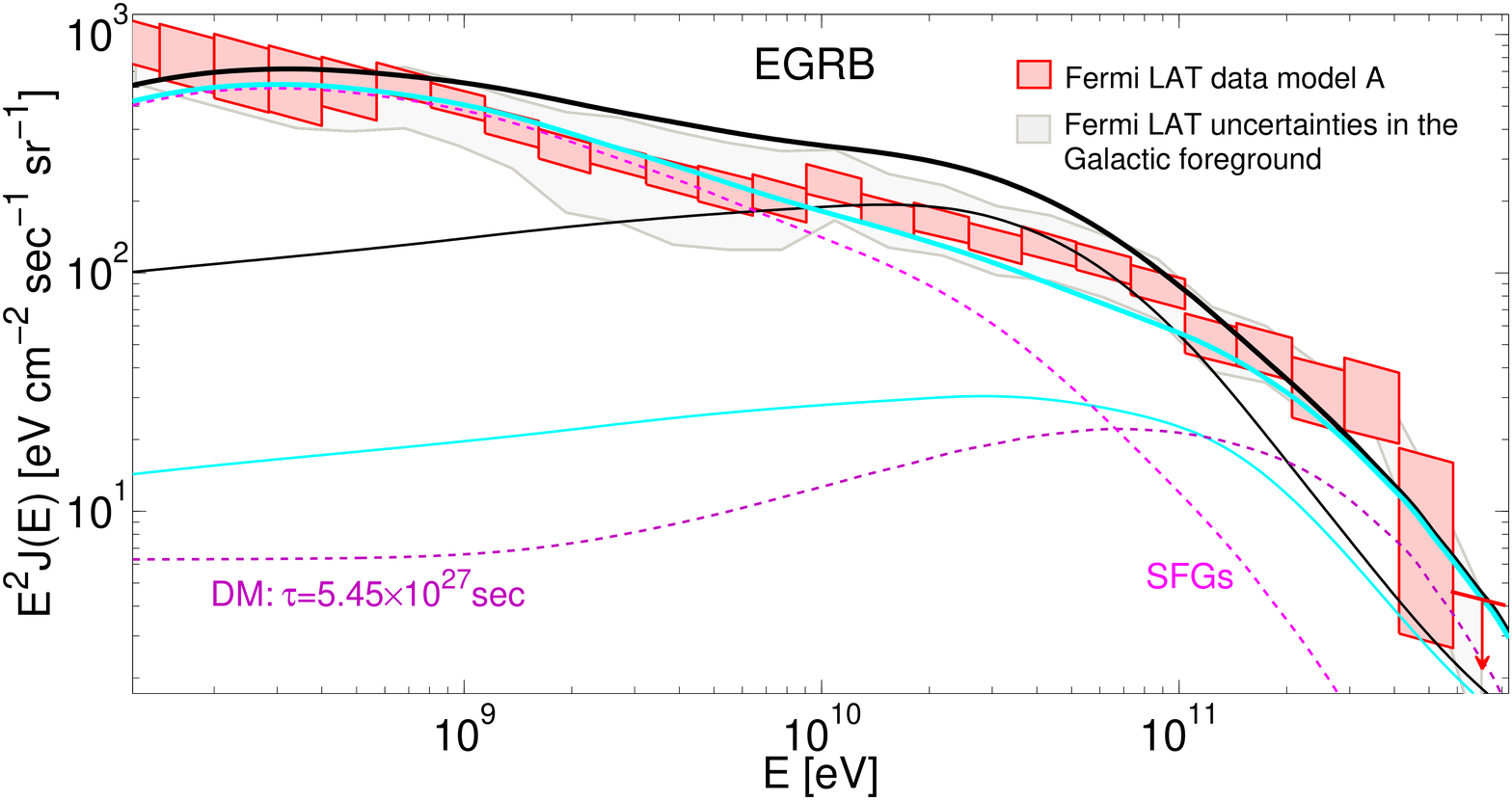}
\caption{The same as in Figure \ref{fig:SFR_HiRes},  but the UHECRs are evolving as MHLAGNs (except one curve that is corresponding to the GRB model). The DM lifetime here is $\tau= 5.45\times10^{27}\mbox{sec}$.}
\label{fig:AGN_HiRes}
\end{figure}

\begin{figure}
\centering
\includegraphics[width=0.49\textwidth]{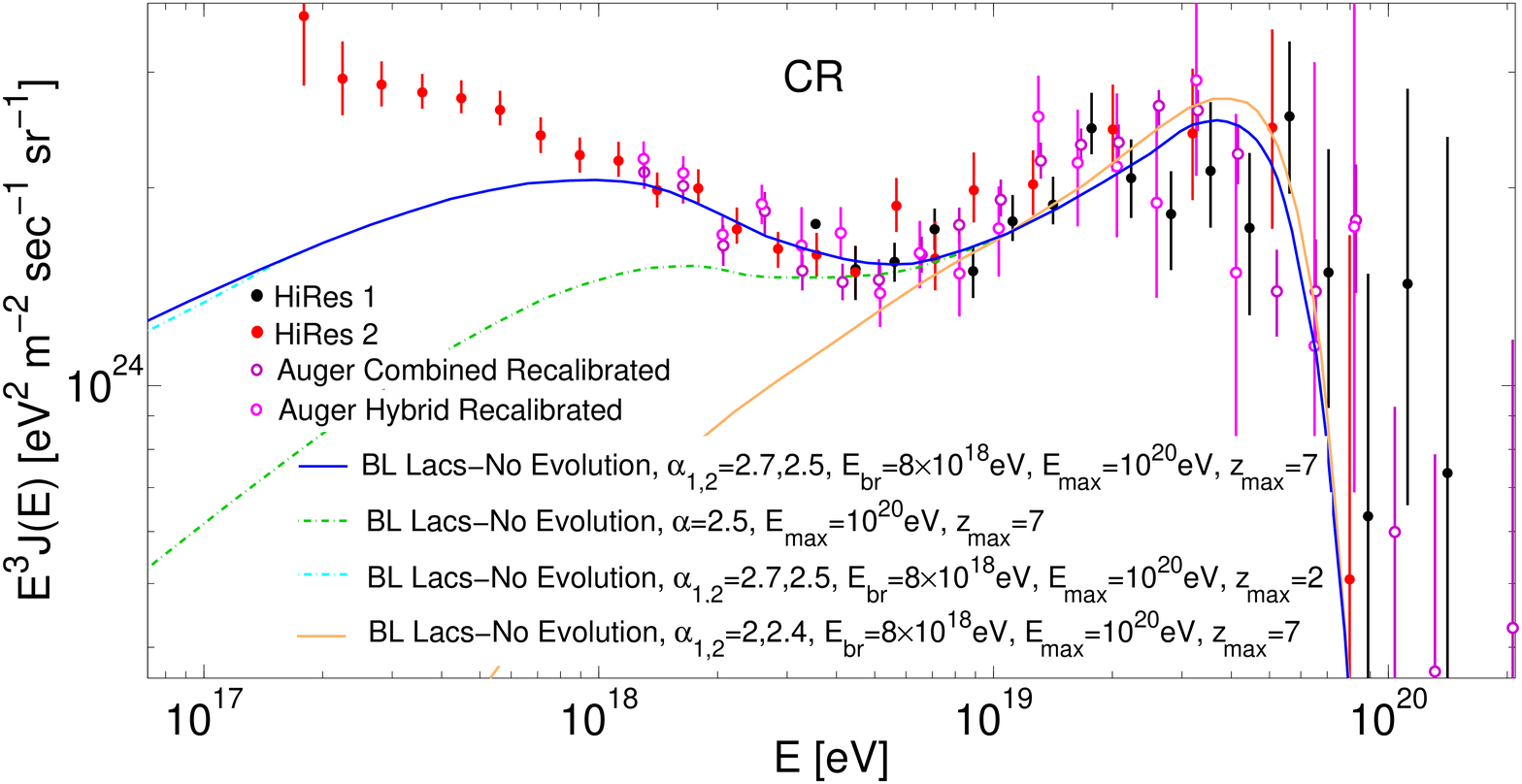}
\includegraphics[width=0.49\textwidth]{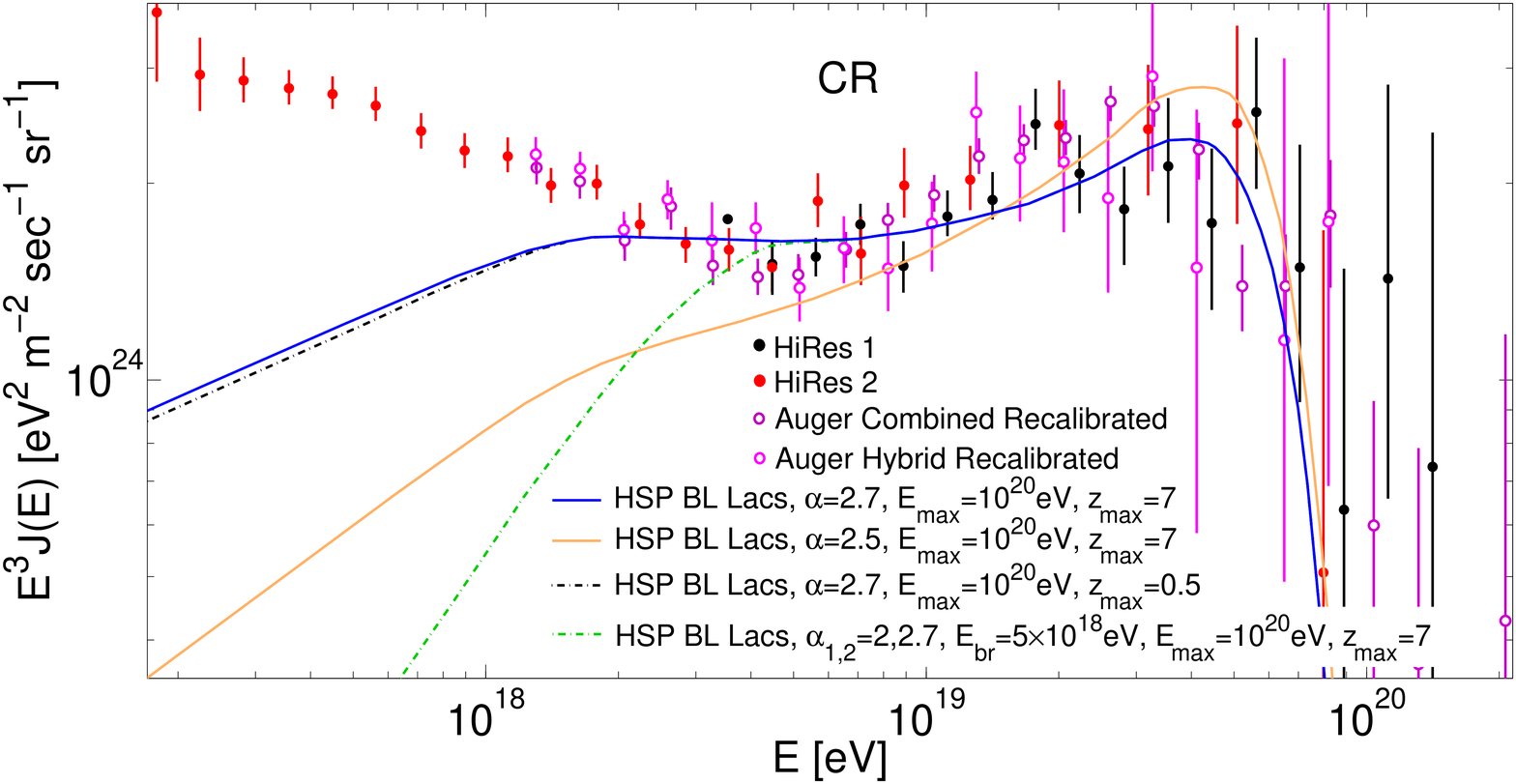}
\includegraphics[width=0.49\textwidth]{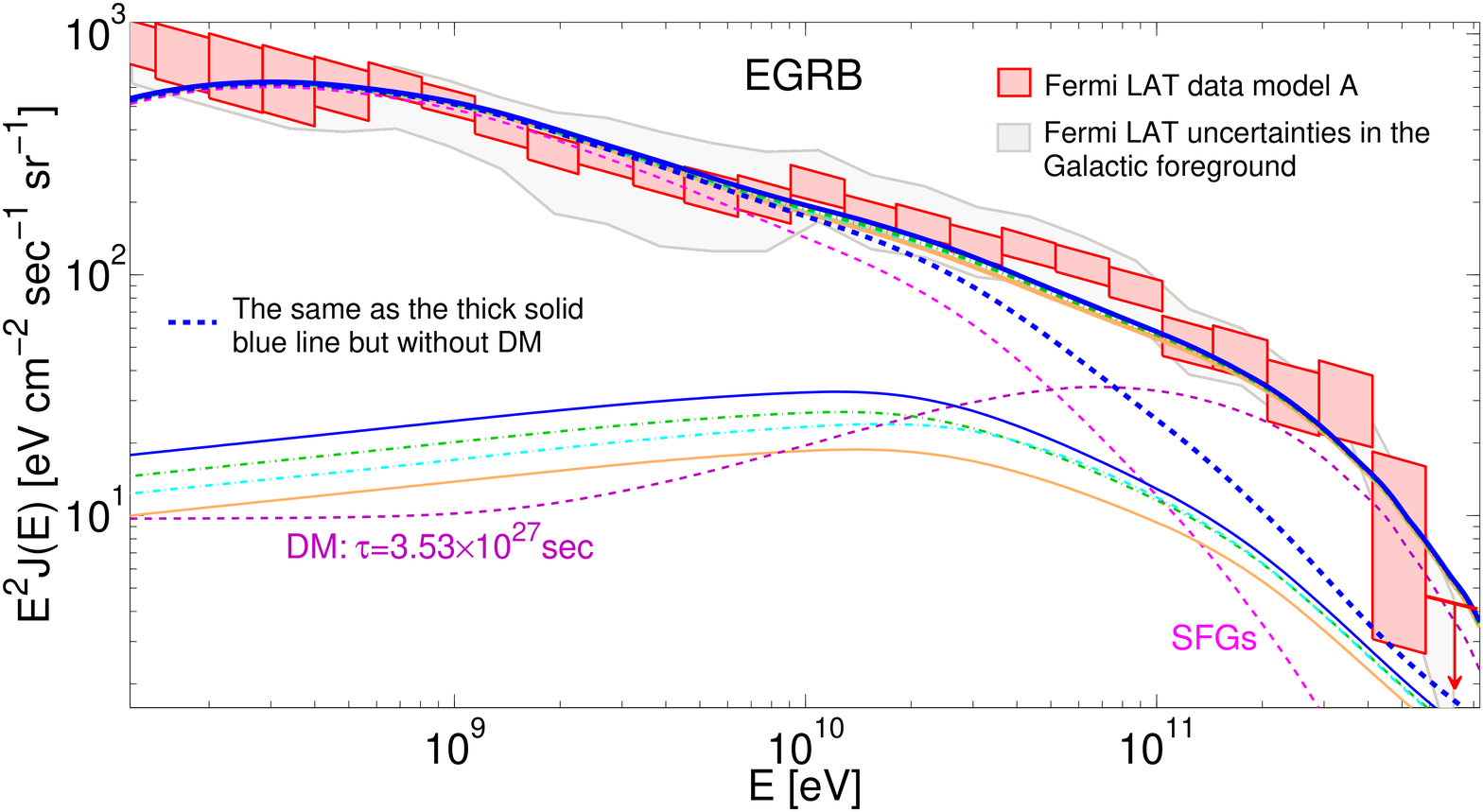}
\includegraphics[width=0.49\textwidth]{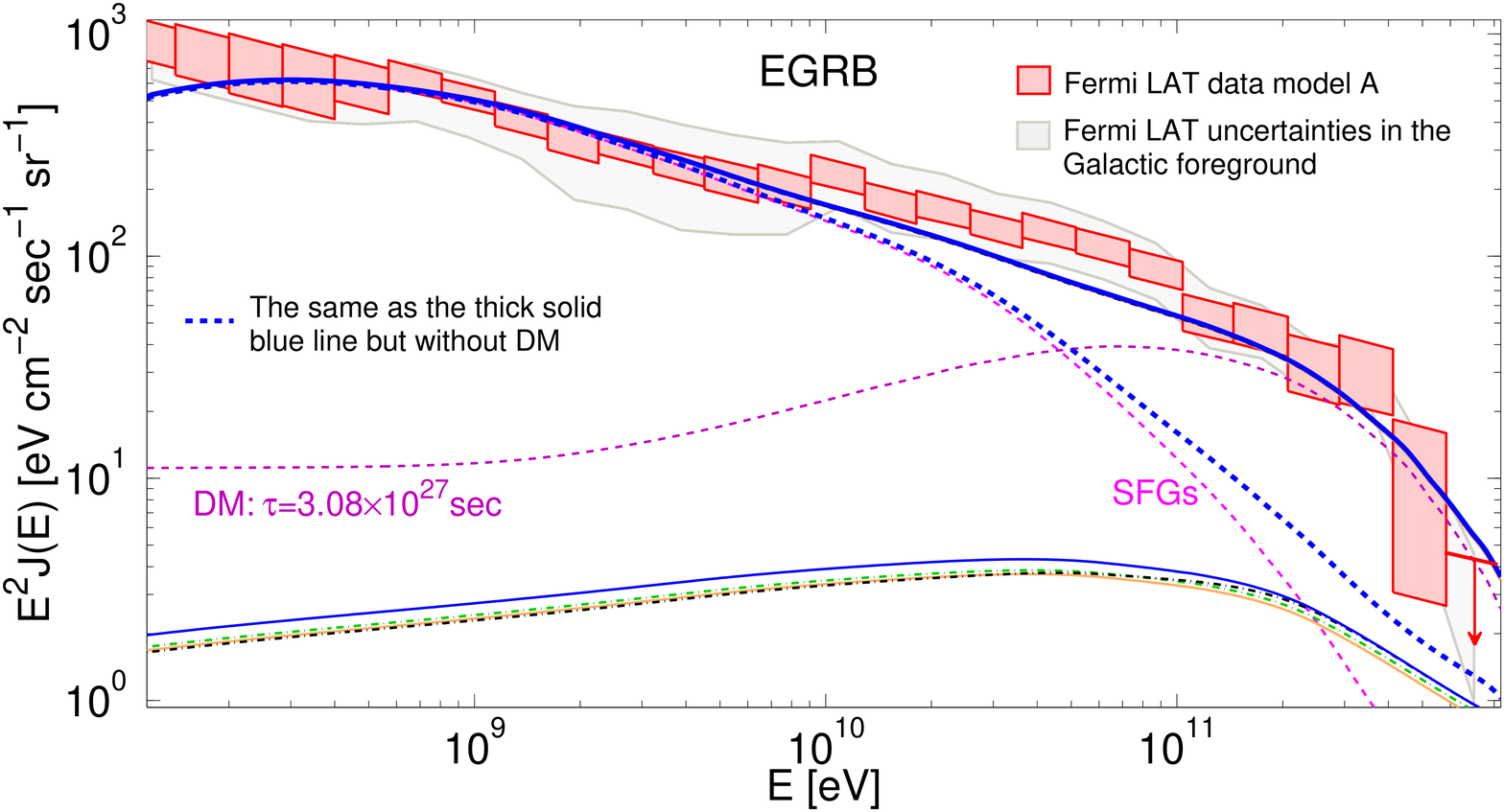}
\caption{The same as in Figure \ref{fig:SFR_HiRes},  but the UHECRs are evolving as BL Lacs. In the left panels are the non-evolving BL Lacs and in the right panels are the HSP BL Lacs which evolve as $(1+z)^m$ with a very negative $m$. The DM lifetimes here are $\tau= 3.53\times10^{27}\mbox{sec}$ for the non-evolving  BL Lacs and $\tau= 3.08\times10^{27}\mbox{sec}$ for the HSP BL Lacs.}
\label{fig:BL_Lac}
\end{figure}

\begin{figure}
\centering
\includegraphics[width=0.8\textwidth]{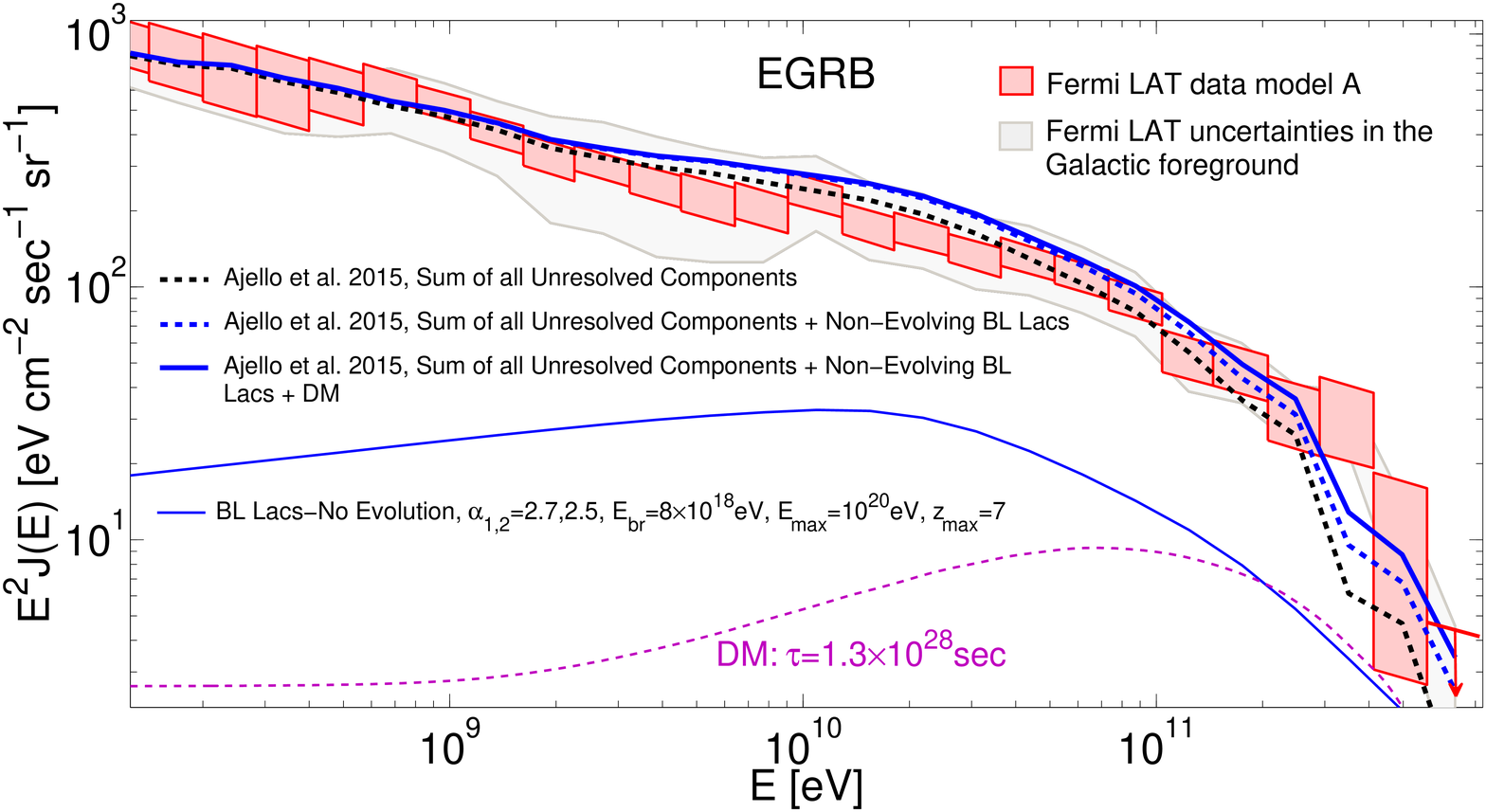}
\includegraphics[width=0.8\textwidth]{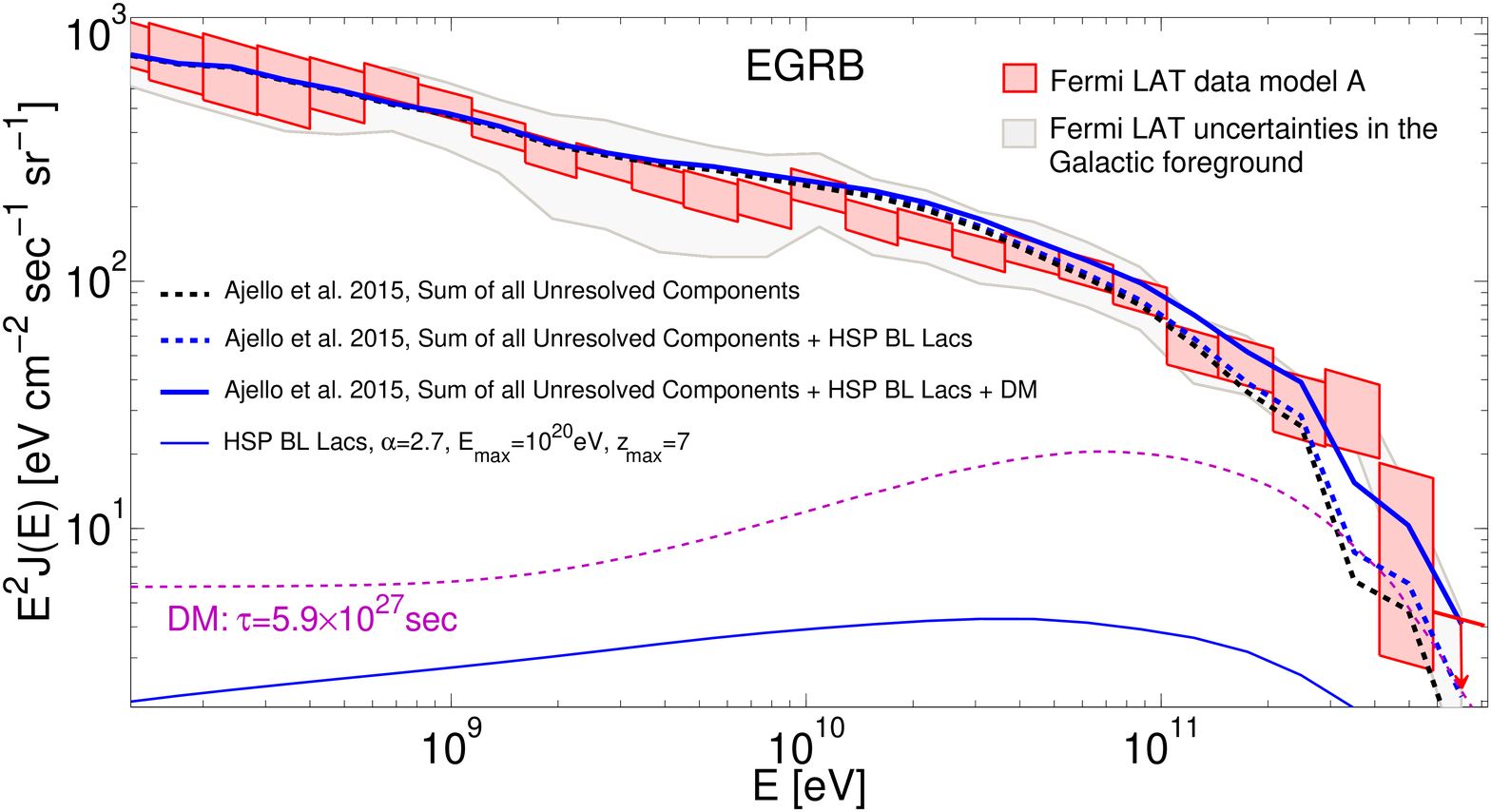}
\caption{The total flux of $\gamma$-rays originating from UHECRs evovling as BL Lacs and from blazars. \textbf{Upper panel:} Non-evolving BL Lacs. The thin blue solid line is corresponding to $\gamma$-rays originating from UHECRs with the parameters: $\alpha_{1,2}=2.7,2.5$, $E_{br}=8\times10^{18}\mbox{eV}$, $E_{max}=10^{20}\mbox{eV}$, and $z_{max}=7$. The thick dashed black line is the sum of all unresolved components (blazars, SFGs, and radio galaxies) in \citet{2015ApJ...800L..27A}, Figure 3 (resolved point sources have been removed). The thick dashed blue line is the sum of the dashed black line and the thin solid blue line. The thick solid blue line is the sum of the dashed black line, the thin solid blue line, and $3\mbox{TeV}$ $W^+W^-$ decay DM with $\tau=1.3\times10^{28}\mbox{eV}$. \textbf{Lower panel:} The same as in the upper panel, but for HSP BL Lacs with $\alpha=2.7$, $E_{max}=10^{20}\mbox{eV}$, and $z_{max}=7$. And for DM with $\tau=5.9\times10^{27}\mbox{sec}$.}
\label{fig:BL_Lac_Blazars}
\end{figure}

In figures \ref{fig:SFR_HiRes}-\ref{fig:BL_Lac_Blazars}, we present our results. Blazars are included only in figures \ref{fig:blazars} and \ref{fig:BL_Lac_Blazars}. In figures \ref{fig:SFR_HiRes}, \ref{fig:SFR_Auger}, \ref{fig:GRB_HiRes}, \ref{fig:GRB_Auger}, \ref{fig:MLLAGN}, \ref{fig:MLLAGN_Auger}, \ref{fig:AGN_HiRes}, and \ref{fig:BL_Lac} we show UHECR spectra for different parameters, the corresponding $\gamma$-ray fluxes (thin lines), and the total flux from the three sources (thick lines): SFGs, UHECRs, and DM decay. In these figures, the dashed-dotted curves differ from the solid blue line in one parameter, in order to show the sensitivity of the spectra to the chosen parameters. We show cases where the UHECR spectra are normalized to the HiRes data as well as cases where the spectra are normalized to the Auger data. The Auger data have a statistical uncertainty of $22\%$ \citep{2010PhRvL.104i1101A, 2014NIMPA.742...22B}. In the plots where we present the HiRes data, we also present the Auger data with energy increased by $16\%$. The recalibrated Auger data and the HiRes data agree well.

In Figure \ref{fig:SFR_HiRes} we show fits to the Fermi LAT data with UHECRs that are adjusted to the HiRes data. The evolution model is SFR, except for one curve that is corresponding to the MHLAGNs model. The DM in this plot is assumed to have a lifetime of $\tau=4.61\times10^{27}\mbox{sec}$. In the upper panels of this plot we show the UHECR spectra and in the lower panels we show the corresponding $\gamma$-ray fluxes (the same lines as in the upper panels), with the SFGs (magenta dashed line) and DM (violet dashed line) contributions, and the sum of the three components (thick lines).   In the left panels we show the sensitivity of the spectra to the chosen parameters, as the dashed-dotted lines differ from the blue line in one parameter. The thick dashed blue line in the lower left panel is the same as the thick solid blue line but without the DM contribution.  In addition, in the lower right panel of this figure we show how our estimate for the SFGs spectrum is compared to the spectrum in other works. The thin dotted black line and the thick dotted black line are SFGs spectra calculated by  \citet{2014JCAP...09..043T} and by \citet{2014ApJ...786...40L} respectively. Our line is in a good agreement with both these spectra.

The maximum $\gamma$-ray contribution from UHECRs evolving as SFR is obtained at redshift $z\sim 7$, and in fact the contribution to the $\gamma$-ray flux from sources with redshifts above $z=4$ is only a few percent of the total flux. Above $z\sim7$, the contribution is negligible.

In Figure \ref{fig:EGRB_DM_limits} we show the total contribution from SFGs, UHECRs, and DM for different DM lifetimes. The blue line in this figure is the same as the blue line in Figure   \ref{fig:SFR_HiRes}. As can be seen from the figure, for this set of UHECR parameters, the shortest DM lifetime allowed in order to respect the bounds set by the Fermi LAT data is $\tau=3.75\times 10^{27}\mbox{sec}$. DM with shorter lifetime (such as $2.5\times10^{27}\mbox{sec}$ presented in the figure) will cause a violation of the limits imposed by the data.

In the upper panel of Figure \ref{fig:shading} we show the uncertainties related to Figure \ref{fig:SFR_HiRes} for SFR evolution, with a maximum redshift of $z_{max}=7$ and a maximum acceleration energy of $E_{max}=10^{20}\mbox{eV}$. The blue band in Figure \ref{fig:shading} is the possible range of $\gamma$-ray fluxes related to these parameters. Its lower limit is the thin solid orange curve in the lower right panel of Figure \ref{fig:SFR_HiRes} and its upper limit is the thin solid blue curve in the lower left panel of Figure \ref{fig:SFR_HiRes}. The violet band represents DM contribution for lifetime values in the range $3.8\mbox{-}5.5\times10^{27}\mbox{sec}$. Magenta dashed line is the SFGs contribution. The green band is the total of the three components: UHECRs, DM, and SFGs. In order to show the importance of the DM contribution to the fit at high energies, we also show the yellow band, which is the sum of the contributions from UHECRs and SFGs (without DM). As can be seen, the high energy part of Fermi LAT data cannot be fitted with only UHECRs and SFGs.

In the lower panel of Figure \ref{fig:shading} we show a fit to the Fermi LAT data, using DM decay in the $\mu^+\mu^-$ channel, with mass $mc^2=30\mbox{TeV}$ and a lifetime of $\tau=1.1\times10^{28}\mbox{sec}$. The $\mbox{DM}\rightarrow\mu^+\mu^-$ spectrum was taken from \citet{2012JCAP...10..043M} and the lifetime was adjusted (\citet{2012JCAP...10..043M} used $\tau=2\times10^{27}\mbox{sec}$). The thin dashed green line in this figure is the $\mbox{DM}\rightarrow\mu^+\mu^-$ contribution. The thick dashed green line is the sum of $\mbox{DM}\rightarrow\mu^+\mu^-$, SFGs, and UHECRs (with parameters written in the plot). We compare this fit to the one with the contribution from the $W^+W^-$ channel (thick violet dashed line). As can be seen, The $W^+W^-$ channel gives a much better fit at high energies. Using the $\mu^+\mu^-$ DM, we cannot fit  all the high energy data points while keeping the $\gamma$-ray flux below the upper limit of the highest energy data point.

In Figure \ref{fig:SFR_Auger}, we normalize the SFR curves to the unrecalibrated Auger data. The DM lifetime in this fit is $\tau=3.87\times10^{27}\mbox{sec}$. The lowest DM lifetime possible in this case is $3.06\times 10^{27}\mbox{sec}$. A lower lifetime will give a too high flux. The $\gamma$-ray fluxes here are lower than in Figure \ref{fig:SFR_HiRes}, since the Auger data have a lower flux than the HiRes data. This is why we need a higher flux at high energies, from DM decay, in order to fit the Fermi LAT data. Here, as in  Figure \ref{fig:SFR_HiRes}, the dashed-dotted lines differ from the solid blue line in one parameter and the thick lines are the sum of the three contributions (UHECRs, SFGs, abd DM). Note that in the left panels, the line corresponding to a maximum acceleration energy of $E_{max}=10^{21}\mbox{eV}$ (dashed-dotted green) and the line corresponding to the spectral index of $\alpha_{1,2}=2.5,2.2$ (dashed-dotted dark red) have almost identical $\gamma$-ray spectra. This is why the total of the $\alpha_{1,2}=2.5,2.2$ line, SFGs, and DM cannot be seen behind the thick dashed-dotted green line.

In Figure \ref{fig:blazars} we show the total contribution of $\gamma$-rays from blazars and from UHECRs evolving as SFR. The unresolved blazar spectrum as extrapolated by \citet{2015ApJ...800L..27A} from the resolved blazar data is represented in Figure \ref{fig:blazars} by the thin black dashed line. We consider the maximum and minimum $\gamma$-ray fluxes from UHECRs that are adjusted to HiRes data and from UHECRs that are adjusted to Auger data. The maximum redshift is $z_{max}=7$ and the maximum acceleration energy is assumed to be $E_{max}=10^{20}\mbox{eV}$. In all cases, the total flux of $\gamma$-rays from blazars and from UHECRs evolving as SFR is too high. In the HiRes case, the sum of blazars and the maximum contribution from UHECRs is already violating the limits imposed by the Fermi LAT data. The total with the minimum contribution from UHECRs is reaching the edge of the Fermi uncertainties, leaving no room for high or low energy components. In the case of Auger data, the totals are lower, but still too high and it is very unlikely that a fit to the entire data will be possible without exceeding the boundaries. Also in this figure we show the sum of all unresolved components (blazars, SFGs, and radio galaxies) in \citet{2015ApJ...800L..27A}, Figure 3 (resolved point sources have been removed), with an additional contribution from $3\mbox{TeV}$ $W^+W^-$ decay DM with lifetime of $4.61\times10^{27}\mbox{sec}$. For $3\mbox{TeV}$ $W^+W^-$ DM, this value of lifetime is the minimum possible to add to \citet{2015ApJ...800L..27A} model, in order to respect the limits set by Fermi LAT data.

In Figure \ref{fig:GRB_HiRes} we do the same as in Figure \ref{fig:SFR_HiRes}, but for the GRB evolution model. The lifetime of the DM is $\tau=5\times10^{27}\mbox{sec}$. As opposed to the SFR cases, in the GRB model, the $\gamma$-rays are violating the limits imposed by the Fermi LAT data, unless we cut off the maximum redshift. This violation can be seen in Figure \ref{fig:GRB_HiRes} in sources with maximum redshift of $z_{max}=4$. The $\gamma$-ray spectrum from UHECRs with a spectral index of $\alpha_{1,2}=2.34,2.22$, break energy $E_{br}=8\times10^{18}\mbox{eV}$, maximum acceleration energy $E_{max}=10^{20}\mbox{eV}$, and maximum redshift $z_{max}=4$ (thin dashed-dotted orange line) is exceeding the boundaries set by Fermi LAT even without the contributions from SFGs and DM. The thin solid black line, corresponding to the parameters $\alpha_{1,2}=2,2.35$, $E_{br}=8\times10^{18}\mbox{eV}$, $E_{max}=10^{20}\mbox{eV}$, and $z_{max}=4$ is not violating the limits imposed by the data, but the sum of it and the contributions from SFGs and DM is violating them. The main difference between these two lines is in the energy of transition from Galactic to extragalactic CRs. While in the orange line case, the transition is below the second Knee at $\sim 0.3\mbox{EeV}$, the transition in the case of the black line is at the Ankle at $\sim 5\mbox{EeV}$. For two UHECR spectra with the same $E_{max}$ and $z_{max}$, a lower energy of transition means a higher flux of $\gamma$-rays. The $\gamma$-ray energy density corresponding to the orange curve is $37\%$ higher than the energy density corresponding to the black curve.

 In Figure \ref{fig:GRB_Auger}, the UHECR spectra are normalized to the unrecalibrated Auger data. The DM contribution has a lifetime of $\tau=4\times10^{27}\mbox{sec}$.  The evolution model is GRB. Here, as in Figure \ref{fig:GRB_HiRes}, the maximum redshift of the UHECR sources has to be cut off in order to respect the boundaries set by Fermi LAT data.

In figures \ref{fig:MLLAGN} and \ref{fig:MLLAGN_Auger}, the UHECR sources are assumed to be MLLAGNs. In Figure \ref{fig:MLLAGN}, the spectra are normalized to the HiRes data and in Figure \ref{fig:MLLAGN_Auger} they are normalized to the Auger data. The MLLAGN sources give almost the same $\gamma$-ray spectra as the SFR and there is no need in a cutoff in the redshift. As in the SFR case,  the $\gamma$-ray flux from both UHECRs and blazars is too high to fit the Fermi LAT data. In Figure \ref{fig:MLLAGN} the DM lifetime is $4.29\times 10^{27}\mbox{sec}$. In Figure \ref{fig:MLLAGN_Auger} the DM lifetimes is $3.75\times 10^{27}\mbox{sec}$ and it can be as low as $2.86\times 10^{27}\mbox{sec}$ without violating the limits imposed by the data.

In Figure \ref{fig:AGN_HiRes}, the curves are corresponding to the MHLAGNs model, except for one curve which is corresponding to the GRBs model. The spectra are normalized to the HiRes data. In the MHLAGNs case, the cut off in redshift needs to be very low if we do not want to violate the limits imposed by the data. As can be seen in the figure, the limits are violated even for a $z_{max}=1.5$ spectrum. In the case of HLAGNs as UHECR sources, the $\gamma$-ray flux is higher and the required cutoff in the redshift is even lower. It is unlikely then, that HLAGNs would be the sources of UHECRs, unless they are in the nearby universe.

In Figure \ref{fig:BL_Lac}, the UHECRs are adjusted to the HiRes data and their assumed sources are non-evolving BL Lacs (left panels) and HSP BL Lacs (right panels). The HSP BL Lacs evolve as $(1+z)^m$ with a very negative $m$ and thus have both a) a very low secondary $\gamma$-ray contribution and b) a relatively high cutoff energy.   It can be seen from the figure that a fit to the Fermi LAT data is marginal with $\gamma$-rays from UHECRs evolving as HSP, $\gamma$-rays from SFGs, and $\gamma$-rays from DM. The total flux at $(4-9)\times10^{10}\mbox{eV}$ may be slightly too low while the total flux at high energies is the highest possible in order to respect the bounds set by Fermi LAT. The non-evolving BL Lacs give a higher $\gamma$-ray flux than the HSPs. Some sets of UHECR parameters (such as $\alpha_{1,2}=2.7,2.5$, $E_{br}=8\times10^{18}\mbox{eV}$, $E_{max}=10^{20}\mbox{eV}$, and $z_{max}=7$) provide better fits to the Fermi LAT data than other sets of parameters (such as $\alpha_{1,2}=2,2.4$, $E_{br}=8\times10^{18}\mbox{eV}$, $E_{max}=10^{20}\mbox{eV}$, and $z_{max}=7$).

In Figure \ref{fig:BL_Lac_Blazars} we show the total $\gamma$-ray contribution from UHECRs evolving as BL Lacs (non-evolving BL Lacs in the upper panel and HSP BL Lacs in the lower panel), the sum of all unresolved components (blazars, SFGs, and radio galaxies), as extrapolated by \citet{2015ApJ...800L..27A} from the resolved blazar data, and DM.  As opposed to the other evolution models (see Figure \ref{fig:blazars}), there is enough room for $\gamma$-rays originating both from blazars and from UHECRs that evolve as BL Lacs. The UHECR parameters in this figure are $\alpha_{1,2}=2.7,2.5$, $E_{br}=8\times10^{18}\mbox{eV}$, $E_{max}=10^{20}\mbox{eV}$, $z_{max}=7$ (upper panel) and $\alpha=2.7$, $E_{max}=10^{20}\mbox{eV}$, $z_{max}=7$ (lower panel). The unresolved part of the sum of all components in \citet{2015ApJ...800L..27A} is represented by the thick dashed black line, the sum of it and the $\gamma$-rays from UHECRs is represented by the thick dashed blue line, and the sum of it and the $\gamma$-rays from UHECRs and from DM is represented by the thick solid blue line. The DM lifetimes are $1.3\times10^{28}\mbox{sec}$ (non-evolving BL Lacs) and $5.9\times10^{27}\mbox{sec}$ (HSP BL Lacs). It can be seen from the figure that the DM contribution improves the fit at high energies because we consider here only astrophysical sources that are too soft. However, blazars come closest to fitting the Fermi LAT data without DM.

\section{Conclusions}
GRB, AGN, and star formation were all more common in the past  $z\gtrsim 1$, with a comoving density now varying as $(1+z)^m$ with $m\gtrsim 3$.
Had UHECR sources been active at $z\sim 1$, photons from these backgrounds would pair produce with these UHECRs and the pairs would ultimately make secondary $\gamma$-radiation.
$\gamma$-rays originating as primaries from SFGs and as secondaries from UHECRs, with an additional high energy contribution (e.g from DM decay), can provide a good fit to the EGRB measured by Fermi LAT. We found that between the evolution models: SFR, GRBs, MLLAGNs, MHLAGNs, HLAGNs, and BL Lacs, the preferable ones for UHECR sources are SFR, MLLAGNs, and BL Lacs. $\gamma$-rays from UHECRs, whose sources evolve in time as SFR, as MLLAGNs, or as BL Lacs, do not violate the bounds set by Fermi LAT measurements. A hypothetical class of UHECR sources that evolve as SFR or as MLLAGNs, is in fact found to be quite robust and most choices of free parameters (provided that the spectral index of the UHECRs was below $\sim2.5$) gave good fits to both the EGRB and the  UHECR spectra. This is consistent with findings of other authors \citep{2015MNRAS.451..751G,2015PhRvD..92b1302G}. The secondary $\gamma$-rays from these UHECRs are softer than the diffuse  high energy $\gamma$-ray background observed by the Fermi LAT,  and these evolutionary models all give a better fit if a contribution from decaying DM particles with masses of $\sim 3 $ TeV are included.

In the case of BL Lacs whose comoving density does not decline or even increases with time (i.e. with decreasing $z$), good fits to the Fermi LAT data could be achieved with a decaying DM contribution. But, the contribution of secondary $\gamma$-rays from these UHECRs is not a major contribution, so other astrophysical theories for the origin of the diffuse EGRB are not preempted by the hypothesis that  UHECRs come from  extragalactic sources with a non-declining comoving density. The DM contribution here appears much more significant than in the other (stronger) evolution scenarios, because the other major contributor, SFG, are assumed to give a softer spectrum than secondary $\gamma$-rays from  UHECRs. However, the possibility that the entire EGRB can be explained by ultimately resolvable blazars has already been suggested, and, as these blazars become better resolved, the lower limits on the lifetime of TeV DM particles will rise. The power in BL Lac objects is about $8\cdot10^{37}\mbox{erg} \ \mbox{sec}^{-1} \ \mbox{Mpc}^{-3}$ \citep{2014ApJ...780...73A}, as compared to $1.5\cdot 10^{36}$, $ 1.7\cdot 10^{37}$, and $6.9\cdot 10^{37}$ for HLAGNs, MHLAGNs, and MLLAGNs respectively \citep{2005A&A...441..417H}, so it is not surprising that BL Lac objects would dominate UHECR production at present. However, other types of AGN were more active in the past,  and their past contribution to the EGRB may have competed with that of BL Lac objects. One might then wonder whether the EGRB imposes a limit on their being as efficient in producing UHECRs. Conceivably, there  could be a physical reason why UHECR production becomes more efficient with cosmic time, e.g. because galactic magnetic fields grow and $E_{max}$ therefore  increases.

In this work we assumed a pure proton composition. A mixed composition produces a somewhat lower EGRB, but a good fit to the data requires a very hard injection spectrum with a spectral index of $\alpha \sim 1\mbox{-}1.6$ \citep{2014JCAP...10..020A}. Such a hard spectrum, might not be achieved within the standard acceleration mechanisms that predict a steeper spectrum, with $\alpha \geq 2$. Thus, the UHECR spectrum, to its highest energies, may have a large proton component.

\begin{figure}
\centering
\includegraphics[width=0.9\textwidth]{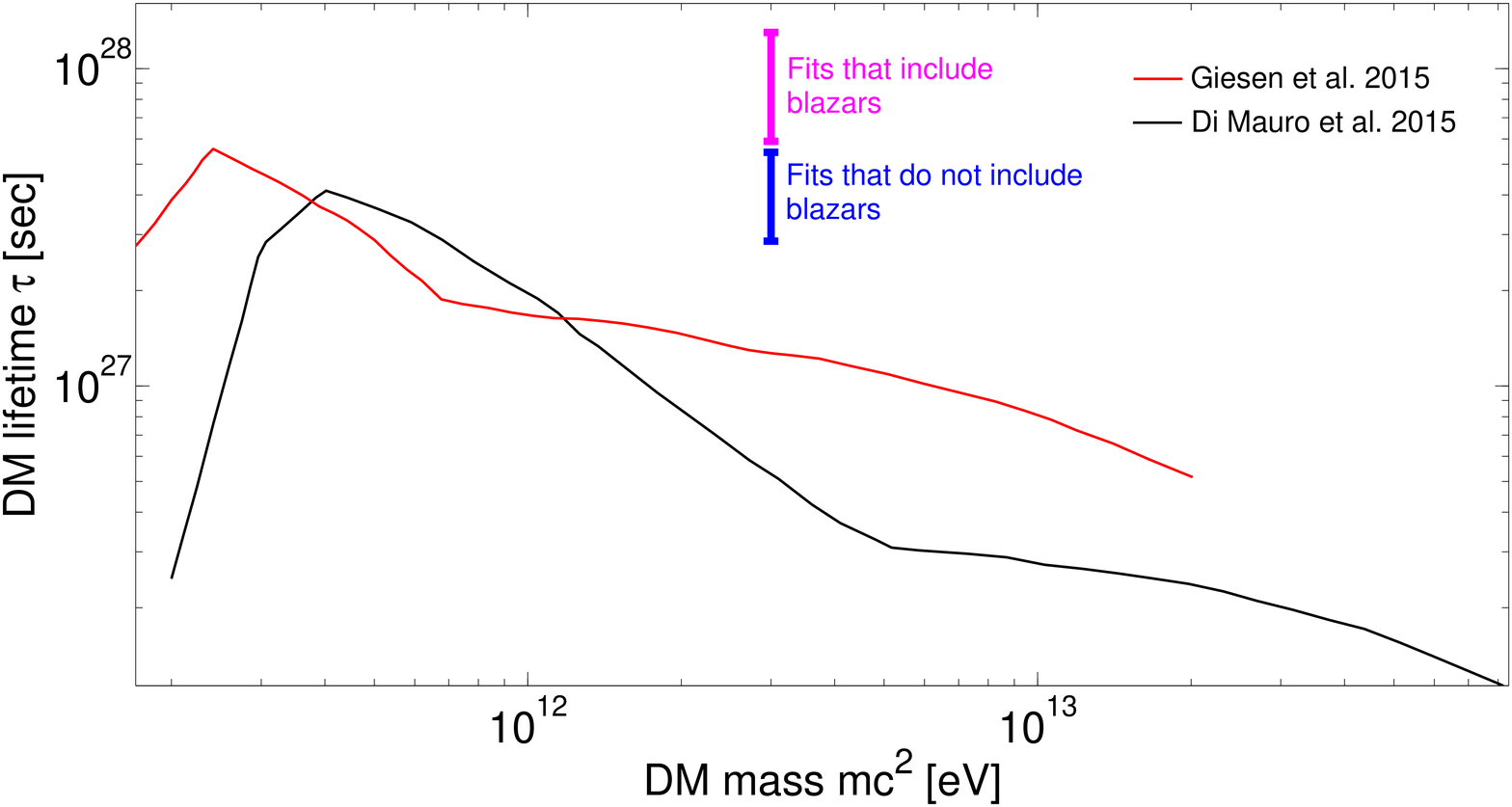}
\caption{Lower limits on the DM lifetimes for a $DM \rightarrow W^+W^-$ decay as a function of the DM mass as obtained by \citet{2015arXiv150707001D} (black line) and by \citet{2015JCAP...09..023G} (red line). The vertical blue line is the possible values of lifetimes for the fits that do not include blazars. The vertical magenta line is the DM lifetimes for the fits that include blazars.}
\label{fig:DM_constraints}
\end{figure}

In fitting the Fermi LAT data, we made use in $\gamma$-rays from DM of mass $mc^2=3\mbox{TeV}$, decaying into $W^+W^-$. The lifetimes we used in the fits that did not include blazars were $\tau=3.08\mbox{-}5.45\times10^{27}\mbox{sec}$. In the fits where blazars were included (Figure \ref{fig:BL_Lac_Blazars}), longer lifetimes were needed: $\tau=5.9\times10^{27}\mbox{sec}$ and $\tau=1.3\times10^{28}\mbox{sec}$. The shortest lifetime possible (and still provide a good fit) in this work  was $2.86\times 10^{27}\mbox{sec}$ and it was obtained in the MLLAGNs model for UHECRs normalized to the Auger data. A lower limit of $\sim6.2\times10^{26}\mbox{sec}$ was found recently by \citet{2015arXiv150707001D} for DM of the same properties. \citet{2015arXiv150707001D} assumed that DM decay produces a $e^+e^-$ flux that contributes to the AMS-02 experiment data \citep{2014PhRvL.113l1102A, 2014PhRvL.113v1102A, 2014PhRvL.113l1101A}. A more strict lower limit on the DM lifetime was obtained by \citet{2015JCAP...09..023G} by fitting the AMS-02 data of anti-proton to proton ratio. The lower limit that was obtained (for DM of the same parametes as we used in this work) by \citet{2015JCAP...09..023G} is $\tau\sim1.3\times10^{27}\mbox{sec}$. In this work we obtained somewhat stricter lower limits on the DM lifetimes, especially in the fits that included blazars (Figure \ref{fig:BL_Lac_Blazars}). In Figure \ref{fig:DM_constraints} we show the lower limits on the DM lifetimes for the $W^+W^-$ channel obtained by  \citet{2015arXiv150707001D} and \citet{2015JCAP...09..023G}, and the range of limfetimes possible for the different models in this work.

If the blazar contribution is as high as is claimed \citep[e.g][]{2015ApJ...800L..27A,2015MNRAS.450.2404G}, then it would not leave  enough room for producing extragalactic UHECRs with AGN-like, GRB-like or even SFR-like source evolution.  The diffuse gamma ray background that is  expected for these evolutionary scenarios to accompany the UHECR production, when added to the blazar contribution, sticks out above the Fermi LAT measurements of the diffuse EGRB. %On the other hand UHECR production is a very recent phenomenon and did not operate  at medium to high redshifts  despite the existence of the same class of sources back then, e.g. because the galactic magnetic fields were still growing. 
By contrast, UHECRs that evolve in time as BL Lacs, produce low enough $\gamma$-ray flux that have enough room to fit the Fermi LAT data together with blazars. The addition of a high energy component from DM decay improves the fit even here, but the DM plays less of a role, so the improvement should not be taken as a strong evidence for the existence of the decaying DM component.

Another, unforced possibility, is that the UHECRs, even at the highest energy, are Galactic \citep{2016GalacticUHECRs}. Here the energy requirements are greatly reduced and Galactic GRBs could easily provide enough energy \citep{1993ApJ...418..386L, 2011ApJ...738L..21E}.

\section*{Acknowledgements}
This work was supported by the Joan and Robert Arnow Chair of Theoretical Astrophysics, the Israel-U.S. Binational Science Foundation, and the Israel Science Foundation, including an ISF-UGC grant. We thank Dr. Noemi Globus for useful discussions.

\bibliography{UHECR}{}

\begin{thebibliography}{}
\expandafter\ifx\csname natexlab\endcsname\relax\def\natexlab#1{#1}\fi

\bibitem[{{Abbasi} {et~al.}(2008){Abbasi}, {Abu-Zayyad}, {Allen}, {Amman},
  {Archbold}, {Belov}, {Belz}, {Ben Zvi}, {Bergman}, {Blake}, {Brusova},
  {Burt}, {Cannon}, {Cao}, {Connolly}, {Deng}, {Fedorova}, {Finley}, {Gray},
  {Hanlon}, {Hoffman}, {Holzscheiter}, {Hughes}, {H{\"u}ntemeyer}, {Jones},
  {Jui}, {Kim}, {Kirn}, {Loh}, {Maestas}, {Manago}, {Marek}, {Martens},
  {Matthews}, {Matthews}, {Moore}, {O'Neill}, {Painter}, {Perera}, {Reil},
  {Riehle}, {Roberts}, {Rodriguez}, {Sasaki}, {Schnetzer}, {Scott}, {Sinnis},
  {Smith}, {Sokolsky}, {Song}, {Springer}, {Stokes}, {Thomas}, {Thomas},
  {Thomson}, {Tupa}, {Westerhoff}, {Wiencke}, {Zhang}, \&
  {Zech}}]{2008PhRvL.100j1101A}
{Abbasi}, R.~U., {Abu-Zayyad}, T., {Allen}, M., {et~al.} 2008, Physical Review
  Letters, 100, 101101

\bibitem[{{Abbasi} {et~al.}(2010){Abbasi}, {Abu-Zayyad}, {Al-Seady}, {Allen},
  {Amman}, {Anderson}, {Archbold}, {Belov}, {Belz}, {Bergman}, {Blake},
  {Brusova}, {Burt}, {Cannon}, {Cao}, {Deng}, {Fedorova}, {Finley}, {Gray},
  {Hanlon}, {Hoffman}, {Holzscheiter}, {Hughes}, {H{\"u}ntemeyer}, {Jones},
  {Jui}, {Kim}, {Kirn}, {Loh}, {Liu}, {Lundquist}, {Maestas}, {Manago},
  {Marek}, {Martens}, {Matthews}, {Matthews}, {Moore}, {O'Neill}, {Painter},
  {Perera}, {Reil}, {Riehle}, {Roberts}, {Rodriguez}, {Sasaki}, {Schnetzer},
  {Scott}, {Sinnis}, {Smith}, {Sokolsky}, {Song}, {Springer}, {Stokes},
  {Stratton}, {Thomas}, {Thomas}, {Thomson}, {Tupa}, {Zech}, \&
  {Zhang}}]{2010PhRvL.104p1101A}
{Abbasi}, R.~U., {Abu-Zayyad}, T., {Al-Seady}, M., {et~al.} 2010, Physical
  Review Letters, 104, 161101

\bibitem[{{Abdo} {et~al.}(2010){Abdo}, {Ackermann}, {Ajello}, {Allafort},
  {Antolini}, {Atwood}, {Axelsson}, {Baldini}, {Ballet}, {Barbiellini},
  {Bastieri}, {Baughman}, {Bechtol}, {Bellazzini}, {Berenji}, {Blandford},
  {Bloom}, {Bogart}, {Bonamente}, {Borgland}, {Bouvier}, {Bregeon}, {Brez},
  {Brigida}, {Bruel}, {Buehler}, {Burnett}, {Buson}, {Caliandro}, {Cameron},
  {Cannon}, {Caraveo}, {Carrigan}, {Casandjian}, {Cavazzuti}, {Cecchi}, {{\c
  C}elik}, {Celotti}, {Charles}, {Chekhtman}, {Chen}, {Cheung}, {Chiang},
  {Ciprini}, {Claus}, {Cohen-Tanugi}, {Conrad}, {Costamante}, {Cotter},
  {Cutini}, {D'Elia}, {Dermer}, {de Angelis}, {de Palma}, {De Rosa}, {Digel},
  {Silva}, {Drell}, {Dubois}, {Dumora}, {Escande}, {Farnier}, {Favuzzi},
  {Fegan}, {Ferrara}, {Focke}, {Fortin}, {Frailis}, {Fukazawa}, {Funk},
  {Fusco}, {Gargano}, {Gasparrini}, {Gehrels}, {Germani}, {Giebels},
  {Giglietto}, {Giommi}, {Giordano}, {Giroletti}, {Glanzman}, {Godfrey},
  {Grandi}, {Grenier}, {Grondin}, {Grove}, {Guiriec}, {Hadasch}, {Harding},
  {Hayashida}, {Hays}, {Healey}, {Hill}, {Horan}, {Hughes}, {Iafrate}, {Itoh},
  {J{\'o}hannesson}, {Johnson}, {Johnson}, {Johnson}, {Johnson}, {Kamae},
  {Katagiri}, {Kataoka}, {Kawai}, {Kerr}, {Kn{\"o}dlseder}, {Kuss}, {Lande},
  {Latronico}, {Lavalley}, {Lemoine-Goumard}, {Llena Garde}, {Longo},
  {Loparco}, {Lott}, {Lovellette}, {Lubrano}, {Madejski}, {Makeev}, {Malaguti},
  {Massaro}, {Mazziotta}, {McConville}, {McEnery}, {McGlynn}, {Michelson},
  {Mitthumsiri}, {Mizuno}, {Moiseev}, {Monte}, {Monzani}, {Morselli},
  {Moskalenko}, {Murgia}, {Nolan}, {Norris}, {Nuss}, {Ohno}, {Ohsugi},
  {Omodei}, {Orlando}, {Ormes}, {Ozaki}, {Paneque}, {Panetta}, {Parent},
  {Pelassa}, {Pepe}, {Pesce-Rollins}, {Piranomonte}, {Piron}, {Porter},
  {Rain{\`o}}, {Rando}, {Razzano}, {Reimer}, {Reimer}, {Reposeur}, {Ripken},
  {Ritz}, {Rodriguez}, {Romani}, {Roth}, {Ryde}, {Sadrozinski}, {Sanchez},
  {Sander}, {Saz Parkinson}, {Scargle}, {Sgr{\`o}}, {Shaw}, {Siskind}, {Smith},
  {Spandre}, {Spinelli}, {Starck}, {Stawarz}, {Strickman}, {Suson}, {Tajima},
  {Takahashi}, {Takahashi}, {Tanaka}, {Taylor}, {Thayer}, {Thayer}, {Thompson},
  {Tibaldo}, {Torres}, {Tosti}, {Tramacere}, {Ubertini}, {Uchiyama}, {Usher},
  {Vasileiou}, {Vilchez}, {Villata}, {Vitale}, {Waite}, {Wallace}, {Wang},
  {Winer}, {Wood}, {Yang}, {Ylinen}, \& {Ziegler}}]{2010ApJ...715..429A}
{Abdo}, A.~A., {Ackermann}, M., {Ajello}, M., {et~al.} 2010, \apj, 715, 429

\bibitem[{{Abraham} {et~al.}(2010){Abraham}, {Abreu}, {Aglietta}, {Ahn},
  {Allard}, {Allekotte}, {Allen}, {Alvarez-Mu{\~n}iz}, {Ambrosio},
  {Anchordoqui}, \& et~al.}]{2010PhRvL.104i1101A}
{Abraham}, J., {Abreu}, P., {Aglietta}, M., {et~al.} 2010, Physical Review
  Letters, 104, 091101

\bibitem[{{Accardo} {et~al.}(2014){Accardo}, {Aguilar}, {Aisa}, {Alvino},
  {Ambrosi}, {Andeen}, {Arruda}, {Attig}, {Azzarello}, {Bachlechner}, \&
  et~al.}]{2014PhRvL.113l1101A}
{Accardo}, L., {Aguilar}, M., {Aisa}, D., {et~al.} 2014, Physical Review
  Letters, 113, 121101

\bibitem[{{Ackermann} {et~al.}(2015){Ackermann}, {Ajello}, {Albert}, {Atwood},
  {Baldini}, {Ballet}, {Barbiellini}, {Bastieri}, {Bechtol}, {Bellazzini},
  {Bissaldi}, {Blandford}, {Bloom}, {Bottacini}, {Brandt}, {Bregeon}, {Bruel},
  {Buehler}, {Buson}, {Caliandro}, {Cameron}, {Caragiulo}, {Caraveo},
  {Cavazzuti}, {Cecchi}, {Charles}, {Chekhtman}, {Chiang}, {Chiaro}, {Ciprini},
  {Claus}, {Cohen-Tanugi}, {Conrad}, {Cuoco}, {Cutini}, {D'Ammando}, {de
  Angelis}, {de Palma}, {Dermer}, {Digel}, {Silva}, {Drell}, {Favuzzi},
  {Ferrara}, {Focke}, {Franckowiak}, {Fukazawa}, {Funk}, {Fusco}, {Gargano},
  {Gasparrini}, {Germani}, {Giglietto}, {Giommi}, {Giordano}, {Giroletti},
  {Godfrey}, {Gomez-Vargas}, {Grenier}, {Guiriec}, {Gustafsson}, {Hadasch},
  {Hayashi}, {Hays}, {Hewitt}, {Ippoliti}, {Jogler}, {J{\'o}hannesson},
  {Johnson}, {Johnson}, {Kamae}, {Kataoka}, {Kn{\"o}dlseder}, {Kuss},
  {Larsson}, {Latronico}, {Li}, {Li}, {Longo}, {Loparco}, {Lott}, {Lovellette},
  {Lubrano}, {Madejski}, {Manfreda}, {Massaro}, {Mayer}, {Mazziotta},
  {McEnery}, {Michelson}, {Mitthumsiri}, {Mizuno}, {Moiseev}, {Monzani},
  {Morselli}, {Moskalenko}, {Murgia}, {Nemmen}, {Nuss}, {Ohsugi}, {Omodei},
  {Orlando}, {Ormes}, {Paneque}, {Panetta}, {Perkins}, {Pesce-Rollins},
  {Piron}, {Pivato}, {Porter}, {Rain{\`o}}, {Rando}, {Razzano}, {Razzaque},
  {Reimer}, {Reimer}, {Reposeur}, {Ritz}, {Romani}, {S{\'a}nchez-Conde},
  {Schaal}, {Schulz}, {Sgr{\`o}}, {Siskind}, {Spandre}, {Spinelli}, {Strong},
  {Suson}, {Takahashi}, {Thayer}, {Thayer}, {Tibaldo}, {Tinivella}, {Torres},
  {Tosti}, {Troja}, {Uchiyama}, {Vianello}, {Werner}, {Winer}, {Wood}, {Wood},
  {Zaharijas}, \& {Zimmer}}]{2015ApJ...799...86A}
{Ackermann}, M., {Ajello}, M., {Albert}, A., {et~al.} 2015, \apj, 799, 86

\bibitem[{{Aguilar} {et~al.}(2014{\natexlab{a}}){Aguilar}, {Aisa}, {Alvino},
  {Ambrosi}, {Andeen}, {Arruda}, {Attig}, {Azzarello}, {Bachlechner}, {Barao},
  \& et~al.}]{2014PhRvL.113l1102A}
{Aguilar}, M., {Aisa}, D., {Alvino}, A., {et~al.} 2014{\natexlab{a}}, Physical
  Review Letters, 113, 121102

\bibitem[{{Aguilar} {et~al.}(2014{\natexlab{b}}){Aguilar}, {Aisa}, {Alpat},
  {Alvino}, {Ambrosi}, {Andeen}, {Arruda}, {Attig}, {Azzarello}, {Bachlechner},
  \& et~al.}]{2014PhRvL.113v1102A}
{Aguilar}, M., {Aisa}, D., {Alpat}, B., {et~al.} 2014{\natexlab{b}}, Physical
  Review Letters, 113, 221102

\bibitem[{{Ajello} {et~al.}(2014){Ajello}, {Romani}, {Gasparrini}, {Shaw},
  {Bolmer}, {Cotter}, {Finke}, {Greiner}, {Healey}, {King}, {Max-Moerbeck},
  {Michelson}, {Potter}, {Rau}, {Readhead}, {Richards}, \&
  {Schady}}]{2014ApJ...780...73A}
{Ajello}, M., {Romani}, R.~W., {Gasparrini}, D., {et~al.} 2014, \apj, 780, 73

\bibitem[{{Ajello} {et~al.}(2015){Ajello}, {Gasparrini}, {S{\'a}nchez-Conde},
  {Zaharijas}, {Gustafsson}, {Cohen-Tanugi}, {Dermer}, {Inoue}, {Hartmann},
  {Ackermann}, {Bechtol}, {Franckowiak}, {Reimer}, {Romani}, \&
  {Strong}}]{2015ApJ...800L..27A}
{Ajello}, M., {Gasparrini}, D., {S{\'a}nchez-Conde}, M., {et~al.} 2015, \apjl,
  800, L27

\bibitem[{{Allard} {et~al.}(2007){Allard}, {Parizot}, \&
  {Olinto}}]{2007APh....27...61A}
{Allard}, D., {Parizot}, E., \& {Olinto}, A.~V. 2007, Astroparticle Physics,
  27, 61

\bibitem[{{Aloisio} {et~al.}(2014){Aloisio}, {Berezinsky}, \&
  {Blasi}}]{2014JCAP...10..020A}
{Aloisio}, R., {Berezinsky}, V., \& {Blasi}, P. 2014, \jcap, 10, 20

\bibitem[{{Aloisio} {et~al.}(2007){Aloisio}, {Berezinsky}, {Blasi}, {Gazizov},
  {Grigorieva}, \& {Hnatyk}}]{2007APh....27...76A}
{Aloisio}, R., {Berezinsky}, V., {Blasi}, P., {et~al.} 2007, Astroparticle
  Physics, 27, 76

\bibitem[{{Aloisio} {et~al.}(2012){Aloisio}, {Berezinsky}, \&
  {Gazizov}}]{2012APh....39..129A}
{Aloisio}, R., {Berezinsky}, V., \& {Gazizov}, A. 2012, Astroparticle Physics,
  39, 129

\bibitem[{{Aloisio} {et~al.}(2013{\natexlab{a}}){Aloisio}, {Berezinsky}, \&
  {Grigorieva}}]{2013APh....41...73A}
{Aloisio}, R., {Berezinsky}, V., \& {Grigorieva}, S. 2013{\natexlab{a}},
  Astroparticle Physics, 41, 73

\bibitem[{{Aloisio} {et~al.}(2013{\natexlab{b}}){Aloisio}, {Berezinsky}, \&
  {Grigorieva}}]{2013APh....41...94A}
---. 2013{\natexlab{b}}, Astroparticle Physics, 41, 94

\bibitem[{{Berezhko} {et~al.}(2012){Berezhko}, {Knurenko}, \&
  {Ksenofontov}}]{2012APh....36...31B}
{Berezhko}, E.~G., {Knurenko}, S.~P., \& {Ksenofontov}, L.~T. 2012,
  Astroparticle Physics, 36, 31

\bibitem[{{Berezinskii} \& {Smirnov}(1975)}]{1975Ap&SS..32..461B}
{Berezinskii}, V.~S., \& {Smirnov}, A.~I. 1975, \apss, 32, 461

\bibitem[{{Berezinsky}(2014)}]{2014APh....53..120B}
{Berezinsky}, V. 2014, Astroparticle Physics, 53, 120

\bibitem[{{Berezinsky} {et~al.}(2006){Berezinsky}, {Gazizov}, \&
  {Grigorieva}}]{2006PhRvD..74d3005B}
{Berezinsky}, V., {Gazizov}, A., \& {Grigorieva}, S. 2006, \prd, 74, 043005

\bibitem[{{Bhattacharjee}(2000)}]{2000PhR...327..109B}
{Bhattacharjee}, P. 2000, \physrep, 327, 109

\bibitem[{{Biermann}(1997)}]{1997JPhG...23....1B}
{Biermann}, P.~L. 1997, Journal of Physics G Nuclear Physics, 23, 1

\bibitem[{{Bird} {et~al.}(1993){Bird}, {Corbat{\'o}}, {Dai}, {Dawson},
  {Elbert}, {Gaisser}, {Green}, {Huang}, {Kieda}, {Ko}, {Larsen}, {Loh}, {Luo},
  {Salamon}, {Smith}, {Sokolsky}, {Sommers}, {Stanev}, {Tang}, {Thomas}, \&
  {Tilav}}]{1993PhRvL..71.3401B}
{Bird}, D.~J., {Corbat{\'o}}, S.~C., {Dai}, H.~Y., {et~al.} 1993, Physical
  Review Letters, 71, 3401

\bibitem[{{Blandford} \& {Eichler}(1987)}]{1987PhR...154....1B}
{Blandford}, R., \& {Eichler}, D. 1987, \physrep, 154, 1

\bibitem[{{Blumenthal} \& {Gould}(1970)}]{1970RvMP...42..237B}
{Blumenthal}, G.~R., \& {Gould}, R.~J. 1970, Reviews of Modern Physics, 42, 237

\bibitem[{{Boncioli}(2014)}]{2014NIMPA.742...22B}
{Boncioli}, D. 2014, Nuclear Instruments and Methods in Physics Research A,
  742, 22

\bibitem[{{Caccianiga} {et~al.}(2002){Caccianiga}, {Maccacaro}, {Wolter},
  {Della Ceca}, \& {Gioia}}]{2002ApJ...566..181C}
{Caccianiga}, A., {Maccacaro}, T., {Wolter}, A., {Della Ceca}, R., \& {Gioia},
  I.~M. 2002, \apj, 566, 181

\bibitem[{{de Marco} \& {Stanev}(2005)}]{2005PhRvD..72h1301D}
{de Marco}, D., \& {Stanev}, T. 2005, \prd, 72, 081301

\bibitem[{{Decerprit} \& {Allard}(2011)}]{2011A&A...535A..66D}
{Decerprit}, G., \& {Allard}, D. 2011, \aap, 535, A66

\bibitem[{{Di Mauro} {et~al.}(2015){Di Mauro}, {Donato}, {Fornengo}, \&
  {Vittino}}]{2015arXiv150707001D}
{Di Mauro}, M., {Donato}, F., {Fornengo}, N., \& {Vittino}, A. 2015, ArXiv
  e-prints, arXiv:1507.07001

\bibitem[{{Eichler} {et~al.}(2016){Eichler}, {Globus}, {Kumar}, \&
  {Gavish}}]{2016GalacticUHECRs}
{Eichler}, D., {Globus}, N., {Kumar}, R., \& {Gavish}, E. 2016, \apj, submitted

\bibitem[{{Eichler} \& {Pohl}(2011)}]{2011ApJ...738L..21E}
{Eichler}, D., \& {Pohl}, M. 2011, \apjl, 738, L21

\bibitem[{{Engel} {et~al.}(2001){Engel}, {Seckel}, \&
  {Stanev}}]{2001PhRvD..64i3010E}
{Engel}, R., {Seckel}, D., \& {Stanev}, T. 2001, \prd, 64, 093010

\bibitem[{{Fichtel} {et~al.}(1977){Fichtel}, {Hartman}, {Kniffen}, {Thompson},
  {Ogelman}, {Ozel}, \& {Tumer}}]{1977ApJ...217L...9F}
{Fichtel}, C.~E., {Hartman}, R.~C., {Kniffen}, D.~A., {et~al.} 1977, \apjl,
  217, L9

\bibitem[{{Fichtel} {et~al.}(1978){Fichtel}, {Simpson}, \&
  {Thompson}}]{1978ApJ...222..833F}
{Fichtel}, C.~E., {Simpson}, G.~A., \& {Thompson}, D.~J. 1978, \apj, 222, 833

\bibitem[{{Giesen} {et~al.}(2015){Giesen}, {Boudaud}, {G{\'e}nolini}, {Poulin},
  {Cirelli}, {Salati}, \& {Serpico}}]{2015JCAP...09..023G}
{Giesen}, G., {Boudaud}, M., {G{\'e}nolini}, Y., {et~al.} 2015, \jcap, 9, 23

\bibitem[{{Giommi} \& {Padovani}(2015)}]{2015MNRAS.450.2404G}
{Giommi}, P., \& {Padovani}, P. 2015, \mnras, 450, 2404

\bibitem[{{Globus} {et~al.}(2015{\natexlab{a}}){Globus}, {Allard},
  {Mochkovitch}, \& {Parizot}}]{2015MNRAS.451..751G}
{Globus}, N., {Allard}, D., {Mochkovitch}, R., \& {Parizot}, E.
  2015{\natexlab{a}}, \mnras, 451, 751

\bibitem[{{Globus} {et~al.}(2015{\natexlab{b}}){Globus}, {Allard}, \&
  {Parizot}}]{2015PhRvD..92b1302G}
{Globus}, N., {Allard}, D., \& {Parizot}, E. 2015{\natexlab{b}}, \prd, 92,
  021302

\bibitem[{{Gould} \& {Schr{\'e}der}(1967)}]{1967PhRv..155.1404G}
{Gould}, R.~J., \& {Schr{\'e}der}, G.~P. 1967, Physical Review, 155, 1404

\bibitem[{{Greisen}(1966)}]{1966PhRvL..16..748G}
{Greisen}, K. 1966, Physical Review Letters, 16, 748

\bibitem[{{Hasinger} {et~al.}(2005){Hasinger}, {Miyaji}, \&
  {Schmidt}}]{2005A&A...441..417H}
{Hasinger}, G., {Miyaji}, T., \& {Schmidt}, M. 2005, \aap, 441, 417

\bibitem[{{Jauch} \& {Rohrlich}(1955)}]{1955jauch}
{Jauch}, J.~M., \& {Rohrlich}, F. 1955, {The theory of photons and electrons}
  ({Addison-Wesley Publishing Company, Inc., Reading, Massachusetts})

\bibitem[{{Jui} \& {the Telescope Array
  Collaboration}(2012)}]{2012JPhCS.404a2037J}
{Jui}, C.~C.~H., \& {the Telescope Array Collaboration}. 2012, Journal of
  Physics Conference Series, 404, 012037

\bibitem[{{Kneiske} {et~al.}(2004){Kneiske}, {Bretz}, {Mannheim}, \&
  {Hartmann}}]{2004A&A...413..807K}
{Kneiske}, T.~M., {Bretz}, T., {Mannheim}, K., \& {Hartmann}, D.~H. 2004, \aap,
  413, 807

\bibitem[{{Lacki} {et~al.}(2014){Lacki}, {Horiuchi}, \&
  {Beacom}}]{2014ApJ...786...40L}
{Lacki}, B.~C., {Horiuchi}, S., \& {Beacom}, J.~F. 2014, \apj, 786, 40

\bibitem[{{Levinson} \& {Eichler}(1993)}]{1993ApJ...418..386L}
{Levinson}, A., \& {Eichler}, D. 1993, \apj, 418, 386

\bibitem[{{Murase} \& {Beacom}(2012)}]{2012JCAP...10..043M}
{Murase}, K., \& {Beacom}, J.~F. 2012, \jcap, 10, 43

\bibitem[{{Norman} {et~al.}(1995){Norman}, {Melrose}, \&
  {Achterberg}}]{1995ApJ...454...60N}
{Norman}, C.~A., {Melrose}, D.~B., \& {Achterberg}, A. 1995, \apj, 454, 60

\bibitem[{{Padovani} {et~al.}(2007){Padovani}, {Giommi}, {Landt}, \&
  {Perlman}}]{2007ApJ...662..182P}
{Padovani}, P., {Giommi}, P., {Landt}, H., \& {Perlman}, E.~S. 2007, \apj, 662,
  182

\bibitem[{{Sreekumar} {et~al.}(1998){Sreekumar}, {Bertsch}, {Dingus},
  {Esposito}, {Fichtel}, {Hartman}, {Hunter}, {Kanbach}, {Kniffen}, {Lin},
  {Mayer-Hasselwander}, {Michelson}, {von Montigny}, {Muecke}, {Mukherjee},
  {Nolan}, {Pohl}, {Reimer}, {Schneid}, {Stacy}, {Stecker}, {Thompson}, \&
  {Willis}}]{1998ApJ...494..523S}
{Sreekumar}, P., {Bertsch}, D.~L., {Dingus}, B.~L., {et~al.} 1998, \apj, 494,
  523

\bibitem[{{Stanev} {et~al.}(1993){Stanev}, {Biermann}, \&
  {Gaisser}}]{1993A&A...274..902S}
{Stanev}, T., {Biermann}, P.~L., \& {Gaisser}, T.~K. 1993, \aap, 274, 902

\bibitem[{{Tamborra} {et~al.}(2014){Tamborra}, {Ando}, \&
  {Murase}}]{2014JCAP...09..043T}
{Tamborra}, I., {Ando}, S., \& {Murase}, K. 2014, \jcap, 9, 43

\bibitem[{{Torres} \& {Anchordoqui}(2004)}]{2004RPPh...67.1663T}
{Torres}, D.~F., \& {Anchordoqui}, L.~A. 2004, Reports on Progress in Physics,
  67, 1663

\bibitem[{{Waxman}(2000)}]{2000NuPhS..87..345W}
{Waxman}, E. 2000, Nuclear Physics B Proceedings Supplements, 87, 345

\bibitem[{{Waxman}(2004)}]{2004NJPh....6..140W}
---. 2004, New Journal of Physics, 6, 140

\bibitem[{{Wdowczyk} {et~al.}(1972){Wdowczyk}, {Tkaczyk}, \&
  {Wolfendale}}]{1972JPhA....5.1419W}
{Wdowczyk}, J., {Tkaczyk}, W., \& {Wolfendale}, A.~W. 1972, Journal of Physics
  A Mathematical General, 5, 1419

\bibitem[{{Y{\"u}ksel} \& {Kistler}(2007)}]{2007PhRvD..75h3004Y}
{Y{\"u}ksel}, H., \& {Kistler}, M.~D. 2007, \prd, 75, 083004

\bibitem[{{Y{\"u}ksel} {et~al.}(2008){Y{\"u}ksel}, {Kistler}, {Beacom}, \&
  {Hopkins}}]{2008ApJ...683L...5Y}
{Y{\"u}ksel}, H., {Kistler}, M.~D., {Beacom}, J.~F., \& {Hopkins}, A.~M. 2008,
  \apjl, 683, L5

\bibitem[{{Zatsepin} \& {Kuz'min}(1966)}]{1966JETPL...4...78Z}
{Zatsepin}, G.~T., \& {Kuz'min}, V.~A. 1966, Soviet Journal of Experimental and
  Theoretical Physics Letters, 4, 78

\end{thebibliography}
\end{document}